\RequirePackage{ifpdf}
\documentclass[11pt,hyper,letterpaper]{JHEP3}
\usepackage{amsmath,amssymb,amsfonts,cases}
\usepackage{graphicx,caption,subcaption}
\usepackage{dcolumn}
\usepackage{bm}
\usepackage{cite}
\usepackage{bbm}
\usepackage{slashed}
\usepackage{wrapfig}

\newcommand{\beq}{\begin{equation}}
\newcommand{\eeq}{\end{equation}}
\newcommand{\bea}{\begin{eqnarray}}
\newcommand{\eea}{\end{eqnarray}}
\newcommand{\bi}{\begin{itemize}}
\newcommand{\ei}{\end{itemize}}
\newcommand{\ben}{\begin{enumerate}}
\newcommand{\een}{\end{enumerate}}

\newcommand{\Txx}{\langle T^{xx}\rangle}
\newcommand{\Jt}{\langle J^t \rangle}

\renewcommand{\a}{\alpha}

\newcommand{\e}{\epsilon}
\newcommand{\G}{\Gamma}

\newcommand{\m}{\mu}

\renewcommand{\t}{\theta}

\newcommand{\sdbi}{S_{\textrm{DBI}}}
\newcommand{\seh}{S_{\textrm{EH}}}
\newcommand{\talpha}{\tilde{\alpha}}
\newcommand{\qext}{Q_{\textrm{ext}}}

\def\a{\alpha}

\def\m{\mu}

\def\o{\omega}  
\def\t{\tau}

\def\G{\Gamma}

\title{ \LARGE Holographic Zero Sound from Spacetime-Filling Branes}
\author{Nikola I. Gushterov,\!$^1$\footnotemark[1]\, Andy O'Bannon,\!$^2$\footnotemark[2]\, and Ronnie Rodgers,\!$^2$\footnotemark[3]
\\
$^1$Rudolf Peierls Centre for Theoretical Physics, University of Oxford \\ Department of Physics, Clarendon Laboratory, Oxford OX1 3NU, United Kingdom
\\
$^2$STAG Research Centre, Physics and Astronomy, University of Southampton \\ Highfield, Southampton SO17 1BJ, United Kingdom}

\footnotetext[1]{E-mail address: \email{nikola.gushterov@physics.ox.ac.uk}}
\footnotetext[2]{E-mail address: \email{a.obannon@soton.ac.uk}}
\footnotetext[3]{E-mail address: \email{r.j.rodgers@soton.ac.uk}}

\abstract{We use holography to study sound modes of strongly-interacting conformal field theories with non-zero temperature, $T$, and $U(1)$ chemical potential, $\mu$. Specifically, we consider charged black brane solutions of Einstein gravity in $(3+1)$-dimensional Anti-de Sitter space coupled to a $U(1)$ gauge field with Dirac-Born-Infeld action, representing a spacetime-filling brane. The brane action has two free parameters: the tension and the non-linearity parameter, which controls higher-order terms in the field strength. For all values of the tension, non-linearity parameter, and $T/\mu$, and at sufficiently small momentum, we find sound modes with speed given by the conformal value and attenuation constant of hydrodynamic form. In particular we find sound at arbitrarily low $T/\mu$, outside the usual hydrodynamic regime, but in the regime where a Fermi liquid exhibits Landau's ``zero'' sound. In fact, the sound attenuation constant as a function of $T/\mu$ qualitatively resembles that of a Fermi liquid, including a maximum, which in a Fermi liquid signals the collisionless to hydrodynamic crossover. We also explore regimes of the tension and non-linearity parameter where two other proposed definitions of the crossover are viable, via pole collisions in Green's functions or peak movement in the charge density spectral function.}

\keywords{AdS/CFT correspondence, Gauge/gravity correspondence, AdS/CMT}
\preprint{OUTP-18-04P}

\begin{document}

\section{Introduction}
\label{intro}

\subsection{Background and Motivation}
\label{bg}

Many systems involve strongly-interacting degrees of freedom with non-zero chemical potential, $\mu \neq 0$. Examples include neutron stars, cold atoms at unitarity, graphene, and more. Such systems can exhibit remarkable properties, such as cold atoms' extremely low ratio of shear viscosity, $\eta$, to entropy density, $s$~\cite{Adams:2012th}. However, few reliable techniques exist to derive these properties from first principles. Perturbation theory is manifestly unreliable, and when $\mu\neq 0$ the ``sign problem'' renders numerical techniques, such as quantum Monte Carlo, practically useless. As a result, the origins of such remarkable properties remain mysterious. 

The Anti-de Sitter/Conformal Field Theory (AdS/CFT) correspondence, also called gauge-gravity duality or holography, offers an alternative approach. AdS/CFT is the statement that certain strongly-interacting CFTs in $d$ spacetime dimensions are equivalent to Einstein gravity in $(d+1)$-dimensional AdS space, $AdS_{d+1}$~\cite{Maldacena:1997re,Gubser:1998bc,Witten:1998qj}. The CFTs are typically non-Abelian gauge theories in the 't Hooft large-$N$ limit~\cite{Aharony:1999ti}. The CFT stress-energy tensor, $T^{\mu\nu}$ (with $\mu,\nu=0,1,\ldots,d-1$), is dual to the metric, $g_{MN}$ (with $M,N=0,1,\ldots,d$), and a $U(1)$ current $J^{\mu}$ is dual to a $U(1)$ gauge field $A_M$. A CFT with non-zero temperature $T$ and entropy density $s\propto N^2$ is dual to a black hole with Hawking temperature $T$ and Bekenstein-Hawking entropy density $s$~\cite{Witten:1998zw}. AdS/CFT thus allows us to study strongly-interacting CFTs with non-zero $T$ and $\mu$ by studying charged black holes in AdS.

AdS/CFT cannot yet describe any real system. Nevertheless, AdS/CFT has the potential to reveal universal principles applicable to real systems. Indeed, AdS/CFT already has several success stories. For example, all rotationally-invariant holographic fluids have the \textit{same} value of $\eta/s$, namely $\eta/s = 1/(4\pi) \approx 0.08$~\cite{Policastro:2001yc,Kovtun:2003wp,Buchel:2003tz,Kovtun:2004de,Starinets:2008fb}, which is surprisingly close to the $\eta/s$ estimated for cold atoms and the quark-gluon plasma~\cite{Adams:2012th}. In other words, AdS/CFT revealed that strongly-interacting fluids have characteristically small $\eta/s \sim 0.1$. AdS/CFT has also revealed universality in second-order transport~\cite{Haack:2008xx,Shaverin:2012kv,Grozdanov:2014kva,Kleinert:2016nav,Grozdanov:2016fkt}, anomalies in transport~\cite{Erdmenger:2008rm,Banerjee:2008th,Son:2009tf}, and more. 

In particular, evidence has accumulated for the possible universality of \textit{sound modes} in holographic compressible quantum matter~\cite{Karch:2008fa,Kulaxizi:2008kv,Kulaxizi:2008jx,Kim:2008bv,Karch:2009zz,Kaminski:2009dh,Edalati:2010pn,HoyosBadajoz:2010kd,Nickel:2010pr,Lee:2010ez,Bergman:2011rf,Ammon:2011hz,Davison:2011ek,Davison:2011uk,Jokela:2012vn,Goykhman:2012vy,Brattan:2012nb,Jokela:2012se,Pang:2013ypa,Dey:2013vja,Edalati:2013tma,Brattan:2013wya,Davison:2013uha,DiNunno:2014bxa,Jokela:2015aha,Itsios:2016ffv,Jokela:2016nsv,Hartnoll:2016apf,Roychowdhury:2017oed,Chen:2017dsy}. ``Compressible'' means the charge density $\langle J^t \rangle \neq 0$ is a smooth function of $\mu \neq 0$ with $d\langle J^t \rangle /d\mu \neq 0$, and ``quantum'' means $T=0$, so that quantum, rather than thermal, effects determine the ground state~\cite{Sachdev:2011wg}. ``Sound modes'' means poles in the longitudinal channel of $T^{\mu\nu}$ and/or $J^{\mu}$'s retarded two-point functions with dispersion relation $\omega(k) = \pm v k +\ldots$, with frequency $\omega$, momentum $k$, speed $v$, and $\ldots$ stands for terms with higher powers of $k$, and where $\textrm{Im}\left(\omega\right)$ determines the mode's attenuation.

To be more specific, $T=0$ sound modes have been found in two classes of holographic models. In both classes the bulk action includes an Einstein-Hilbert term,
\beq
\label{seh}
\seh = \frac{1}{16\pi G} \int d^{d+1}x \, \sqrt{-g} \left ( R + \frac{d(d-1)}{L_0^2}\right),
\eeq
with Newton's constant $G$, $g = \textrm{det}\left(g_{MN}\right)$, Ricci scalar $R$ of $g_{MN}$, and $AdS_{d+1}$ radius $L_0$. The two classes of models differ in $A_M$'s dynamics. The first class is ``probe brane'' models~\cite{Karch:2008fa,Kulaxizi:2008kv,Kulaxizi:2008jx,Kim:2008bv,Karch:2009zz,Kaminski:2009dh,HoyosBadajoz:2010kd,Nickel:2010pr,Lee:2010ez,Bergman:2011rf,Ammon:2011hz,Davison:2011ek,Jokela:2012vn,Goykhman:2012vy,Brattan:2012nb,Jokela:2012se,Pang:2013ypa,Dey:2013vja,Edalati:2013tma,Brattan:2013wya,DiNunno:2014bxa,Jokela:2015aha,Itsios:2016ffv,Jokela:2016nsv,Hartnoll:2016apf,Roychowdhury:2017oed,Chen:2017dsy}, in which $A_M$ has a Dirac-Born-Infeld (DBI) action,
\beq
\label{sdbi}
\sdbi = - T_D \int d^{d+1}x \, \sqrt{-\textrm{det}\left(g_{MN}+ \alpha \, F_{MN}\right)},
\eeq
with tension $T_D$, constant $\alpha$ of dimension $\left(\textrm{length}\right)^2$, and $F_{MN}=\partial_M A_N - \partial_N A_M$. These models employ the probe limit: expand solutions for $g_{MN}$ and $A_M$ in $G T_D \ll 1$ to leading non-trivial order. In the probe limit, $A_M$'s stress-energy tensor is neglected in Einstein's equation, and $A_M$'s equation of motion reduces to that in the ``unperturbed'' background $g_{MN}$. We will consider only spacetime-filling branes~\cite{Tarrio:2013tta}, \textit{i.e.} the integral in eq.~\eqref{sdbi} is over all $(d+1)$ bulk dimensions, although defect branes, of non-zero co-dimension, can also give rise to $T=0$ sound modes~\cite{Karch:2008fa,Karch:2009zz}.

In field theory terms, the probe limit is justified when the charged fields comprise a negligibly small fraction of the total degrees of freedom. For example, in string theory a D-brane action includes a DBI term~\cite{Polchinski:1998rr}. In holography, a D-brane that reaches the $AdS_{d+1}$ boundary is typically dual to ``flavor fields,'' meaning fields in the gauge group's fundamental representation, just like quarks in Quantum Chromodynamics (QCD)~\cite{Karch:2002sh}. The $U(1)$ is then a flavor symmetry, analogous to QCD's quark number symmetry. In such cases, typically $T_D \propto N$ whereas $G \propto 1/N^2$, so that $G T_D \propto 1/N \ll 1$. In other words, the order $N^2$ adjoint fields (gluons) vastly outnumber the order $N$ flavor fields (quarks), which are thus negligible.

The second class of models is Einstein-Maxwell theory~\cite{Edalati:2010pn,Davison:2011uk}, possibly coupled to an uncharged scalar ``dilaton'' field~\cite{Davison:2013uha}, with no probe limit, \textit{i.e.} $A_M$'s stress-energy tensor is not neglected in Einstein's equation. The gauge field thus back-reacts on the metric, hence we will also call these models ``back-reacted.'' In field theory terms, in back-reacted models the charged fields comprise a non-negligible fraction of the total number of degrees of freedom. Moreover, a DBI action truncated at second order in $\alpha F_{MN}$ is a Maxwell action. From that perspective, using a Maxwell action means discarding certain all-orders corrections in $\alpha$.

In both classes of models, sound modes appear in extremal solutions where $A_M$'s only non-zero component is $A_t$, and both $g_{MN}$ and $A_t$ depend only on the holographic radial coordinate. For example, in Einstein-Maxwell theory sound modes appear in the extremal $AdS_{d+1}$-Reissner-Nordstr\"om (AdS-RN) charged black brane solution~\cite{Edalati:2010pn,Davison:2011uk}.

The physical origin of these sound modes in holographic compressible quantum matter is mysterious. To see why, consider the three most familiar forms of compressible quantum matter, each characterized by symmetry breaking, and each supporting a sound mode~\cite{Landau,PhysRevLett.17.74,Pines,LP1,LP2,Negele,Sachdev:2011wg}. In solids, translational symmetry breaking produces a phonon. In Bose-Einstein condensates, spontaneous breaking of the particle number $U(1)$ produces a superfluid phonon. In a Landau Fermi liquid (LFL), no symmetries are necessarily broken, but fluctuations of the Fermi surface's shape produce Landau's ``zero sound'' excitation~\cite{Landau,PhysRevLett.17.74,Pines,LP1,LP2,Negele}, a longitudinal excitation with a dispersion relation of the form of a hydrodynamic sound mode, $\omega = \pm v k - i \Gamma k^2 + \ldots$, with attenuation constant $\Gamma$ and $\ldots$ representing powers of $k$ greater than $k^2$.

In holographic compressible quantum matter the sound modes appear in states that preserve the translational and $U(1)$ symmetries, hence they cannot be (superfluid) phonons. Moreover, they almost certainly cannot be zero sound either, because the effective theories describing holographic quantum compressible matter differ dramatically from LFL theory.

In LFL theory, the ground state is a degenerate system of interacting fermionic quasi-particles, producing a Fermi surface, and fluctuations about the ground state are either quasi-particles/holes or collective excitations, such as zero sound. In contrast, probe brane models show no sign of a Fermi surface~\cite{Karch:2008fa,Kulaxizi:2008kv,Kulaxizi:2008jx,Kim:2008bv,Karch:2009zz,Kaminski:2009dh,HoyosBadajoz:2010kd,Nickel:2010pr,Lee:2010ez,Bergman:2011rf,Ammon:2011hz,Davison:2011ek,Jokela:2012vn,Goykhman:2012vy,Brattan:2012nb,Jokela:2012se,Pang:2013ypa,Dey:2013vja,Edalati:2013tma,Brattan:2013wya,DiNunno:2014bxa,Jokela:2015aha,Itsios:2016ffv,Jokela:2016nsv,Hartnoll:2016apf,Roychowdhury:2017oed,Chen:2017dsy}, although they do exhibit spectral weight at $\omega=0$ for $k$ up to some finite value, similar to a smeared Fermi-Dirac distribution~\cite{Anantua:2012nj}. The equations of their effective description have the same form as hydrodynamics with weak momentum relaxation, but with momentum replaced by $\langle J^t \rangle$~\cite{Chen:2017dsy,Davison:2014lua}.

Einstein-Maxwell models can have a Fermi surface~\cite{Liu:2009dm,Cubrovic:2009ye,Faulkner:2009wj}, but violate Luttinger's theorem: the Fermi surface volume is smaller than $\langle J^t \rangle$ by powers of $N$~\cite{Edalati:2010pn,Davison:2011uk,Davison:2013uha,Hartnoll:2016apf}. In these models, the effective description remains mysterious, primarily because extremal AdS-RN has a near-horizon $AdS_2$, indicating some $(0+1)$-dimensional CFT among the light modes~\cite{Faulkner:2009wj}. Indeed, correlators of $T^{\mu\nu}$ and $J^{\mu}$ exhibit branch cuts due to these light modes, in addition to the sound modes~\cite{Edalati:2010hk,Edalati:2010pn}. The effective description is thus neither LFL theory nor hydrodynamics, but rather some kind of ``semi-local quantum liquid''~\cite{Iqbal:2011in} wherein space divides into ``patches'' of size $\ell\propto1/\mu$, such that correlators at separations $<\ell$ exhibit $(0+1)$-dimensional scale invariance, and at separations $>\ell$ exhibit exponential decay.

Although the sound modes in holographic compressible quantum matter are almost certainly not LFL zero sound, following convention we will call them ``holographic zero sound'' (HZS)~\cite{Karch:2008fa,Karch:2009zz}, where ``zero sound'' is chosen mainly because they are not phonons,\footnote{HZS can however be interpreted as a Goldstone boson arising from the breaking of an abstract symmetry, namely two $U(1)$'s at different values of the $AdS_{d+1}$ radial coordinate broken to the diagonal~\cite{Nickel:2010pr}.} while ``holographic'' emphasizes that they are probably not LFL zero sound.

\begin{figure}[t!]
\begin{center}
\includegraphics[width=0.6\textwidth]{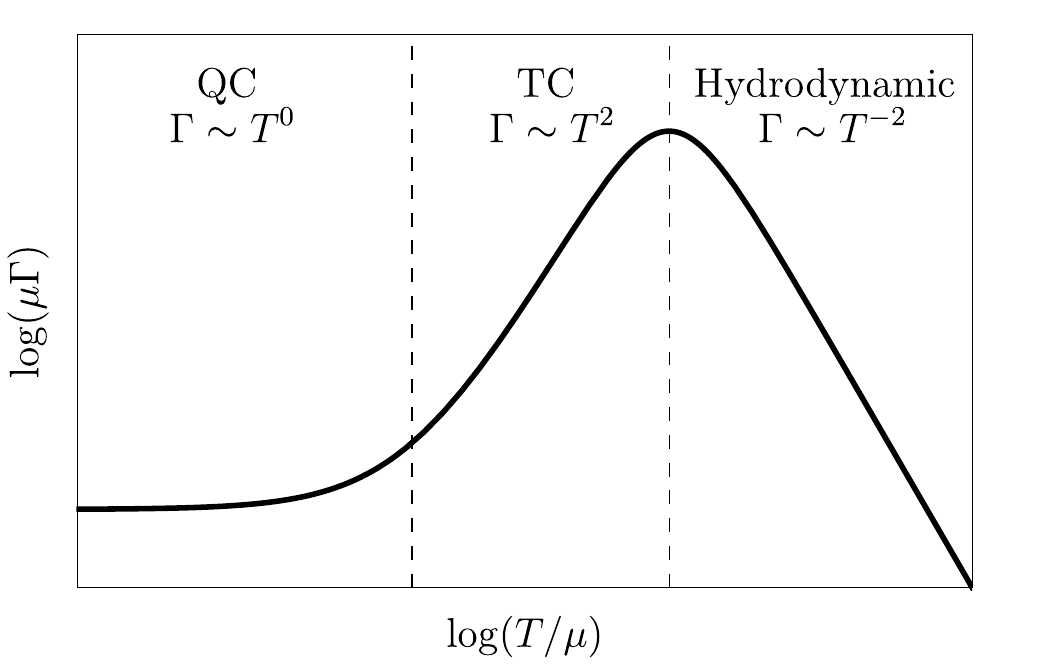}
\caption{\label{fig:attenuation_cartoon} Schematic depiction of the LFL theory form of $\ln\left(\mu\Gamma\right)$, with sound attenuation constant $\Gamma$, as a function of $\ln\left(T/\mu\right)$ at fixed frequency $\omega$ and momentum $k$. The two vertical dashed black lines represent $\pi T /\mu = \omega/\mu$ (left) and $\sqrt{\omega/\mu}$ (right). In the quantum collisionless (QC) regime $\Gamma \propto T^0$, in the thermal collisionless (TC) regime $\Gamma \propto T^2$, and in the hydrodynamic regime $\Gamma \propto T^{-2}$. A maximum appears between the thermal collisionless and hydrodynamic regimes, signaling the collisionless-to-hydrodynamic crossover.}
\end{center}
\end{figure}

Remarkably, however, in probe models the fate of HZS when $T>0$ is strikingly similar to that of LFL zero sound~\cite{Davison:2011ek,Davison:2011uk}. LFL theory is an expansion in $\omega$ about the Fermi energy~\cite{Landau,PhysRevLett.17.74,Pines,LP1,LP2,Negele}, so the LFL zero sound dispersion relation is typically expressed as $k(\omega)$, with real-valued $\omega$ and complex-valued $k$. When $T/\mu=0$, $|\textrm{Im}(k)| \propto \omega^2/\mu$ at leading order in $\omega$. As $T/\mu$ increases with $\mu$ and $\omega$ fixed, LFL theory predicts a three-stage ``collisionless-to-hydrodynamic'' crossover, characterized by changes to $\textrm{Im}(k)$ due to collisions with thermally-excited quasi-particles. Fig.~\ref{fig:attenuation_cartoon} is a schematic depiction of the crossover. The LFL prediction for the crossover has been confirmed experimentally in liquid Helium 3~\cite{PhysRevLett.17.74}. 

First, in the ``quantum collisionless'' regime, $0 \leq \pi T/\mu < \omega/\mu$, the collisions are too weak and infrequent to change zero sound's dispersion from the $T/\mu=0$ form, that is, $|\textrm{Im}(k)| \propto \omega^2/\mu$ persists. Second, in the ``thermal collisionless'' regime, $\omega/\mu < \pi T/\mu < \sqrt{\omega/\mu}$, the collisions become sufficiently strong and frequent that $|\textrm{Im}(k)|$ increases at a rate $\propto \left(\pi T\right)^2/\mu$. Third, in the ``hydrodynamic'' regime, the collisions are so strong and frequent as to destroy zero sound, however the thermal excitations now support the usual hydrodynamic (``first'') sound mode, whose attenuation decreases at a rate $\propto \mu \,\omega^2/T^2$. The transition from thermal collisionless scaling, $|\textrm{Im}(k)|\propto T^2$, to hydrodynamic scaling, $|\textrm{Im}(k)|\propto T^{-2}$, is thus marked by a maximum of $\textrm{Im}(k)$, which provides a definition for a precise moment (value of $T/\mu$) of crossover from collisionless to hydrodynamic regimes. For more details on the collisionless-to-hydrodynamic crossover in LFLs, see for example refs.~\cite{Davison:2011ek,Davison:2011uk,Pines,Negele}.

\begin{figure}[t!]
	\begin{center}
		\begin{subfigure}{0.49\textwidth}
			\includegraphics[width=\textwidth]{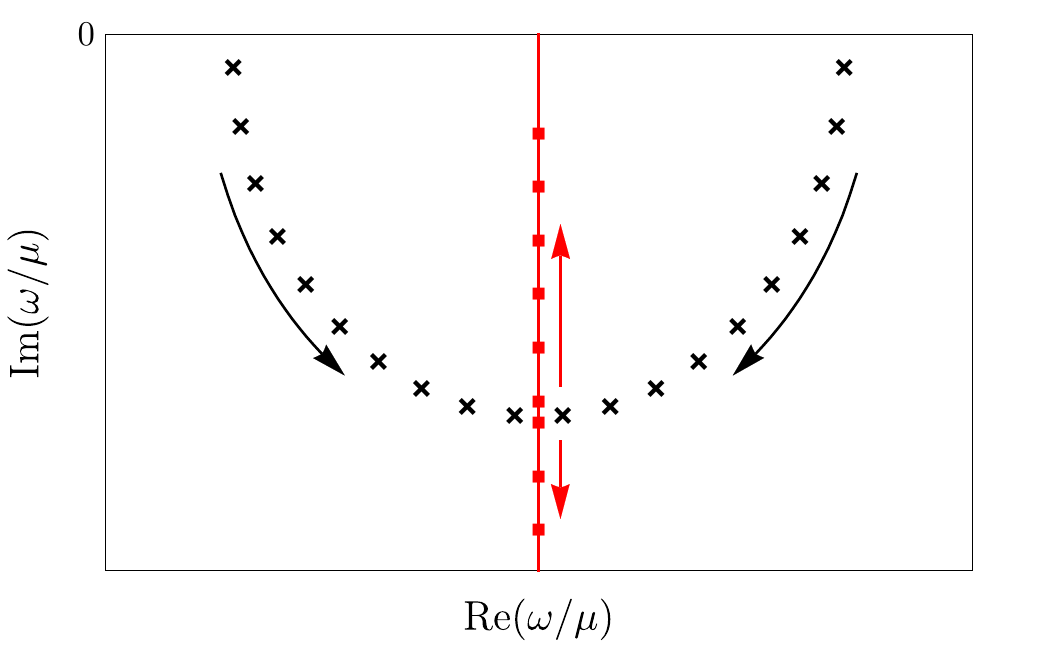}
			\caption{\label{fig:probe_cartoon} Probe brane models}
		\end{subfigure}
		\begin{subfigure}{0.49\textwidth}
			\includegraphics[width=\textwidth]{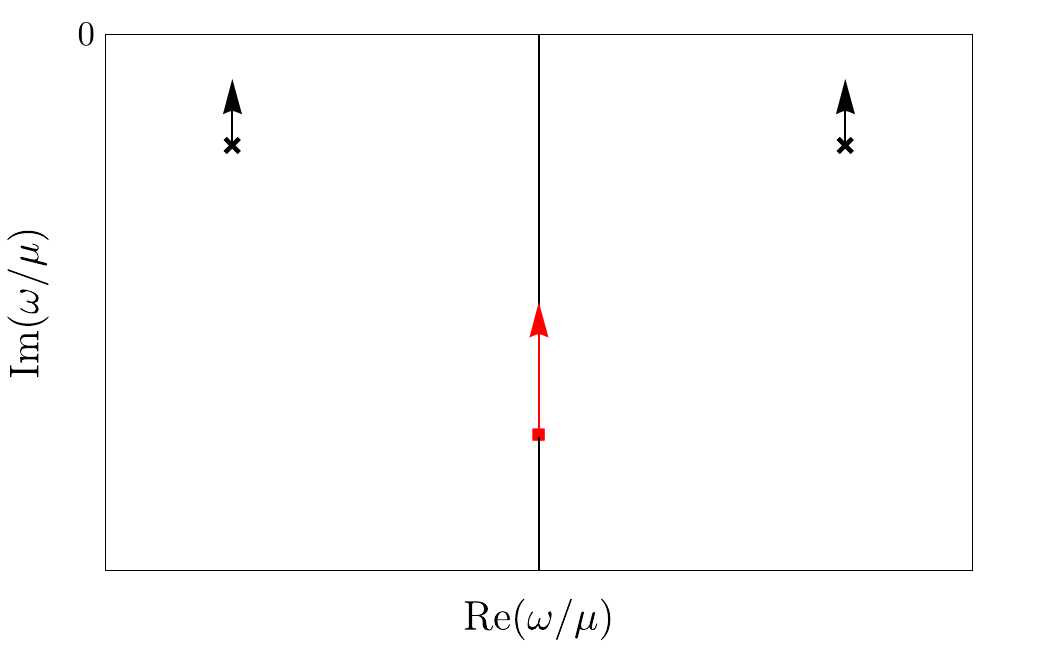}
			\caption{\label{fig:rn_cartoon} Einstein-Maxwell models}
		\end{subfigure}
		\caption{\label{fig:cartoons} Schematic depictions of the movement of poles in the complex $\omega/\mu$ plane in sound-channel $J^{\mu}$ and/or $T^{\mu\nu}$ correlators in HZS models. The arrows indicate the motion of poles as $T/\mu$ increases. The black crosses represent sound poles, while the (upper) red square represents the charge diffusion pole. (a) In spacetime-filling probe brane models, the two HZS poles move down, approximately tracing semi-circles, collide on the imaginary axis, and split into two purely imaginary poles, one moving up and one moving down, where the former is the charge diffusion pole. (b) In Einstein-Maxwell models, all three poles simply move up towards the real axis, with the sound poles' real parts constant.}
	\end{center}
\end{figure}

In probe brane models the HZS attenuation behaves identically to LFL zero sound in the quantum and thermal collisionless regimes~\cite{Davison:2011ek}. However, in the probe limit the HZS pole appears only in correlators of $J^{\mu}$, and not those of $T^{\mu\nu}$, so when $T/\mu > \sqrt{\omega/\mu}$, HZS crosses over to charge diffusion, not hydrodynamic sound: returning to complex-valued $\omega$ and real-valued $k$, the dispersion becomes $\omega = - i D k^2 + \ldots$, with charge diffusion constant $D$. As a result, the sound attenuation exhibits no maximum. Nevertheless, a precise moment of crossover can be defined from the pole movement in the complex $\omega/\mu$ plane as $T/\mu$ increases with fixed $k$ and $\mu$~\cite{Davison:2011ek}, as depicted schematically in fig.~\ref{fig:probe_cartoon}. This pole movement is in fact identical to that of a harmonic oscillator evolving from under- to over-damped~\cite{Chen:2017dsy}. First, the two HZS poles move down, approximately along semi-circles, and eventually meet on the imaginary axis, subsequently splitting into two purely imaginary poles, one that descends down the imaginary axis and one that rises to become the charge diffusion pole. The meeting point provides a precise definition for the exact moment of crossover~\cite{Davison:2011ek}.

However, in Einstein-Maxwell models  the crossover is qualitatively different from both LFL and probe brane models~\cite{Davison:2011uk}. At low $T/\mu$ the sound attenuation scales as $|\textrm{Im}(k)|\propto T^0$, similar to the LFL quantum collisionless regime, but at intermediate $T/\mu$ it scales as a power of $T$ less than the $T^2$ of the LFL thermal collisionless regime. At higher $T/\mu$ a hydrodynamic regime emerges where $|\textrm{Im}(k)|\propto T^{-1}$, unlike the $T^{-2}$ of a LFL, but as expected for a CFT: for $T/\mu$ high enough that all scales besides $T$ are negligible, dimensional analysis requires $|\textrm{Im}\left(\omega\right)| \propto T^{-1}$, the AdS-Schwarzschild (AdS-SCH) result~\cite{Herzog:2003ke,Kovtun:2005ev}. Nevertheless, for sufficiently small $k$ the sound attenuation exhibits a maximum before the hydrodynamic regime, so the LFL definition of the crossover remains viable.

Moreover, in Einstein-Maxwell models the pole movement differs dramatically from probe brane models. In the complex $\omega/\mu$ plane, the sound-channel correlators of $J^{\mu}$ and $T^{\mu\nu}$ exhibit both sound and charge diffusion poles for all $T/\mu$, which simply move up, closer to the real axis, as $T/\mu$ increases, as depicted schematically in fig.~\ref{fig:rn_cartoon}. Indeed, a crossover is apparent only in the charge density's spectral function, which we denote $\rho_J$, where as $T/\mu$ increases, a peak produced by the sound poles is suppressed, and a peak produced by the charge diffusion pole rises. A second definition of the crossover is then possible, as the $T/\mu$ where the charge diffusion peak first becomes taller than the sound peak~\cite{Davison:2011uk}. No crossover is apparent in the energy density spectral function, which we denote $\rho_{tt}$, where only a single peak produced by the sound poles is apparent for all $T/\mu$. Equivalently, this crossover occurs as a transfer of dominance in the~\textit{residues} of the poles in the charge density's retarded Green's function, which partly determine the corresponding spectral weights in $\rho_J$.

In short, the LFL and holographic results present us with three possible definitions for a precise $T/\mu$ of crossover. The LFL definition is the sound attenuation maximum. The probe limit definition is the collision of poles on the imaginary axis in fig.~\ref{fig:probe_cartoon}. The AdS-RN definition is the transfer of dominance in $\rho_J$ from the sound peak to the charge diffusion peak. A natural question is how common each of these behaviors is, and whether a ``universal'' definition exists, applicable to all cases above, and more generally to all quantum compressible matter.

More broadly, the accumulating evidence from holography suggests that compressible quantum matter supports sound modes typically, and perhaps universally, and can be characterized by the crossover behavior of such sound. Furthermore, holography shows that, unlike a LFL, the crossover can (and sometimes must) be characterized not only by sound attenuation, but also by pole movement or spectral functions. These results raise many crucial questions. What classes of effective theories of compressible quantum matter support sound modes? What does the crossover of such sound modes reveal about these effective theories? Does any real strongly-interacting quantum compressible matter support sound modes, and if so, what do they reveal about the underlying degrees of freedom?

\subsection{The Model}
\label{model}

As a small step towards answering these questions, and to provide some larger context for the existing holographic results, in this paper we consider a simple model that allows us to interpolate continuously between the two classes of models described above. Specifically, we consider a back-reacted DBI model, with bulk action
\beq
\label{action}
S=\seh + \sdbi,
\eeq
and study the collisionless-to-hydrodynamic crossover as a function of two dimensionless parameters. First is the ``effective tension'' or ``back-reaction'' parameter,
\beq
\label{taudef}
\tau \equiv \left(8 \pi G\right) T_D L^2
\eeq
which appears in Einstein's equation, controlling $\sdbi$'s back-reaction (the back-reacted AdS radius $L$ depends on $T_D$: see eq.~\eqref{ldef}). In particular, the probe limit is an expansion in $GT_D\propto\tau\ll1$ to leading non-trivial order. As suggested above, $\tau$ measures the ratio of the number of charged degrees of freedom to total degrees of freedom, and $\tau \ll 1$ simply means the number of charged degrees of freedom is $\ll N^2$. Second is the ``non-linearity'' parameter,
\beq
\label{talphadef}
\talpha \equiv \alpha/L^2,
\eeq
which controls the strength of higher-order terms in $F_{MN}$. In particular, we can recover a Maxwell action from $\sdbi$ by sending $\talpha \to 0$ with $\tau\talpha^2$ fixed. In probe D-brane models, $\talpha$ is proportional to the string length squared, and is holographically dual to an \textit{inverse} power of the 't Hooft coupling, so that $\sdbi$ includes an infinite sum of finite-coupling corrections.

Of course, $\alpha$ appears in $\sdbi$ only as $F_{MN}$'s pre-factor, so in fact we can absorb $\alpha$ into $F_{MN}$ by a simple re-scaling. To be more precise, the gravity theory's action is invariant under the re-scaling $\alpha \to \lambda \, \alpha$ and $F_{MN} \to \lambda^{-1} \, F_{MN}$ with constant $\lambda \in \mathbb{R}^+$. We could use this re-scaling symmetry to absorb $\alpha$ into $F_{MN}$, which would then be dimensionless, however we will retain $\alpha$ for various reasons: to facilitate comparison to the existing literature, to keep track of powers of the 't Hooft coupling, to facilitate the Maxwell limit of $\sdbi$, etc.

For the theory with action in eq.~\eqref{action}, an exact, closed-form charged black brane solution is known for all values of $\tau$ and $\talpha$~\cite{Fernando:2003tz,Dey:2004yt,Cai:2004eh,Pal:2012zn,Tarrio:2013tta}. The solution is analogous to AdS-RN, and indeed shares many qualitative features with AdS-RN. For example, for any $\tau \neq 0$, the extremal solutions have near-horizon $AdS_2$, so the effective theory is a semi-local quantum liquid.

To be concrete, we restrict to $d=3$ and $T/\mu>0$ (never $T/\mu=0$), and numerically compute the positions of sound-channel poles in $J^{\mu}$ and $T^{\mu\nu}$ correlators in the complex $\omega/\mu$ plane, as a function of either $T/\mu$ or, to determine dispersion relations, $k/\mu$. In holography, the poles in retarded Green's functions are dual to normalizable in-going solutions of the linearized fluctuation equations, \textit{i.e.}\ quasi-normal modes (QNMs) of the charged black brane~\cite{Son:2002sd,Policastro:2002se,Starinets:2002br,Policastro:2002tn,Kovtun:2005ev}. For any $\tau>0$ these poles are shared by all sound-channel correlators of $J^{\mu}$ and $T^{\mu\nu}$, because the dual linearized metric fluctuations are coupled. We also numerically compute $\rho_J$ and $\rho_{tt}$ holographically, from the on-shell action of the bulk fluctuations~\cite{Son:2002sd,Policastro:2002se,Policastro:2002tn}.

\subsection{Summary of Results}
\label{results}

We explore the two-dimensional space parameterized by $\tau$ and $\talpha$ in two steps. First we fix $\talpha$ and increase $\tau$, starting  from the probe limit $\tau=0$. Second, for certain $\tau$ values we scan through decreasing values of $\talpha$. In each case we calculate three things: the spectrum of poles closest to the origin of the complex $\omega/\mu$ plane, the spectral functions $\rho_{tt}$ and $\rho_J$, and the sound dispersion. Our results are summarized as follows.

\textbf{Pole Movement:} At low $T/\mu$ and small $k/\mu$ we always find two HZS poles and a few other poles, which depending on the values of $\tau$ and $\talpha$, may be propagating (non-zero real part) or dissipative (zero real part). As we increase $T/\mu$ the motion of these poles is more complicated than either case in fig.~\ref{fig:cartoons}, and indeed depends sensitively on the values of $\tau$ and $\talpha$. We leave the details to sec.~\ref{sec:poles}, and here just sketch some key general features.

When we fix $\talpha$ and $k/\mu$ and increase $\tau$, purely imaginary poles begin moving up the imaginary axis and ``interfering'' with the poles closer to the origin, producing various complicated pole collisions and splittings as $T/\mu$ increases. However, for $\tau$ below a critical value, two poles eventually emerge at high enough $T/\mu$ that move similarly to the probe limit of fig.~\ref{fig:probe_cartoon}, i.e. they collide on the imaginary axis and produce the charge diffusion pole. On the other hand, for $\tau$ above the critical value the three poles closest to the origin are similar to those of AdS-RN, namely two sound poles and a purely imaginary pole, which move similarly to the AdS-RN case in fig.~\ref{fig:rn_cartoon}, unaffected by the complicated collisions and splittings occurring lower in the complex $\omega/\mu$ plane. Notice that we do not have to take the AdS-RN limit to make the three poles closest to the origin behave similarly to those of AdS-RN: we merely increase $\tau$. For fixed $\talpha$ and $k/\mu$ and increasing $\tau$, the probe limit definition of the crossover thus remains viable only for $\tau$ below a critical value.

Fixing $\tau$ and $k/\mu$ and increasing $\talpha$ actually has the same effect, qualitatively, that is, for fixed $\tau$ and $k/\mu$ the probe limit definition of crossover is viable only for $\talpha$ below a critical value. To see why, suppose $\talpha$ is small, so that the higher-order terms in $F_{MN}$ are suppressed. The leading Maxwell term has coefficient proportional to the product $\tau \talpha^2$, so indeed we expect that fixing one and changing the other should be qualitatively similar to the converse.

In short, when the DBI action back-reacts the probe limit definition of the crossover can remain viable, but only for sufficiently small $\tau$ or $\talpha$, at fixed $k/\mu$.

The gravity theory's scaling symmetry $\talpha \to \lambda \, \talpha$ and $F_{MN} \to \lambda^{-1}\, F_{MN}$ allows for another interpretation of our results for changing $\talpha$ at fixed $T/\mu$, $\omega/\mu$, and $k/\mu$. In an appropriate gauge, at the AdS boundary $A_t \to \mu$, so in the CFT the scaling acts as $\mu \to \lambda^{-1}\,\mu$. Changing $\talpha \to \lambda \, \talpha$ with $T/\mu$, $\omega/\mu$, and $k/\mu$ fixed is thus equivalent to fixing $\talpha$ and changing $T/\mu \to \lambda \, T/\mu$, and similarly for $\omega/\mu$ and $k/\mu$. In particular, changing $\talpha$ at fixed $k/\mu$ is equivalent to fixing $\talpha$ and changing $k/\mu$, which thus provides information about dispersion relations. Occasionally such an interpretation will be useful in what follows, though primarily we will stick to our interpretation of changing $\talpha$ with fixed $k/\mu$.

For all $\talpha$ and $\tau$ (outside of the probe limit) and fixed $k/\mu$ we find sound poles for all $T/\mu$, representing HZS at low $T/\mu$ and hydrodynamic sound at high $T/\mu$. The HZS poles do not always cross over directly to hydrodynamic sound, but instead for small $\talpha$ or $\tau$ they collide on the imaginary axis, as in fig.~\ref{fig:probe_cartoon}, while other poles evolve into hydrodynamic sound. In any case, HZS appears to be ubiquitous in this class of models.

\textbf{Spectral Functions:} For all $\talpha$ and $\tau$ that we consider, with fixed $k/\mu$, the energy density spectral function $\rho_{tt}$ as a function of $\omega/\mu$ always exhibits only a single peak for all $T/\mu$, arising from the sound pole, whether HZS or hydrodynamic.

For fixed $\talpha$ and all $\tau$ we consider, and small $k/\mu$, the charge density spectral function $\rho_J$ at low $T/\mu$ exhibits a peak from the HZS pole. As $T/\mu$ increases a second peak rises closer to $\omega/\mu=0$, due to the charge diffusion pole. The charge diffusion peak eventually grows taller than the sound peak, so the AdS-RN definition of crossover thus remains viable in these cases. However, we suspect that for $\tau$ non-zero but smaller than we could access numerically the AdS-RN definition could eventually fail, because in the probe limit, $\tau=0$, $\rho_J$ always exhibits only a single peak, from either HZS (before the HZS poles collide) or charge diffusion (after the HZS poles collide). In that case no transfer of dominance is possible. Instead, the single peak simply moves towards $\omega/\mu=0$ and broadens as $T/\mu$ increases.

As mentioned above, fixing one of $\talpha$ and $\tau$ and changing the other should have the same qualitative effect as the converse, so long as $k/\mu$ and the higher-order terms in $F_{MN}$ remain sufficiently small. We thus expect that if we fix $\tau$ and decrease $\talpha$ with small $k/\mu$ then $\rho_J$ should eventually behave qualitatively similar to the probe limit. Our results confirm that expectation. In particular, if we fix $\tau$ and decrease $\talpha$ with small $k/\mu$, then we find that the peak in $\rho_J$ due to HZS is eventually overwhelmed by a taller and broader peak, and indeed for $\talpha$ below a critical value $\rho_J$ exhibits only a single peak that moves similarly to the probe limit. The gravity theory's scaling symmetry then implies that fixing $\talpha$ and increasing $k/\mu$ will produce only a single peak in $\rho_J$, as occurs in AdS-RN with increasing $k/\mu$~\cite{Davison:2011uk}.

In short, when the DBI action back-reacts the AdS-RN definition of the crossover can become viable for sufficiently large $\tau$ or $\talpha$, when $k/\mu$ is fixed.

Additionally, we compare our numerical results for $\rho_{tt}$ and $\rho_J$ to a simple approximation that treats each underlying Green's function as a sum of just a few poles close to the origin of the complex $\omega/\mu$ plane. This approximation turns out to work extremely well for many, but not all, values of $\tau$, $\talpha$ and $T/\mu$ that we consider.

\textbf{Sound Dispersion:} For all $\tau$ we consider, with fixed $\talpha$ and sufficiently small $k/\mu$ we always find a sound mode with speed given by (within our numerical accuracy) the conformal value, $v=1/\sqrt{2}$, as in other back-reacted models~\cite{Davison:2011uk,Davison:2013uha}.

If we fix $\talpha$ and $k/\mu$ and increase $\tau$, then the sound pole's $|\textrm{Im}\left(\omega\right)|$ (shown in fig.~\ref{fig:sound_attenuation}) at low $T/\mu$ always scales as $T^0$, similar to a LFL's quantum collisionless regime, and at high $T/\mu$ scales as $T^{-1}$, as expected for a CFT. However, at intermediate $T/\mu$ the power of $T$ decreases as $\tau$ increases, from the $T^2$ of the probe limit down to, but never quite exactly to, $T^0$. An immediate consequence is that a maximum always appears in $|\textrm{Im}\left(\omega\right)|$ at the transition from the intermediate $T/\mu$ scaling to the high $T/\mu$ hydrodynamic scaling.

Furthermore, as $\tau$ increases the maximum's position drifts to higher $T/\mu$. The maximum's height also shrinks, which is perhaps surprising if we recall that $\tau$ effectively measures the number of charged degrees of freedom. In particular, if we increase the number of charged degrees of freedom, and hence increase $\tau$, then na\"ively we expect a larger number of ``decay channels'' for practically any mode, including sound. The na\"ive expectation is thus for $|\textrm{Im}\left(\omega\right)|$ to \textit{increase}, that is for sound to be \textit{dampened}, as $\tau$ increases. Instead we find the opposite: in our holographic model, sound becomes \textit{less} damped as we increase $\tau$.

Fixing $\tau$ and changing $\talpha$ with fixed $k/\mu$ shrinks the overall size of $|\textrm{Im}\left(\omega\right)|$ and shifts it to larger $T/\mu$, but a maximum still appears. In short, for all $\tau$ and $\talpha$ we consider, with fixed $k/\mu$, the sound pole's $|\textrm{Im}\left(\omega\right)|$ as a function of $T/\mu$ is qualitatively similar to that of a LFL in fig.~\ref{fig:attenuation_cartoon}, though quantitatively distinct at intermediate and high $T/\mu$. Most prominently, a maximum always appears in $|\textrm{Im}\left(\omega\right)|$, so the LFL definition of the crossover remains viable.

Finally, for all $\tau$ and $\talpha$ that we consider, we find numerically that the sound attenuation constant takes the hydrodynamic form, $\Gamma=\frac{1}{2} \eta/(\varepsilon+P)$, with shear viscosity $\eta$, energy density $\varepsilon$, and pressure $P$, for all $T/\mu$. In particular, we find this form even at low $T/\mu$, or equivalently for energies $\gg T/\mu$, which is \textit{outside} the usual hydrodynamic regime. The fact that our model, like all (rotationally-invariant) holographic models, has $\eta=s/(4\pi)$ with entropy density $s$~\cite{Policastro:2001yc,Buchel:2003tz,Kovtun:2004de}, then implies that $\Gamma \propto s/(\varepsilon+P)$ is in fact completely determined by thermodynamics. Plugging the Einstein-DBI charged black brane's values of $s$, $\varepsilon$, and $P$  into $\Gamma = \frac{1}{2} s/(\varepsilon+P)$ then enables us to obtain an extremely good approximate expression for the position of the maximum in $|\textrm{Im}\left(\omega\right)|$.

Our paper is a companion to ref.~\cite{Gushterov:2018nht}, which focuses on the shear channel rather than the sound channel, and finds many complementary results. In particular, in hydrodynamics the shear diffusion constant is also $\propto \eta/(\varepsilon+P)$, and a key numerical result of ref.~\cite{Gushterov:2018nht} is that the shear diffusion constant computed numerically from the Einstein-DBI charged black brane also retains the hydrodynamic form down to arbitrarily low $T/\mu$.

These same phenomena occur in other back-reacted models~\cite{Bhattacharyya:2007vs,Davison:2013bxa,Davison:2013uha}, and suggest that in these models the hydrodynamic derivative expansion may be valid even for energies $\gg T/\mu$, outside the normal hydrodynamic regime, so long as $k \ll \mu$ or $T$. More generally, hydrodynamics may be reliable for all $T/\mu$, on length scales larger than a mean free path defined by $\eta/\left(\varepsilon+P\right)$~\cite{Bhattacharyya:2007vs}, giving a mean free path $\propto 1/T$ at high $T/\mu$ but $\propto 1/\mu$ at low $T/\mu$.

Surveying of all the results above makes clear that no definition of the crossover is ``universal.'' At fixed $k/\mu$, the probe limit definition is viable only for sufficiently small $\tau$ or $\talpha$. The AdS-RN definition is viable only for sufficiently large $\tau$ or $\talpha$. The LFL definition is viable for all $\tau$ and $\talpha$ except the probe limit.

This paper is organized as follows. In sec.~\ref{review} we review the charged black brane solutions of the fully back-reacted DBI action. In sec.~\ref{numerical} we present our numerical results for the pole movement, spectral functions, and sound dispersion. We conclude in sec.~\ref{summary} with discussion of our results, including some speculation on the effective theory describing long wavelength excitations, and outlook for future research. The appendix contains the technical details of computing the retarded Green's functions and QNMs.

\section{Charged Black Brane Solutions}
\label{review}

The equations of motion arising from the action in eq.~\eqref{action} with $d=3$ admit the charged black brane solution~\cite{Fernando:2003tz,Dey:2004yt,Cai:2004eh,Pal:2012zn,Tarrio:2013tta},
\begin{subequations}
\label{bgsol}
\beq
ds^2 = g_{MN} \, dx^M dx^N = \frac{L^2}{z^2} \left(\frac{dz^2}{f(z)} - f(z) \, dt^2 + dx^2 + dy^2\right),
\eeq
\beq
f(z) = 1- \frac{z^3}{z_H^3} + \frac{\tau}{3} \left[1 - \frac{z^3}{z_H^3} + \,_2F_1\left(-\frac{1}{2},-\frac{3}{4};\frac{1}{4};-\tilde{\alpha}^2 Q^2\right) \frac{z^3}{z_H^3}- \, _2F_1\left(-\frac{1}{2},-\frac{3}{4};\frac{1}{4};-\tilde{\alpha}^2 Q^2\frac{z^4}{z_H^4}\right)\right], \nonumber
\eeq
\beq
F_{tz} = - F_{zt} = \frac{Q/z_H^2}{\sqrt{1+ \tilde{\alpha}^2 Q^2 z^4/z_H^4}},
\eeq
\end{subequations}
with CFT time coordinate $t$ and spatial coordinates $x$ and $y$, and holographic coordinate $z$. The horizon $z_H$ is the smallest real solution of $f(z_H)=0$, and the asymptotic $AdS_4$ boundary is at $z \to 0$, with $AdS_4$ radius $L$ given by
\beq
\label{ldef}
L^2 = \frac{L_0^2}{1-\left(8 \pi G\right)T_DL_0^2/3}.
\eeq
The brane changes the $AdS_4$ radius from $L_0$ to $L$ because when $F_{MN}=0$ the DBI action is simply the brane's volume, which makes a positive contribution to the cosmological constant. Roughly speaking, $L$ is a measure of the total degrees of freedom in the CFT, for example when $d=4$ the central charges are $L^3/G$~\cite{Ammon:2015wua}. Clearly $L^2\geq 0$ if and only if $\left(8 \pi G\right)T_D \leq 3 L_0^{-2}$. As suggested in sec.~\ref{intro}, $T_D$ is a measure of the number of charged degrees of freedom in the CFT. The bound $\left(8 \pi G\right)T_D \leq 3 L_0^{-2}$ suggests that the model in eq.~\eqref{action} describes a CFT in which the number of charged degrees of freedom can increase while preserving conformal symmetry, \textit{i.e.} zero beta function(s), only up to a limit determined by the number of uncharged degrees of freedom. Indeed, appealing to our intuition from probe branes, generically flavor fields make a positive contribution to the gauge coupling's beta function, hence we expect the flavor fields to preserve conformal symmetry only within some ``conformal window.''

In subsequent sections we use units with $L\equiv 1$. In that case, if we change $\left(8 \pi G\right)T_D$ then implicitly we also change $L_0$ to maintain $L\equiv 1$, or more precisely, to maintain all quantities in units of $L$. As a result, $\left(8 \pi G\right)T_D$, and hence $\tau$, will effectively have no upper limit.

For given $\tau$ and $\tilde{\alpha}$, the dimensionless integration constant $Q$ completely determines the solution in eq.~\eqref{bgsol}. Correspondingly, the CFT's state is determined by the single dimensionless parameter $T/\mu$, hence $Q$ must determine $T/\mu$. For the solution in eq.~\eqref{bgsol},
\begin{subequations}
\beq
\label{teq}
T = \frac{|f'(z_H)|}{4 \pi} = \frac{3+\tau\left(1-\sqrt{1+\tilde{\alpha}^2 Q^2}\right)}{4\pi z_H},
\eeq
\beq
\label{mueq}
\mu =  \int_0^{z_H} dz \, F_{tz} = \frac{Q}{z_H} \,_2F_1\left(\frac{1}{2},\frac{1}{4};\frac{5}{4};-\tilde{\alpha}^2 Q^2\right),
\eeq
\end{subequations}
where $f'(z) \equiv \partial f(z)/\partial z$. The mapping from $Q$ to $T/\mu$ is thus given by
\beq
\label{tmu}
\frac{T}{\mu} = \frac{3+\tau\left(1-\sqrt{1+\tilde{\alpha}^2 Q^2}\right)}{4\pi Q \,_2F_1\left(\frac{1}{2},\frac{1}{4};\frac{5}{4};-\tilde{\alpha}^2 Q^2\right)}.
\eeq
Only the product $\talpha Q$ appears in $g_{MN}$, so invariance of $g_{MN}$ under the gravity theory's scaling symmetry $\talpha \to \lambda \,\talpha$ and $F_{MN} \to \lambda^{-1}\,F_{MN}$ requires $Q \to \lambda^{-1} \, Q$ and hence $T \to T$ and $\mu \to \lambda^{-1} \, \mu$, as mentioned in sec.~\ref{results}.

All thermodynamic quantities can be written as a function of $T/\mu$ only, or equivalently of $Q$ only, times an overall factor of either $T$ or $\mu$ to a power determined by dimensional analysis. For example, using eq.~\eqref{teq} the solution's Bekenstein-Hawking entropy density $s$, namely $1/(4G)$ times the horizon area density, can be written as
\beq
\label{entropy}
s = \frac{L^2}{4G} \frac{1}{z_H^2} = \frac{L^2}{4G} \left(\frac{4 \pi T}{3}\right)^2 \left[1+\frac{\tau}{3}\left(1-\sqrt{1+\talpha^2 Q^2}\right)\right]^{-2}.
\eeq
The on-shell Euclidean gravity action density equals the CFT's free energy density times $1/T$~\cite{Witten:1998zw}. To compute the energy density, $\varepsilon \equiv \langle T^{tt} \rangle$, and pressure, $P\equiv \Txx = \langle T^{yy} \rangle$, we must therefore evaluate the Euclidean version of the action, eq.~\eqref{action}, on the Euclidean version of the solution, eq.~\eqref{bgsol}. The result diverges, and requires holographic renormalization~\cite{deHaro:2000xn,Skenderis:2002wp}, which proceeds similarly to the AdS-RN case.\footnote{To compute correlators via holographic renormalization, we introduce a cutoff surface near the asymptotic $AdS_4$ boundary, $z = \epsilon$, introduce covariant counterterms at $z = \epsilon$, take variational derivatives of the on-shell bulk action plus counterterms, and then send $\epsilon \to 0$. The Einstein-DBI counterterms are identical to those of Einstein-Maxwell, namely the Gibbons-Hawking term, a counterterm proportional to the cutoff surface's volume, a counterterm proportional to the cutoff surface's intrinsic curvature, and a counterterm proportional to a Maxwell term for $F_{MN}$. The latter is actually unnecessary for the solution in eq.~\eqref{bgsol}, consistent with the field theory statement that the vacuum counterterms suffice for renormalization at non-zero $T$ and $\mu$~\cite{Kapusta:2006pm}. The Einstein-Maxwell counterterms appear explicitly for example in ref.~\cite{Edalati:2010pn}.} We thus find
\beq
\label{energy}
\varepsilon = \frac{L^2}{8\pi G}\left(\frac{4 \pi T}{3}\right)^3 \, \frac{1+ \frac{\tau}{3} \left [ 1 -  \,_2F_1\left(-\frac{1}{2},-\frac{3}{4};\frac{1}{4};-\tilde{\alpha}^2 Q^2\right)\right]}{\left[1+\frac{\tau}{3}\left(1-\sqrt{1+\talpha^2 Q^2}\right)\right]^3},
\eeq
and $P = \varepsilon/2$, as required by scale invariance~\cite{LLfluids}. In the hydrodynamic regime, $v^2 = \frac{\partial P}{\partial \varepsilon}=1/(d-1)$~\cite{LLfluids}, which in our case is $v^2 = 1/2$. Remarkably, for both AdS-RN and probe branes in AdS-SCH, HZS also has $v^2 = 1/(d-1)$~\cite{Karch:2008fa,Edalati:2010pn,Davison:2011ek,Davison:2011uk}, as we will see in sec.~\ref{numerical}. In a LFL the speeds of hydrodynamic and zero sound coincide only in the limit of infinite quasi-particle interaction strength~\cite{Pines}. The charge density $\Jt$ of the solution in eq.~\eqref{bgsol} is
\beq
\label{jt}
\Jt = \frac{L^2}{8\pi G} \left(\frac{4 \pi T}{3}\right)^2 \frac{ \tau \talpha^2 \, Q}{\left[1+\frac{\tau}{3}\left(1-\sqrt{1+\talpha^2 Q^2}\right)\right]^2},
\eeq
which obeys $\varepsilon + P = s\,T + \mu \Jt$, as expected. Moreover, we can write $\Jt$ in terms of $s$ as $\Jt = \tau \talpha^2 Q \, s/(2\pi)$, which we will use in sec.~\ref{sec:soundatt}.

The solution in eq.~\eqref{bgsol} admits an extremal limit, $T=0$, with $Q$'s corresponding extremal value, $\qext$, given by
\beq
\label{qext}
\qext^2 = \frac{1}{\tau \tilde{\alpha}^2} \left( 6 + \frac{9}{\tau}\right).
\eeq
We can show that the extremal limit of the solution in eq.~\eqref{bgsol} has near-horizon geometry $AdS_2 \times \mathbb{R}^2$ in the usual way, as follows. We expand $f(z)$ near the horizon, \textit{i.e.} in powers of $\left(z_H-z\right)$, where of course $f(z_H)=0$, and if $Q = \qext$ then also $f'(z_H)=0$. In that case, truncating the expansion at order $(z_H-z)^2$ and defining a new radial coordinate
\beq
\xi \equiv \frac{1}{(z_H-z) \, \frac{1}{2} \left . f''(z_H) \right|_{\qext}},
\eeq
produces the near-horizon metric
\beq
ds^2 = \frac{L^2_{AdS_2}}{\xi^2} \left (d\xi^2 - dt^2\right) + \frac{L^2}{z_H^2} \left( dx^2 + dy^2 \right),
\eeq
which is $AdS_2 \times \mathbb{R}^2$, with $AdS_2$ radius $L_{AdS_2}$ given by
\beq
\label{ads2radius}
L^2_{AdS_2} = \frac{L^2}{z_H^2 \, \frac{1}{2} \left . f''(z_H) \right |_{\qext}},
\eeq
where for the solution in eq.~\eqref{bgsol}
\beq
\label{fderivs}
z_H^2 \left . \frac{1}{2} f''(z_H) \right |_{\qext} = \frac{9+6 \tau}{3+\tau}.
\eeq
As in AdS-RN, the near-horizon $AdS_2 \times \mathbb{R}^2$ indicates that the dual CFT state is a semi-local quantum liquid~\cite{Iqbal:2011in}. In $T^{\mu\nu}$ and $J^{\mu}$'s Green's functions we then expect branch cuts along the imaginary axis~\cite{Edalati:2010hk,Edalati:2010pn}. However, in subsequent sections we will always have $T/\mu>0$, so instead of branch cuts we expect poles along the imaginary axis that grow more and more dense as $T/\mu$ decreases, presumably coalescing into a branch cut when $T/\mu=0$~\cite{Edalati:2010hk,Edalati:2010pn}. In sec.~\ref{numerical} we will not explore $T/\mu$ small enough to see any such dense collection of poles.

\subsection{The Probe Limit}

As mentioned below eq.~\eqref{taudef}, the probe limit is an expansion in $GT_D\propto\tau \ll 1$, with $\talpha$ fixed. More specifically, we expand in $\tau$, and in all field theory quantities retain all terms up to the first non-trivial order in $\tau$. In the holographically dual gravity theory, those leading non-trivial contributions come from the probe DBI action evaluated in the uncorrected background metric. For the $g_{MN}$ in eq.~\eqref{bgsol} we thus set $\tau=0$, in which case $L^2 = L_0^2$ and $f(z)=1-z^3/z_H^3$, that is, $g_{MN}$ becomes that of AdS-SCH in $d=3$ with radius $L_0$. Consequently, the probe limit expressions for $T$, $\mu$, and $T/\mu$ are simply those in eqs.~\eqref{teq},~\eqref{mueq}, and~\eqref{tmu}, respectively, but with $\tau=0$. Moreover, in eq.~\eqref{qext} taking $\tau \to 0$ sends $\qext \to \infty$. In that limit, $g_{MN}$ is that of $AdS_4$, with no horizon and hence no near-horizon $AdS_2 \times \mathbb{R}^2$.

However, in these conformal cases the probe limit breaks down when $T/\mu=0$~\cite{Hartnoll:2009ns,Bigazzi:2013jqa}. To see why, consider for example the probe limit of $s$, or any other quantity obtained from the on-shell action/free energy.\footnote{The entropy density $s$ can be calculated either from the horizon area or from $-\frac{\partial}{\partial T}$ of the free energy density. In the first case, calculating the order $GT_D$ contribution to $s$ requires calculating $\sdbi$'s linearized back-reaction and the corresponding change in $z_H$. The second case requires only calculating the on-shell $\sdbi$ with the un-corrected $g_{MN}$ and then taking $-\frac{\partial}{\partial T}$. In particular, the second calculation requires no back-reaction. The two calculations agree, as required by thermodynamic consistency: see for example refs.~\cite{Mateos:2006yd,Mateos:2007vn,Bigazzi:2009bk}.} Expanding eq.~\eqref{entropy} to first order in $GT_D$ gives
\beq
\label{sprobe}
s = \frac{L_0^2}{4G} \left(\frac{4 \pi T}{3}\right)^2 \left[1 - \frac{1}{3} \tau + \frac{2}{3} \tau \sqrt{1+\talpha^2 Q^2}+ \mathcal{O}\left(\tau^2\right)\right] ,
\eeq
where now $\tau = \left(8 \pi G\right) T_D L_0^2$ and $\talpha = \alpha/L_0^2$. Following refs.~\cite{Karch:2008fa,Karch:2009zz}, we next replace $Q$ with $T/\mu$, or equivalently $T^2/\Jt$, using the probe limit of eq.~\eqref{jt}
\beq
\label{jtprobe}
\Jt = \frac{L_0^2}{8\pi G} \left(\frac{4 \pi T}{3}\right)^2 \tau \talpha^2 \, Q,
\eeq
where, as in eq.~\eqref{sprobe}, $\tau$ and $\talpha$ now involve $L_0$ rather than $L$. Inserting eq.~\eqref{jtprobe} into eq.~\eqref{sprobe} and expanding in $T^2/\Jt \ll 1$ gives
\beq
\label{sprobelowt}
s = \frac{L_0^2}{4G} \left(\frac{4 \pi T}{3}\right)^2 \left[1 - \frac{1}{3} \tau \right ] + \frac{4\pi}{3} \frac{\Jt}{\talpha}+ \frac{1}{2} \left(\frac{4\pi}{3}\right)^5 \frac{\tau^2 \talpha L_0^4}{(8\pi G)^2} \, \frac{T^4}{\Jt}+ \mathcal{O}\left(\frac{\tau^4 T^8}{\Jt^3}\right) + \mathcal{O}\left(\tau^2\right).
\eeq
On the right-hand-side of eq.~\eqref{sprobelowt}, the first term is $s$ of $d=3$ AdS-SCH minus the probe's $\Jt$-independent order $\tau$ correction. The second term is $T$-independent, leading to a residual entropy: if $T/\mu=0$ then $s \propto \Jt/\talpha+ \mathcal{O}\left(\tau^2\right)$. In that case the probe limit clearly breaks down because the order $\Jt \propto \tau$ term is larger than the order $\tau^0$ term~\cite{Hartnoll:2009ns,Bigazzi:2013jqa}. As mentioned above, in subsequent sections we will always have $T/\mu>0$, avoiding such probe limit breakdown. The third term on the right-hand-side of eq.~\eqref{sprobelowt} gives the leading $\Jt$-dependent contribution to the heat capacity, $T \partial s/\partial T$, which is $\propto T^4$. For general $d$ that term is $\propto T^{2(d-1)}$, in stark contrast to $T$ for free fermions or $T^{d-1}$ for free bosons~\cite{Karch:2008fa,Karch:2009zz}.

\subsection{The AdS-RN Limit}

As mentioned below eq.~\eqref{talphadef}, to recover Einstein-Maxwell from Einstein-DBI we take $\talpha \to 0$ keeping $\tau \talpha^2$ fixed, so that $\tau$ diverges as $\talpha^{-2}$. Moreover we adjust $L_0$ to keep $L$ fixed. In that limit, $f(z)$, and hence $T/\mu$, takes the AdS-RN form,
\beq
f(z) = 1- \frac{z^3}{z_H^3} - \frac{1}{2} \tau \talpha^2 Q^2 \frac{z^3}{z_H^3} + \frac{1}{2} \tau \talpha^2 Q^2 \frac{z^4}{z_H^4},
\eeq
\beq
\label{tadsrn}
\frac{T}{\mu} = \frac{3-\frac{1}{2}\,\tau \talpha^2 Q^2}{4\pi Q}.
\eeq
In particular, now $\qext^2 = 6/(\tau \talpha^2)$, which is also obvious from taking $\tau \propto \talpha^{-2} \to \infty$ in eq.~\eqref{qext}. That same limit of eq.~\eqref{ads2radius} gives $L_{AdS_2}^2=L^2/6$, as expected. In the AdS-RN limit, we also find the expected form of the entropy density,
\beq
s = \frac{L^2}{4G} \left(\frac{4 \pi T}{3}\right)^2 \left[1-\frac{1}{6} \, \tau \talpha^2 \, Q^2\right]^{-2}.
\eeq

In contrast to the probe limit, for AdS-RN at small $T/\mu$ the heat capacity's leading $\Jt$-dependent term is $\propto T$, similar to free fermions---though other observables differ dramatically from those of free fermions, as discussed in sec.~\ref{intro}. The AdS-RN limit of eq.~\eqref{jt} is
\beq
\Jt = \frac{L^2}{8\pi G} \left(\frac{4 \pi T}{3}\right)^2 \,\tau \talpha^2 \, Q\, \left[1-\frac{1}{6}\tau\talpha^2 Q^2\right]^{-2}.
\eeq
In the limit $T^2/\Jt \ll 1$, we thus find
\beq
s = \frac{2\pi}{\sqrt{6}} \frac{\Jt}{\sqrt{\tau \talpha^2}} + \frac{8\pi^2}{6^{5/4}} \frac{L}{\sqrt{8\pi G}} \frac{T\sqrt{\Jt}}{\left(\tau \talpha^2\right)^{1/4}} + \mathcal{O}\left(T^2\right),
\eeq
where the first term is $T$-independent, leading to a residual entropy $\propto \Jt/\sqrt{\tau \talpha^2}$, while the second term gives a leading contribution to the heat capacity $\propto T$, as advertised.

\section{Numerical Results}
\label{numerical}

For given values of $\tau$ and $\talpha$, we want to know whether a sound pole exists at low $T/\mu$, and how its dispersion changes in the crossover to hydrodynamics as $T/\mu$ increases. More generally we want to know the spectrum of poles in the sound channel of the charge and energy retarded Green's functions, $G_J$ and $G_{tt}$ respectively, at low $T/\mu$ and small $k/\mu$, and how they move as $T/\mu$ increases (the crossover) or as $k/\mu$ increases (the dispersion relations). We also want to know how the poles affect the charge and energy spectral functions, $\rho_J$ and $\rho_{tt}$, respectively. We will focus on the ``highest'' poles, meaning those highest in the complex $\omega/\mu$ plane (closest to the origin), which represent the longest-lived excitations.

In the appendix we explain in detail how we compute $G_J$ and $G_{tt}$, their poles, and $\rho_J$ and $\rho_{tt}$ holographically, by solving for the linearized fluctuations of the gravity fields dual to $J^{\mu}$ and $T^{\mu\nu}$, using the techniques of ref.~\cite{Kaminski:2009dh}. Crucially, in the gravity theory in general the fluctuations couple, implying that $G_J$ and $G_{tt}$ share poles. However, in the probe limit the fluctuations decouple, in which case we can distinguish which poles appear in $G_J$ versus $G_{tt}$.

As mentioned in sec.~\ref{results}, we will sample values of $\tau$ and $\talpha$ in two steps. First we will fix $\talpha=1$ and increase $\tau$, typically starting from the probe limit, $\tau=0$, and then going through $\tau=10^{-5},10^{-4},10^{-3}$, and $10^{-2}$ and in some cases larger $\tau$. Second we will choose representative $\tau$ values, and for each scan through $\talpha$ values.

To stay within the hydrodynamic regime at high $T/\mu$, we fix $k/\mu=10^{-2}$ throughout, except of course when computing dispersion relations. However, as mentioned in secs.~\ref{results} and~\ref{review}, the gravity theory's scaling symmetry $\talpha \to \lambda \,\talpha$ and $F_{MN} \to \lambda^{-1}\,F_{MN}$ acts in the CFT to re-scale the chemical potential, $\mu \to \lambda^{-1}\,\mu$, thus allowing for an alternative interpretation of the effect of changing $\talpha$, as instead fixing $\talpha$ and changing $T/\mu$, $\omega/\mu$, and $k/\mu$. Such an interpretation will be useful in a few cases below.

We present our numerical result for the poles in $G_J$ and $G_{tt}$ in sec.~\ref{sec:poles}, for the spectral functions $\rho_J$ and $\rho_{tt}$ in sec.~\ref{sec:spectral_functions}, and for the sound attenuation in sec.~\ref{sec:soundatt}.

\subsection{Poles and Dispersion Relations}
\label{sec:poles}

In the probe limit with $T/\mu=0$ the metric $g_{MN}$ is that of $AdS_4$, in which case conformal invariance fixes $G_{tt}$ completely, up to an overall constant~\cite{DiFrancesco:1997nk}, whose only non-analyticities are branch points at $\omega = \pm k$ and $\omega= \infty$, connected by an arbitrary contour. However, $G_J$ has no branch points, but rather two highest poles identified as HZS~\cite{Karch:2008fa,Karch:2009zz}, with dispersion
\beq
\label{soundisp}
\omega = \pm v \, k - i \, \Gamma \, k^2 + \mathcal{O}\left(k^3\right),
\eeq
with $v = 1/\sqrt{d-1}$ and attenuation constant
\beq
\label{probehzsatt}
\Gamma = \frac{v^2}{2 \mu} = \frac{\Gamma\left(\frac{1}{2}\right)}{\Gamma\left(\frac{1}{2(d-1)}\right)\Gamma\left(\frac{d-2}{2(d-1)}\right)} \, \Jt^{-\frac{1}{d-1}},
\eeq
both with $d=3$. When $T/\mu>0$, but still in the probe limit, the metric $g_{MN}$ is that of AdS-SCH, so $G_{tt}$ will have the usual hydrodynamic sound poles, with dispersion of the same form as in eq.~\eqref{soundisp}, where scale invariance requires $v = 1/\sqrt{d-1}$ and now
\beq \label{eq:sound_mode_hydro}
\Gamma = \frac{d-2}{d-1} \, \frac{\eta}{\varepsilon + P},
\eeq
with $d=3$. In (rotationally-invariant) holographic QFTs the shear viscosity $\eta = s/(4\pi)$~\cite{Policastro:2001yc,Buchel:2003tz,Kovtun:2004de}. The $s$ and $\varepsilon$ of AdS-SCH in $d=3$ are simply the probe limits of eqs.~\eqref{entropy} and~\eqref{energy}, respectively, where also $P=\varepsilon/2$. These values give $v=1/\sqrt{2}$ and $\Gamma = 1/(8 \pi T)$~\cite{Herzog:2003ke,Kovtun:2005ev}.

As reviewed in sec.~\ref{intro}, in the probe limit with $T/\mu>0$, the HZS survives for $0<\pi T/\mu<\o/\mu$, with dispersion unchanged from the $T/\mu=0$ form~\cite{Davison:2011ek,Chen:2017dsy}, just like the LFL quantum collisionless regime. The HZS also survives for $\o/\mu < \pi T/\mu < \sqrt{\o/\mu}$, still with $v=1/\sqrt{2}$, but now with $\Gamma \propto T^2$, just like the LFL thermal collisionless regime~\cite{Davison:2011ek,Chen:2017dsy}. However, in the hydrodynamic regime, $\pi T/\mu > \sqrt{\o/\mu}$, $J^{\mu}$'s conservation equation dictates that the highest pole in $G_J$ is not that of sound, but rather hydrodynamic charge diffusion, with dispersion
\beq
\label{chargediff}
\omega = - i \, D \, k^2 + \mathcal{O}\left(k^3\right),
\eeq
where a probe DBI action in $d=3$ AdS-SCH gives a charge diffusion constant~\cite{Kim:2008bv,Mas:2008qs}
\beq \label{eq:probe_diffusion}
D = \frac{3}{4\pi T} \, \sqrt{1 + \talpha^2 Q^2}\, _2F_1 \left(\frac{3}{2}, \frac{1}{4}; \frac{5}{4}; - \talpha^2 Q^2 \right).
\eeq

\subsubsection{Changing $\tau$}
\label{tauchange}

\begin{figure}[t!]
	\begin{center}
		\begin{subfigure}{0.49\textwidth}
			\includegraphics[width=\textwidth]{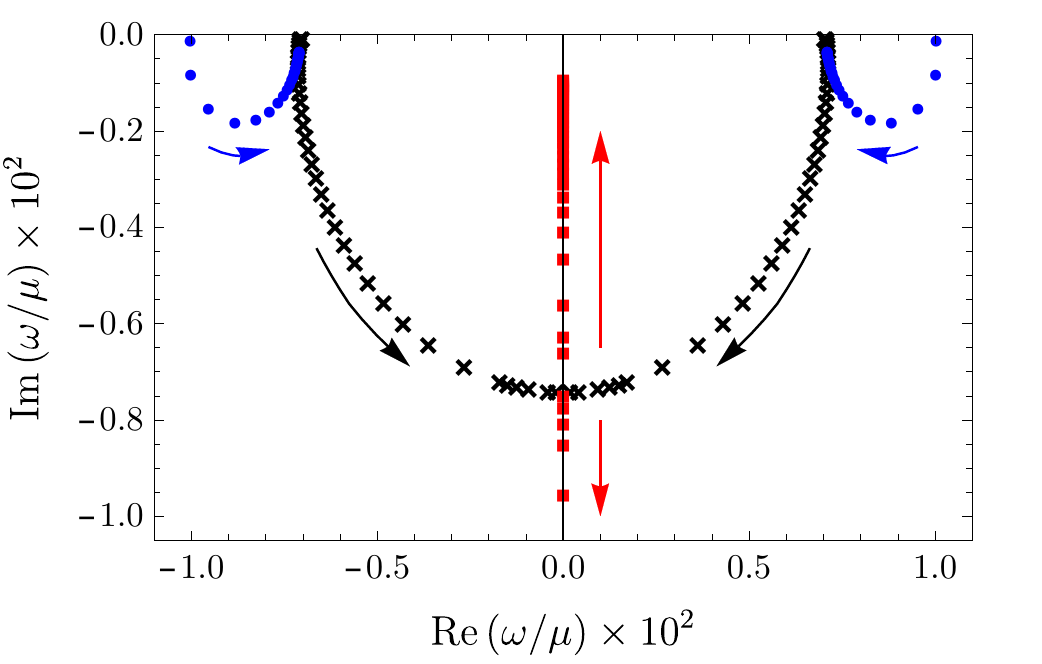}
			\caption{\label{plane_probe}}
		\end{subfigure}
		\begin{subfigure}{0.49\textwidth}
			\includegraphics[width=\textwidth]{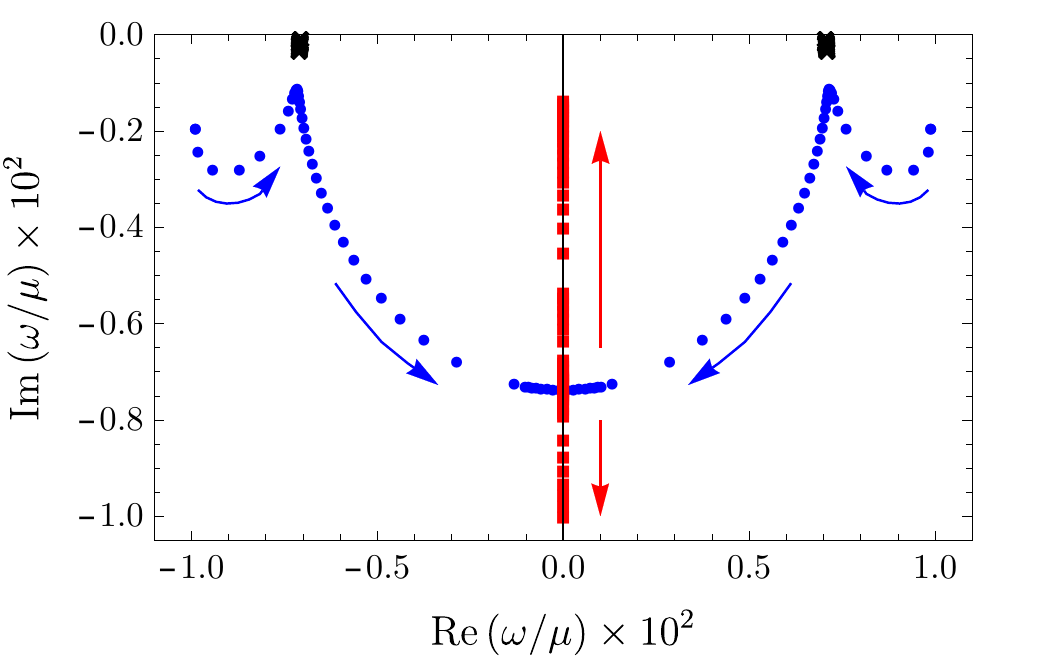}
			\caption{\label{plane_10Em4}}
		\end{subfigure}
		\caption{\label{planesmalltau} Positions of poles of $G_J$ and $G_{tt}$ in the complex $\omega/\mu$ plane for increasing $T/\mu$, with $\talpha=1$ and $k/\mu=10^{-2}$. We have enhanced $\textrm{Re}\left(\omega/\mu\right)$ and $\textrm{Im}\left(\omega/\mu\right)$ by $10^2$ for clarity. The arrows indicate the movement of poles as $T/\mu$ increases. (a) $\tau=0$ and $5\times10^{-4}\leq T/\mu\leq0.1$. At $T/\mu=5\times10^{-4}$ we find four poles, two only in $G_{tt}$, with relativistic dispersion (blue dots), and two only in $G_J$, with dispersion well-approximated by the HZS dispersion in eqs.~\eqref{soundisp} and~\eqref{probehzsatt} (black crosses). As $T/\mu$ increases the blue dots move down and then back up, eventually becoming hydrodynamic sound poles. The black crosses move down and eventually collide and split on the imaginary axis, producing two purely imaginary poles (red squares), one of which moves up and becomes the charge diffusion pole (see also fig.~\ref{fig:probe_cartoon}). (b) $\tau=10^{-4}$ and $10^{-4}\leq T/\mu\leq0.05$. At $T/\mu=10^{-4}$ we again find four poles, similar to $\tau=0$, however now all poles are shared by $G_J$ and $G_{tt}$, and the black crosses denote sound poles which persist mostly unchanged as $T/\mu$ increases, while the poles with relativistic dispersion collide and split on the imaginary axis, producing the charge diffusion pole. (Animated versions of both figures are available on this paper's arxiv page.)}
	\end{center}
\end{figure}

Fig.~\ref{plane_probe} shows our numerical results for the positions of poles in the complex $\omega/\mu$ plane for $\talpha=1$ and $\tau = 0$, i.e. the probe limit. The arrows indicate the motion of the poles as $T/\mu$ increases from $T/\mu=5\times10^{-4}$ to $0.1$. (An animated version of fig.~\ref{plane_probe} is available on this paper's arxiv page.)

Our results are similar to those of refs.~\cite{Policastro:2002tn,Kovtun:2005ev} for $G_{tt}$ and refs.~\cite{Davison:2011ek,Chen:2017dsy} for $G_J$, the main difference being that our spacetime is asymptotically $AdS_4$ rather than $AdS_5$. At low temperature, $T/\mu=5\times10^{-4}$, we find four poles, two in $G_{tt}$, denoted by blue dots in fig.~\ref{plane_probe}, with relativistic dispersion $\omega = \pm k + \ldots$~\cite{Kovtun:2005ev}, and two in $G_J$, denoted by black crosses in fig.~\ref{plane_probe}, with dispersion well-approximated by the $T/\mu=0$ HZS form in eqs.~\eqref{soundisp} and~\eqref{probehzsatt}~\cite{Davison:2011ek,Chen:2017dsy}.

As $T/\mu$ increases the blue dots first descend into the complex $\omega/\mu$ plane before turning around and moving back up, always with decreasing real part. By the time $T/\mu=0.1$ they have become the hydrodynamic sound poles. Similar crossover behavior in $G_{tt}$'s poles from relativistic to sound dispersion was observed in ref.~\cite{Kovtun:2005ev}. Meanwhile the black crosses move as depicted in fig.~\ref{fig:probe_cartoon}: they move down and towards the imaginary axis, approximately tracing semi-circles~\cite{Davison:2011ek,Chen:2017dsy}, and then collide on the imaginary axis at $T/\mu=0.033$, where they split into two purely imaginary poles, one moving up the axis and the other moving down. The one moving up eventually becomes the charge diffusion pole, with dispersion given by eqs.~\eqref{chargediff} and~\eqref{eq:probe_diffusion}. Such crossover behavior in $G_J$ in the probe limit was observed in refs.~\cite{Davison:2011ek,Chen:2017dsy}. As mentioned in sec.~\ref{intro}, in ref.~\cite{Davison:2011ek} the collision of poles on the imaginary axis was used as a definition of the precise moment of crossover (value of $T/\mu$) to the hydrodynamic regime.

\begin{figure}[t!]
\begin{center}
\begin{subfigure}{0.49\textwidth}
\includegraphics[width=\textwidth]{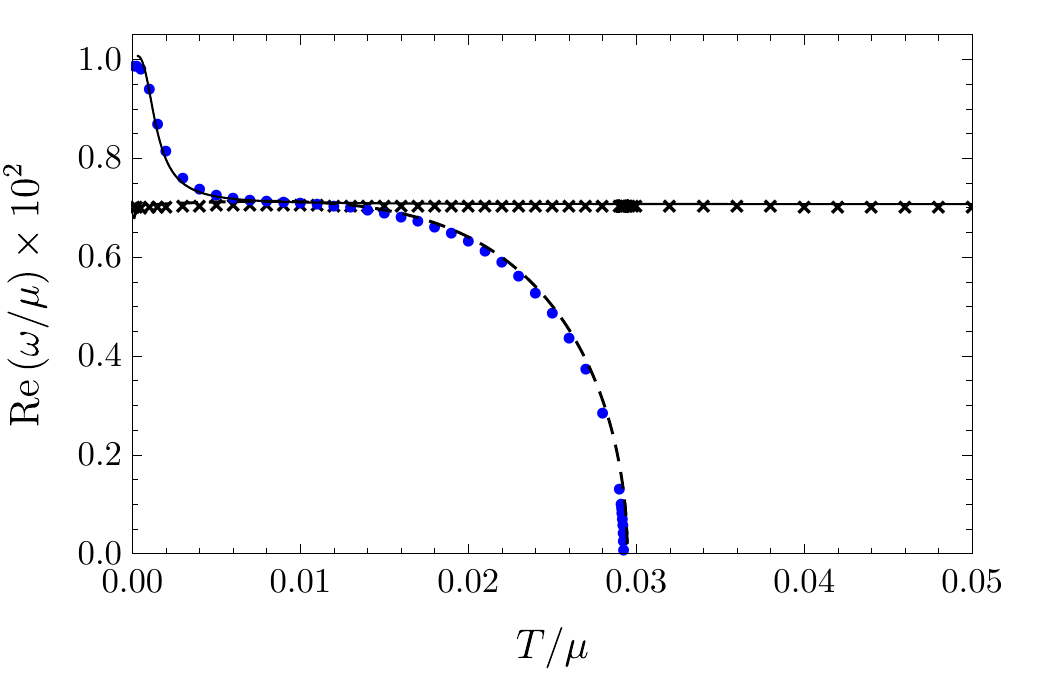}
\caption{\label{realimsmalltaua}}
\end{subfigure}
\begin{subfigure}{0.49\textwidth}
\includegraphics[width=\linewidth]{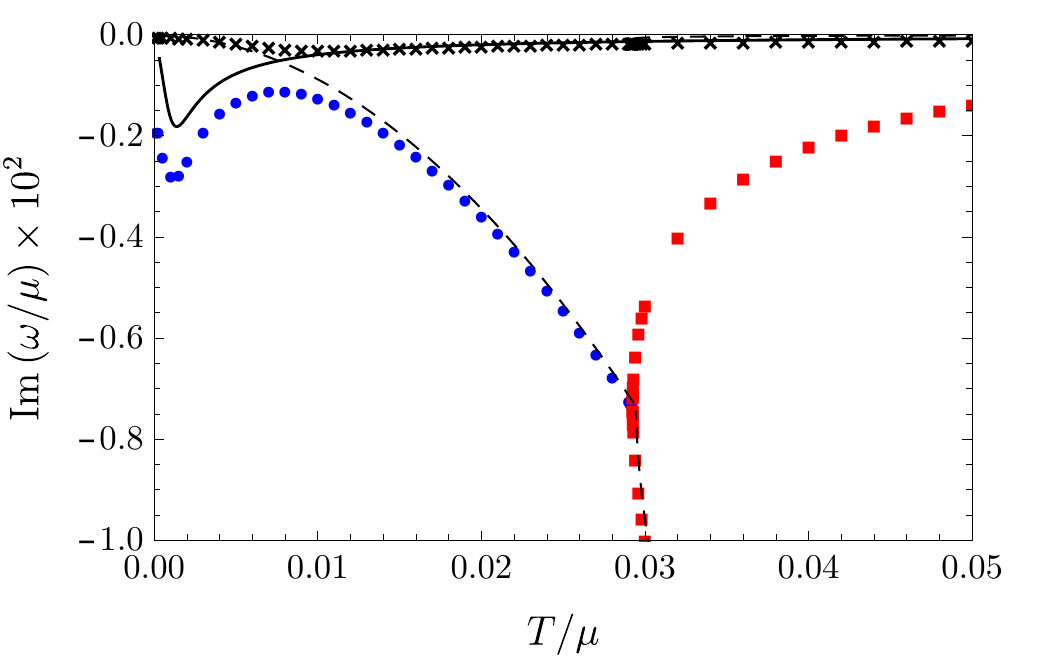}
\caption{\label{realimsmalltaub}}
\end{subfigure}
\caption{\label{realimsmalltau} The same data as fig.~\ref{plane_10Em4} but with separate plots for (a) $\textrm{Re}\left(\omega/\mu\right)$ and (b) $\textrm{Im}\left(\omega/\mu\right)$, each enhanced by $10^2$ for clarity, as functions of $T/\mu$. The color and shape coding are the same as fig.~\ref{plane_10Em4}. The dashed black lines denotes the probe limit HZS dispersion in eqs.~\eqref{soundisp} and~\eqref{probehzsatt} while the solid black line denotes the hydrodynamic sound dispersion. At low $T/\mu$ the black crosses follow the black dashed line, identifying those poles as HZS, and as $T/\mu$ increases they crossover to the solid black line, indicating they have become hydrodynamic sound. The upper branch of red squares eventually approaches the probe limit charge diffusion dispersion in eqs.~\eqref{chargediff} and~\eqref{eq:probe_diffusion} (not shown), identifying that as the diffusion pole.}
\end{center}
\end{figure}

We next introduce small back-reaction, $\tau\neq 0$ but $\ll 1$. We found that the pole movement for $\tau=10^{-5}$ is qualitatively similar to that for $\tau=10^{-4}$, so we will only present results for the latter. Fig.~\ref{plane_10Em4} shows our numerical results for the pole positions for $\talpha=1$ and $\tau = 10^{-4}$, for $10^{-4}\leq T/\mu \leq 0.05$. (An animated version of fig.~\ref{plane_10Em4} is available on this paper's arxiv page.) For clarity, fig.~\ref{realimsmalltau} shows the same data as fig.~\ref{plane_10Em4}, but with $\textrm{Re}\left(\omega/\mu\right)$ and $\textrm{Im}\left(\omega/\mu\right)$ plotted separately versus $T/
\mu$ in figs.~\ref{realimsmalltaua} and~\ref{realimsmalltaub}, respectively.

In fig.~\ref{plane_10Em4} and fig.~\ref{realimsmalltau}, at the low temperature $T/\mu=10^{-4}$, similar to fig.~\ref{plane_probe} we again find four poles, two with relativistic dispersion, again denoted by blue dots, and two with HZS dispersion, again denoted by black crosses. However as $T/\mu$ increases the pole movement has some dramatic qualitative differences from the probe limit. The blue dots again first move down and up while their real part decreases, but then they move down again, still with decreasing real part. Meanwhile the black crosses barely move: fig.~\ref{realimsmalltaua} shows the real part is apparently constant (within our numerical accuracy), with $v=1/\sqrt{2}$, while fig.~\ref{realimsmalltaub} shows the imaginary part changes by at most $10\%$, with the largest deviation at the point of closest approach to the blue dots. However, after that point of closest approach the remaining evolution is similar to the probe limit. The blue dots approximately trace semi-circles and ultimately collide on the imaginary axis at $T/\mu = 0.029$, where they then split into two purely imaginary poles, one moving up the axis and one moving down, where the one moving up eventually becomes the charge diffusion pole. The black crosses eventually become the hydrodynamic sound poles, with $\Gamma = 1/(8 \pi T)$.

\begin{figure}[t!]
\begin{center}
\begin{subfigure}{0.49\textwidth}
	\includegraphics[width=\textwidth]{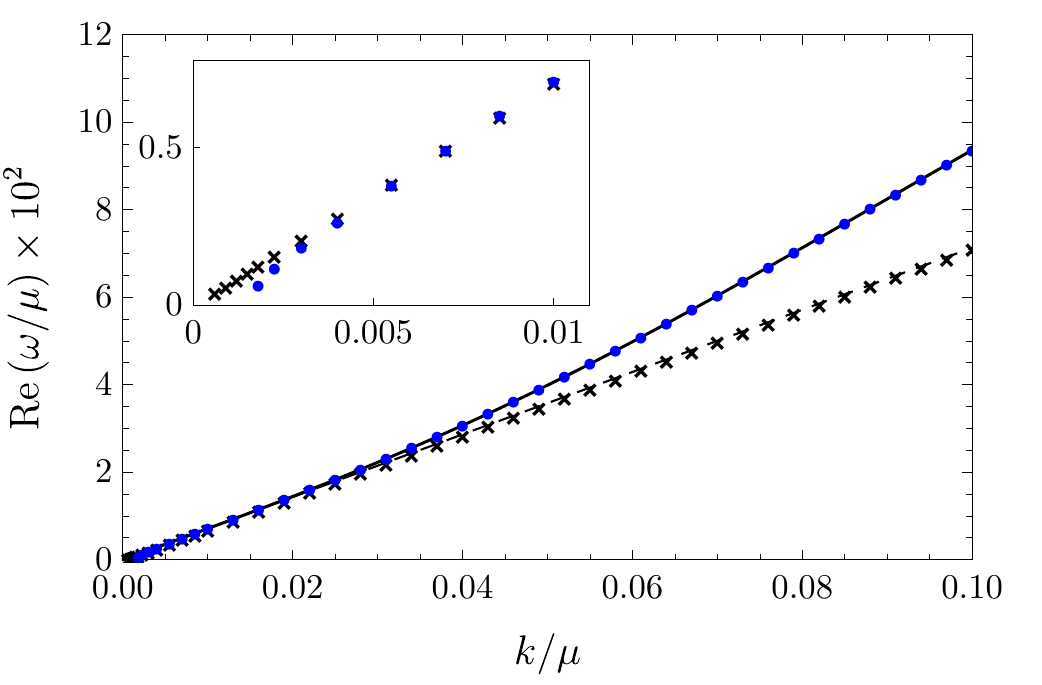}
	\caption{}
\end{subfigure}
\begin{subfigure}{0.49\textwidth}
	\includegraphics[width=\textwidth]{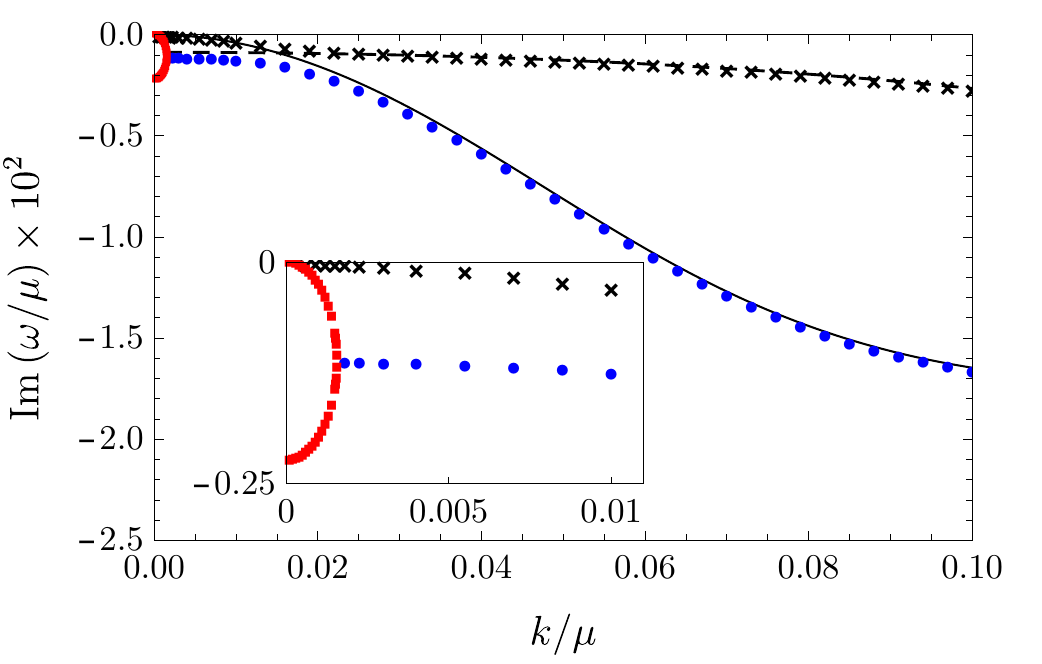}
	\caption{}
\end{subfigure}
\caption{\label{fig:dispersionsmalltau} Dispersion relations of the four highest poles for $\tau=10^{-4}$, $\talpha=1$, $T/\mu = 10^{-2}$, and $10^{-4} \leq k/\mu \leq 0.1$. (a) $\textrm{Re}\left(\omega\right)/\mu$ and (b) $\textrm{Im}\left(\omega\right)/\mu$, each versus $k/\mu$.   The solid and dashed black lines show the poles in $G_{tt}$ and $G_J$ in the probe limit, respectively. The black crosses follow the probe HZS dispersion for large $k/\mu$, with $\Gamma$ from eq.~\eqref{probehzsatt}, and the hydrodynamic sound dispersion for small $k/\mu$, with $\Gamma$ in eq.~\eqref{eq:sound_mode_hydro}. At large $k/\mu$ the blue dots have the dispersion of the poles in $G_{tt}$, with $\textrm{Re}\left(\omega\right) = \pm k$, but at $k/\mu \approx 0.02$ have $\textrm{Re}\left(\omega\right) = \pm k/\sqrt{2}$, and for $k/\mu \lesssim 0.02$ they drop to $\textrm{Re}\left(\omega\right) = 0$ around $k/\mu \approx 2\times 10^{-3}$, as shown in the inset of (a). They then split into two purely imaginary poles, the red squares, as shown in the inset of (b). One of these moves up the imaginary axis and becomes the charge diffusion pole, with the probe limit dispersion in eqs.~\eqref{chargediff} and~\eqref{eq:probe_diffusion}.}
\end{center}
\end{figure}

Fig.~\ref{fig:dispersionsmalltau} shows dispersion relations for $\tau=10^{-4}$, $\talpha=1$, $T/\mu=10^{-2}$, and $10^{-4} \leq k/\mu \leq 0.1$. The two poles with least negative imaginary part (the black crosses) follow the probe HZS dispersion in eqs.~\eqref{soundisp} and~\eqref{probehzsatt} to excellent approximation everywhere in this regime of $k/\mu$. The next two highest poles (the blue dots) have relativistic dispersion $\mathrm{Re}\left(\omega\right) = k$ for large $k/\mu$, but upon decreasing to $k/\mu \approx 0.02$ they have $\mathrm{Re}\left(\omega\right) \approx k/\sqrt{2}$, suggesting they have become an \textit{additional} pair of sound poles. However, as $k/\mu$ continues decreasing to $k/\mu \lesssim 0.02$, these two poles meet on the imaginary axis and split into two purely imaginary poles (the red squares), one of which moves up the imaginary axis and becomes the hydrodynamic diffusion pole, with the probe limit dispersion in eqs.~\eqref{chargediff} and~\eqref{eq:probe_diffusion}.

Fig.~\ref{fig:dispersionsmalltau} will be the only plot of dispersion relations that we present. However, in subsequent cases we have calculated dispersion relations, which we use to identify poles as HZS, relativistic, hydrodynamic sound, or hydrodynamic charge diffusion.\footnote{To clarify terminology: in sec.~\ref{sec:soundatt} we will show that in fact $\Gamma$ takes the hydrodynamic form, $\Gamma = \frac{1}{2} \frac{\eta}{\varepsilon+P}$, for all $T/\mu$, and thus could be called ``hydrodynamic'' for all $T/\mu$. However, throughout the paper we instead use $\Gamma$'s limiting values to distinguish sound as HZS or hydrodynamic. For example, if $\Gamma$ approaches the probe value in eq.~\eqref{probehzsatt} as $T/\mu \to 0$ then we call the poles HZS, whereas if $\Gamma \to 1/(8 \pi T)$ as $T/\mu \to \infty$  then we call the poles hydrodynamic sound. Hopefully the meaning of ``hydrodynamic'' will always be clear by the context.} Crucially, for all $\tau$, $\talpha$, and $T/\mu$, we have found that the speed of sound, whether HZS or hydrodynamic, always takes the conformal value, $v=1/\sqrt{2}$, as in other back-reacted models~\cite{Davison:2011uk,Davison:2013uha}.

The main effect of small back-reaction $\tau=10^{-4}$, compared to the probe limit $\tau=0$, is clearly a ``pole switch'' in the crossover. In the probe limit, the two relativistic poles crossover to the hydrodynamic sound poles, while the two HZS poles trace semicircles and collide on the imaginary axis, producing two purely imaginary poles, one of which becomes the charge diffusion pole. However, with a small amount of back-reaction the two relativistic poles at first move similarly to the probe limit case, but then change direction and become the two poles tracing semicircles and eventually giving rise to the charge diffusion pole. Meanwhile the HZS crosses over directly to the hydrodynamic sound poles, with no aparent change in $\textrm{Re}\left(\omega\right)$ and only slight change in $\textrm{Im}\left(\omega\right)$. Such sound pole behavior is similar to the crossover in AdS-RN~\cite{Davison:2011uk}. Nevertheless, despite the pole switch we could still define a precise moment the crossover occurs in the same way as the probe limit~\cite{Davison:2011ek}, when the two poles collide on the imaginary axis and produce the charge diffusion pole.

\begin{figure}[t!]
\begin{center}
	\begin{subfigure}{0.49\textwidth}
		\includegraphics[width=\textwidth]{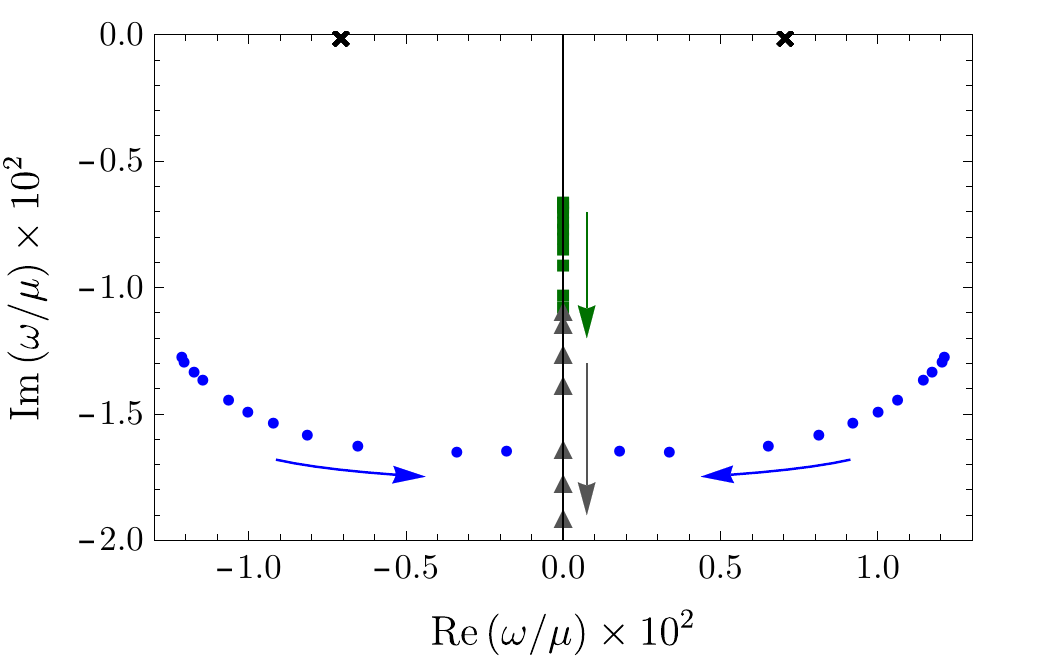}
		\caption{\label{plane_tau_0p001_low_T} $T/\mu=1.25\times10^{-3}$ to $2.23\times10^{-3}$.}
	\end{subfigure}
	\begin{subfigure}{0.49\textwidth}
		\includegraphics[width=\textwidth]{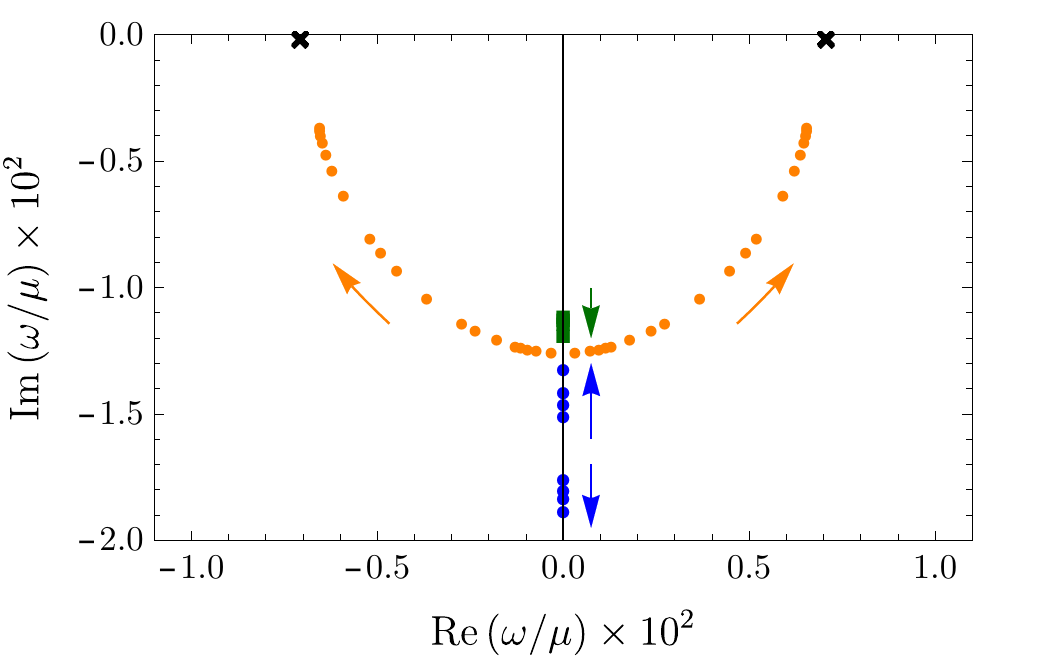}
		\caption{\label{plane_tau_0p001_mid_T} $T/\mu = 2.23\times10^{-3}$ to $10^{-2}$.}
	\end{subfigure}
	\begin{subfigure}{0.49\textwidth}
		\includegraphics[width=\textwidth]{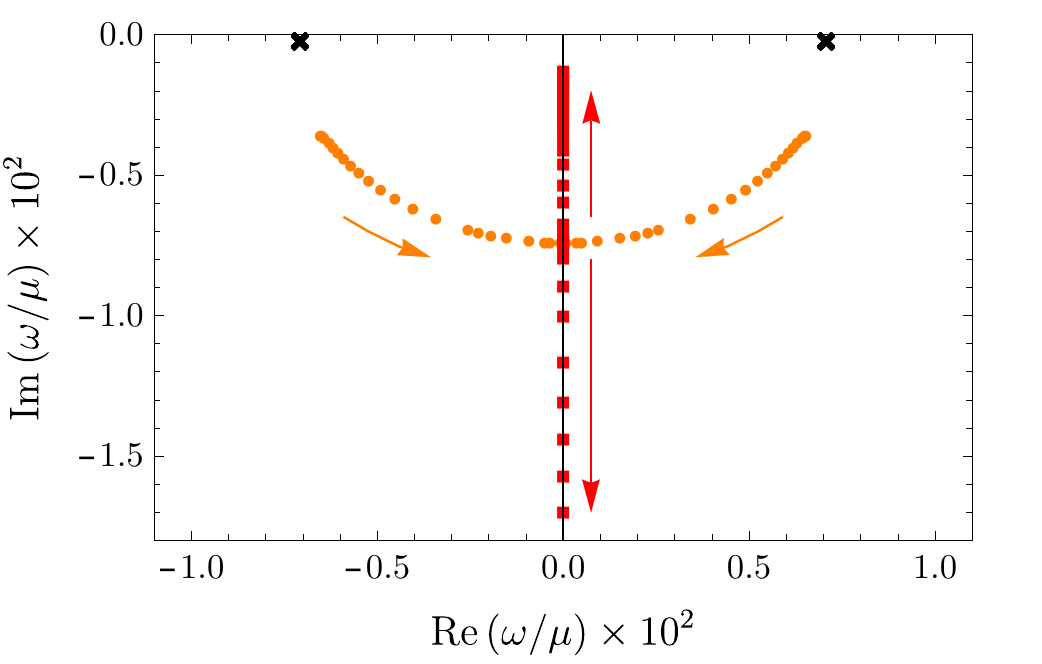}
	\caption{\label{plane_tau_0p001_high_T} \(T/\mu = 0.011\) to \(0.05\).}
	\end{subfigure}
	\caption{\label{plane_tau_0p001} Positions of poles of $G_J$ and $G_{tt}$ in the complex $\omega/\mu$ plane, with $\tau = 10^{-3}$, $\talpha=1$, $k/\mu = 10^{-2}$, and (a) $1.25 \times 10^{-3} \leq T/\mu \leq 2.23 \times 10^{-3}$, (b) $2.23\times10^{-3}\leq T/\mu \leq 10^{-2}$, and (c) $0.011\leq T/\mu \leq 0.05$. We have enhanced $\textrm{Re}\left(\omega/\mu\right)$ and $\textrm{Im}\left(\omega/\mu\right)$ by $10^2$ for clarity. The arrows indicate the movement of poles as $T/\mu$ increases. The pole motion is considerably more complicated than the previous smaller $\tau$ cases, so for detailed descriptions of the poles and their movement, including the color and shape coding, see the accompanying text. (Animated versions of these figures are available on this paper's arxiv page.)}
\end{center}
\end{figure}

\begin{figure}[t!]
\begin{center}
\begin{subfigure}{0.49\textwidth}
\includegraphics[width=\textwidth]{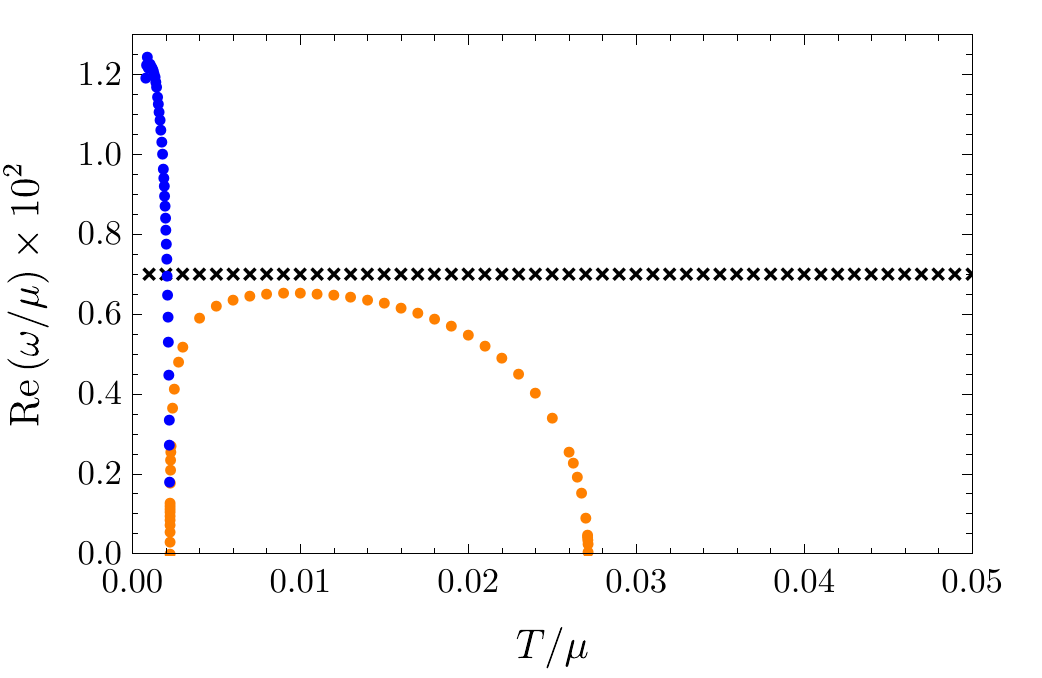}
\caption{\label{realimtau10e3a}}
\end{subfigure}
\begin{subfigure}{0.49\textwidth}
\includegraphics[width=\linewidth]{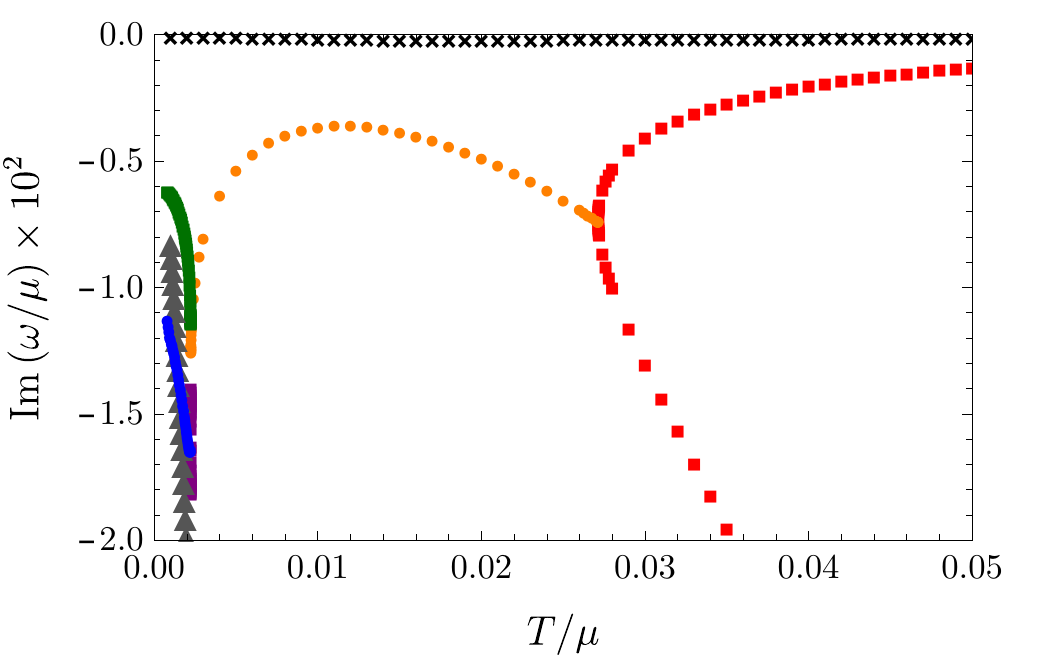}
\caption{\label{realimtau10e3b}}
\end{subfigure}
\caption{\label{realim_tau_0p001} The same data as fig.~\ref{plane_tau_0p001}, but with separate plots for (a) $\textrm{Re}\left(\omega/\mu\right)$ and (b) $\textrm{Im}\left(\omega/\mu\right)$, each enhanced by $10^2$ for clarity, as functions of $T/\mu$. The pole motion is considerably more complicated than the previous smaller $\tau$ cases, so for detailed descriptions of the poles and their movement, including the color and shape coding, see the accompanying text.}
\end{center}
\end{figure}

Fig.~\ref{plane_tau_0p001} shows our numerical results for the poles with larger back-reaction, $\t = 10^{-3}$, still with $\talpha=1$ and $k/\mu = 10^{-2}$, and now for $1.25 \times 10^{-3} \leq T/\mu \leq 0.05$. The arrows again indicate the pole movement as $T/\mu$ increases. (An animated version of fig.~\ref{plane_tau_0p001} is available on this paper's arxiv page.) For clarity, fig.~\ref{realim_tau_0p001} shows the same data as fig.~\ref{plane_tau_0p001}, but with $\textrm{Re}\left(\omega/\mu\right)$ and $\textrm{Im}\left(\omega/\mu\right)$ plotted separately versus $T/\mu$ in figs.~\ref{realimtau10e3a} and~\ref{realimtau10e3b}, respectively.

The crossover with $\tau = 10^{-3}$ is more complicated than with $\tau = 10^{-4}$, so we divide the evolution into three regimes of $T/\mu$. First, fig.~\ref{plane_tau_0p001_low_T} shows the six highest poles for $1.25 \times 10^{-3} \leq T/\mu \leq 2.23 \times 10^{-3}$. At the smallest $T/\mu$ we find two poles with HZS dispersion (black crosses), and then lower in the complex $\omega/\mu$ plane we find two purely imaginary poles (green squares and gray triangles) and two poles with relativistic dispersion (blue dots). As we increase $T/\mu$, the black crosses barely move, while the green squares and gray triangles move down the imaginary axis, and the two blue dots move down and towards the imaginary axis, meeting there at $T/\mu = 2.23\times10^{-3}$. Crucially, they meet below the green square but above the gray triangle. That is a key difference from $\tau =10^{-4}$, where two poles met on the imaginary axis but with no purely imaginary poles above them.

Fig.~\ref{plane_tau_0p001_mid_T} then shows the four highest poles for $2.23\times10^{-3}\leq T/\mu \leq 10^{-2}$. The two poles that met on the imaginary axis split into two purely imaginary poles (still blue dots), one of which moves up while the other moves down. The one moving up collides with the green square at $T/\mu = 2.24\times10^{-3}$ and splits into two poles with non-zero real parts (orange dots), which move away from the imaginary axis and up towards the real axis as $T/\mu$ increases (the U-shape in fig.~\ref{plane_tau_0p001_mid_T}). However at $T/\mu \approx 10^{-2}$ the orange dots stop, reaching their maximum distance from the imaginary axis and highest point in the complex $\omega/\mu$ plane.

Fig.~\ref{plane_tau_0p001_high_T} shows the subsequent evolution for $0.011\leq T/\mu \leq 0.05$ which is in fact similar to the previous cases. The orange dots reverse direction, moving back down into the complex $\omega/\mu$ plane and closer to the imaginary axis, tracing semicircles before colliding on the imaginary axis at $T/\mu\approx 0.027$ and then splitting into two purely imaginary poles (red squares), one of which moves down the imaginary axis while the other moves up and eventually becomes the hydrodynamic charge diffusion pole.

In short, the key difference with $\tau=10^{-3}$, compared to $\tau =10^{-4}$, is that when the two propagating poles (blue dots) hit the imaginary axis a purely imaginary pole is already present on the axis above them. As a result, when they split into two purely imaginary poles, one moving up the axis and one moving down, the one moving up must collide with this ``extra'' imaginary pole. Those two poles then ``pop off'' the imaginary axis and become increasingly long-lived propagating poles (orange dots), until at $T/\mu\approx10^{-2}$ they stop and reverse course. The subsequent evolution is then similar to the previous cases: they trace semicircles until they hit the imaginary axis, producing the charge diffusion pole. As a result, despite the more complicated pole movement at low $T/\mu$, the probe limit definition of the crossover actually remains viable at $\tau=10^{-3}$, and gives a crossover temperature of $T/\mu\approx 0.027$, i.e. the temperature of the \textit{second} pole collision on the imaginary axis.

More generally, we have learned that as $\tau$ increases, purely imaginary poles rise up the imaginary $\omega/\mu$ axis and begin to ``interfere'' with the relativistic poles that collide on the axis. Clearly a critical value of $\tau$ exists, somewhere between $\tau = 10^{-4}$ and $10^{-3}$, where as $\tau$ increases the highest of these purely imaginary poles first has imaginary part equal to that of the colliding poles. We have found this critical value to be $\tau \approx 9 \times 10^{-4}$.

\begin{figure}[h!]
\begin{center}
\begin{subfigure}{0.49\textwidth}
	\includegraphics[width=\textwidth]{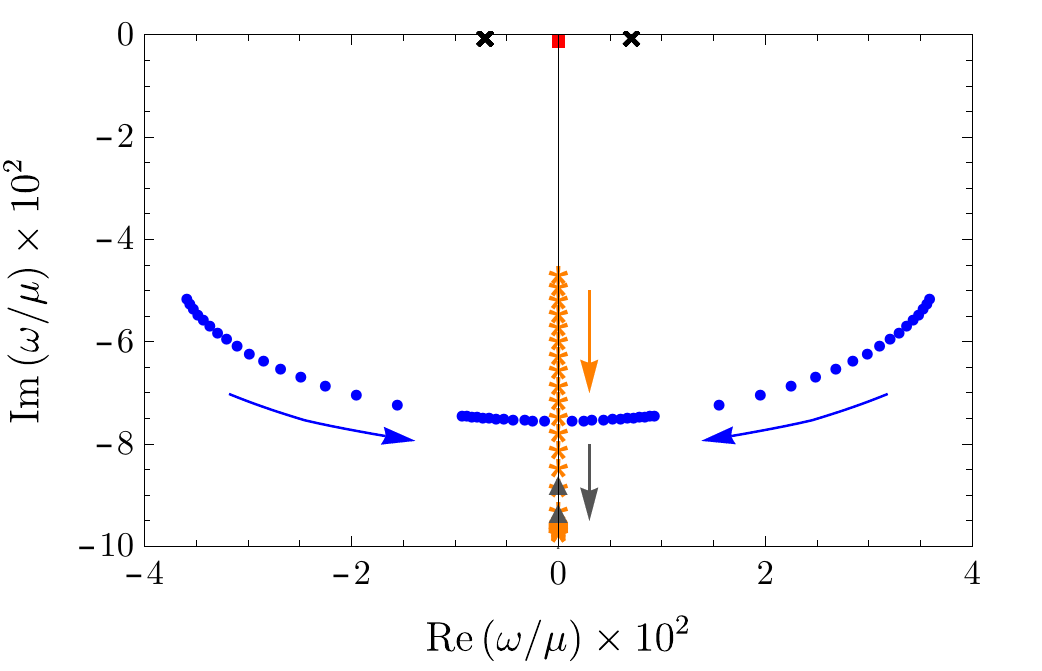}
	\caption{\label{complex_plane_tau0p01_a1_q0p01_t0p005_to_0p0083}$T/\mu = 5\times 10^{-3}$ to $8.3\times 10^{-3}$.}
\end{subfigure}
\begin{subfigure}{0.49\textwidth}
	\includegraphics[width=\textwidth]{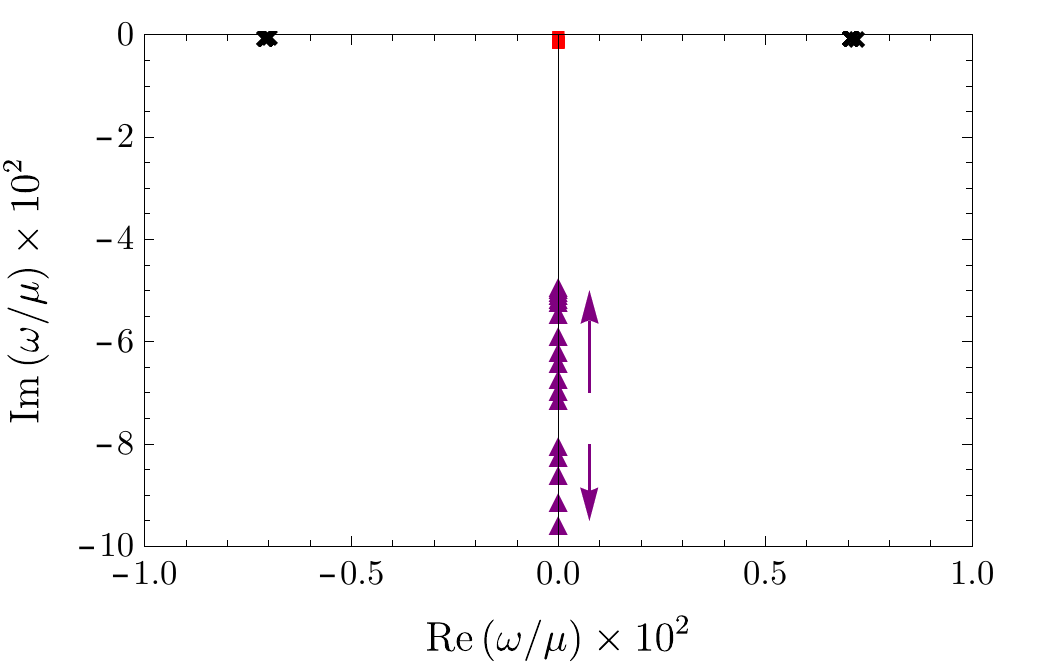}
	\caption{\label{complex_plane_tau0p01_a1_q0p01_t0p0083_to_0p01}$T/\mu =  8.3\times 10^{-3}$ to $10^{-2}$.}
\end{subfigure}
\begin{subfigure}{0.49\textwidth}
	\includegraphics[width=\textwidth]{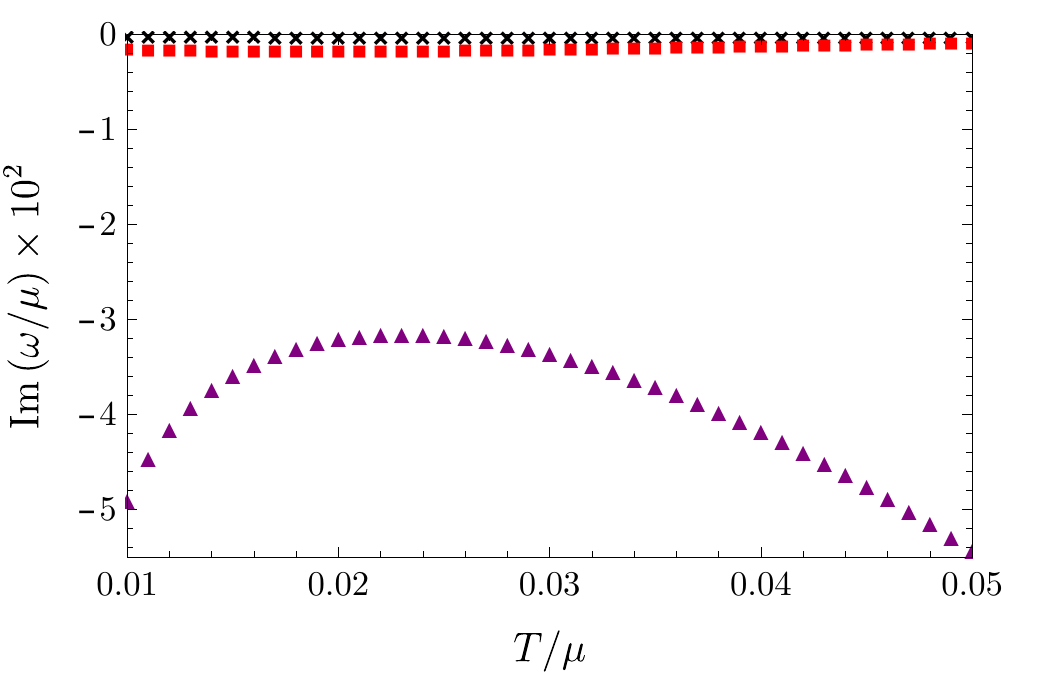}
	\caption{\label{imaginary_tau_0p01}}
\end{subfigure}
\begin{subfigure}{0.49\textwidth}
	\includegraphics[width=\textwidth]{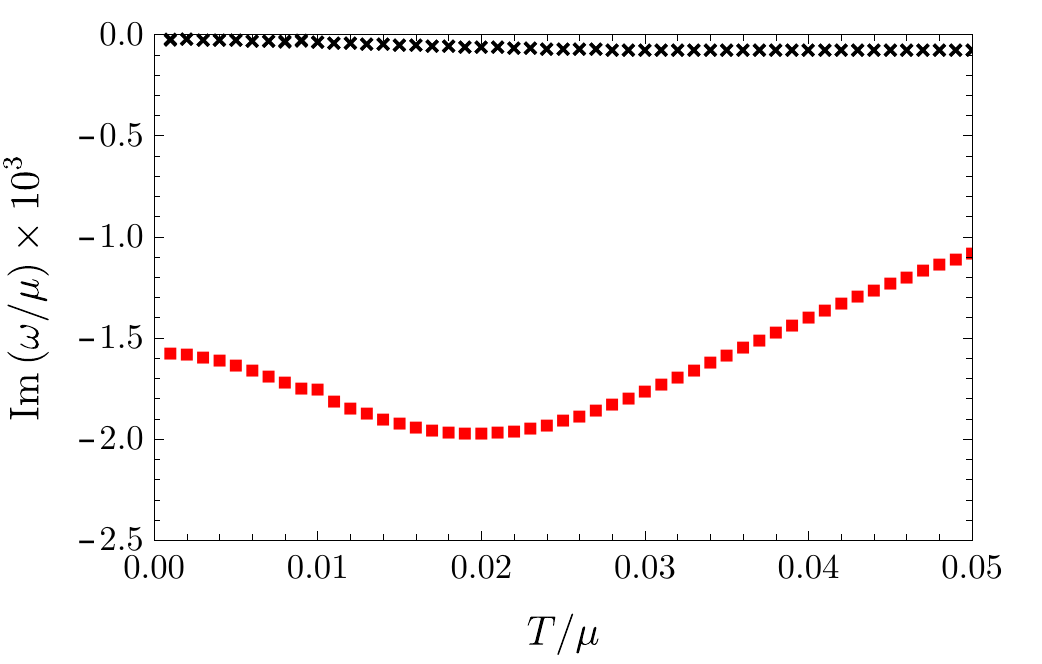}
	\caption{\label{imaginary_tau0p01_a1_q0p01}}
\end{subfigure}
\caption{\label{plane_tau_0p01} Positions of poles of $G_J$ and $G_{tt}$ in the complex $\omega/\mu$ plane with $\tau=10^{-2}$, $\talpha=1$, $k/\mu=10^{-2}$, and (a) $5\times 10^{-3} \leq T/\mu \leq 8.3\times 10^{-3}$ and (b) $8.3\times 10^{-3}\leq T/\mu \leq 10^{-2}$. We have enhanced $\textrm{Re}\left(\omega/\mu\right)$ and $\textrm{Im}\left(\omega/\mu\right)$ by $10^2$ for clarity. Arrows indicate the movement of poles as $T/\mu$ increases. At $T/\mu=5\times 10^{-3}$ we find seven highest poles, the three highest being two HZS poles (black crosses) and a purely imaginary pole (red square), and four lower poles, two purely imaginary (orange and gray triangles), and two with non-zero real parts (blue dots). As $T/\mu$ increases the three highest poles barely move, while the orange and gray triangles move down. The blue dots move down and collide on the imaginary axis, above the orange and gray triangles, and then split into two purely imaginary poles (purple triangles), one moving up the axis and one moving down. However, unlike the previous smaller $\tau$ cases, the one moving up does not become the charge diffusion pole, instead stopping, reversing direction, and moving back down the axis. The three highest poles eventually become the hydrodynamic sound and charge diffusion poles, respectively. (c) $\textrm{Im}\left(\omega/\mu\right)\times 10^2$ versus $T/\mu$ for the four highest poles, showing the upper purple triangle's highest point at $T/\mu \approx 2.4\times 10^{-2}$. (d) Close-up of (c) for the three highest poles, showing how little these move compared to the others. (Animated versions of these figures are available on this paper's arxiv page.)}
\end{center}
\end{figure}

Fig.~\ref{complex_plane_tau0p01_a1_q0p01_t0p005_to_0p0083} shows our numerical results for the pole positions for higher back-reaction, $\tau = 10^{-2}$, still with $\talpha=1$ and $k/\mu=10^{-2}$, and now for $5 \times 10^{-3}\leq T/\mu \leq 8.3\times 10^{-3}$. (An animated version of fig.~\ref{plane_tau_0p01} is available on this paper's arxiv page.) At the smallest $T/\mu$ we again find two poles with HZS dispersion (black crosses) but now also a purely imaginary pole high in the complex $\omega/\mu$ plane (red square). Lower in the complex $\omega/\mu$ plane we find four poles, two purely imaginary (orange and gray triangles) and two with relativistic dispersion (blue dots). As $T/\mu$ increases, the black crosses and red square barely move, while the orange and gray triangles move down the imaginary axis and the blue dots move down and towards the imaginary axis, colliding there at $T/\mu\approx 8.3\times 10^{-3}$, above the orange and gray triangles. Fig.~\ref{complex_plane_tau0p01_a1_q0p01_t0p0083_to_0p01} shows the subsequent movement for $8.3\times 10^{-3}\leq T/\mu \leq 10^{-2}$, where the poles that collided split into two purely imaginary poles (purple triangles), one of which moves up the axis while the other moves down. However, \textit{both} remain below the red square.

Indeed, as $T/\mu$ continues increasing, to $10^{-2}\leq T/\mu \leq 0.05$, fig.~\ref{imaginary_tau_0p01} shows $\textrm{Im}\left(\omega/\mu\right)$ for the black crosses, red square, and purple triangle. The purple triangle reaches a highest point around $T/\mu \approx 0.024$, well below the red square, before turning around and descending back down the imaginary axis. Fig.~\ref{imaginary_tau0p01_a1_q0p01} shows a close-up of $\textrm{Im}\left(\omega/\mu\right)$ for the black crosses and red square for $0\leq T/\mu \leq 0.05$. In that $T/\mu$ range, the black crosses decrease from $\textrm{Im}\left(\omega/\mu\right)\approx-0.05$ only to $\approx-0.1$ while the red square decreases from $\textrm{Im}\left(\omega/\mu\right)\approx-1.6$ down to a minimum of $\approx-2$ at $T/\mu\approx 0.02$ before rising again to $\textrm{Im}\left(\omega/\mu\right)\approx-1.1$. As $T/\mu$ increases, the black crosses and red square eventually become the hydrodynamic sound and charge diffusion poles, respectively.

In short, the evolution with $\tau=10^{-2}$ is qualitatively different from that with smaller $\tau$. With $\tau=10^{-2}$ we find two propagating poles and a single purely imaginary pole relatively high in the complex $\omega/\mu$ plane, and then lower in the complex $\omega/\mu$ plane two poles that collide on the imaginary axis and split into two purely imaginary poles, one moving up the axis and one moving down, where the one moving up eventually stops, turns around, and moves back down, never becoming the highest purely imaginary pole. The two highest propagating poles cross over from HZS to hydrodynamic sound, and the highest purely imaginary pole becomes the hydrodynamic charge diffusion pole at sufficiently high $T/\mu$.

Recalling that as $\tau$ increases purely imaginary poles move farther up the imaginary axis, clearly a second critical value of $\tau$ exists, somewhere between $\tau=10^{-3}$ and $10^{-2}$, where the highest purely imaginary pole no longer moves down and ``interferes'' with the colliding poles, and instead crosses over directly to the hydrodynamic charge diffusion pole. We have found this critical value to be $\t \approx 3.2 \times 10^{-3}$. Moreover, we have sampled various $\t \gtrsim 3.2 \times 10^{-3}$, including values $\tau>10^{-2}$, and found behavior qualitatively similar to $\tau=10^{-2}$.

Clearly for $\tau > 3.2 \times 10^{-3}$ we cannot use the probe limit definition of the crossover, since at no point do poles collide on the imaginary axis and produce the hydrodynamic charge diffusion pole. Instead, the three highest poles behave similarly to the AdS-RN case, namely they move very little as $T/\mu$ increases. In sec.~\ref{sec:spectral_functions} we will show that the AdS-RN definition of the crossover, via a transfer of dominance in peaks of $\rho_J$, is viable for $\tau \gtrsim 3.2 \times 10^{-3}$.

\subsubsection{Changing $\talpha$}
\label{alphachange}

We will now consider $\tau=10^{-4}$ and $\tau=10^{-2}$ and in each case decrease $\talpha$, with $k/\mu=10^{-2}$. In $\sdbi$ decreasing $\talpha$ at fixed $\tau$ suppresses higher-order terms in $F_{MN}$, but is not exactly the Maxwell limit, which requires $\talpha \to 0$ with $\tau \talpha^2$ fixed, so that $\tau \propto \talpha^{-2}$ diverges. Instead, as discussed in sec.~\ref{results}, fixing $\tau$ and decreasing $\talpha$ with fixed $k/\mu$ is more akin to the probe limit: higher-order terms in $F_{MN}$ are suppressed, while the leading Maxwell term has coefficient $\tau \talpha^2$, so fixing $\tau$ and decreasing $\talpha$ should be qualitatively similar to decreasing $\tau$ with fixed $\talpha$. Indeed, that intuition turns out to be correct.

As also mentioned in sec.~\ref{results}, due to the gravity theory's scaling symmetry $\alpha \to \lambda \, \alpha$ and $F_{MN} \to \lambda^{-1}\,F_{MN}$, for a given $\tau$, fixing $k/\mu$ and \textit{decreasing} $\talpha$ is equivalent to fixing $\talpha$ and \textit{increasing} $k/\mu$. For a given $\tau$, the following results thus provide information about dispersion relations at fixed $\talpha=1$. Indeed, as $k/\mu$ increases higher-order terms in $k/\mu$ will alter the sound poles' $\textrm{Re}\left(\omega\right)$ in dramatic ways.

\begin{figure}[h!]
\begin{center}
\begin{subfigure}{0.49\textwidth}
\includegraphics[width=\textwidth]{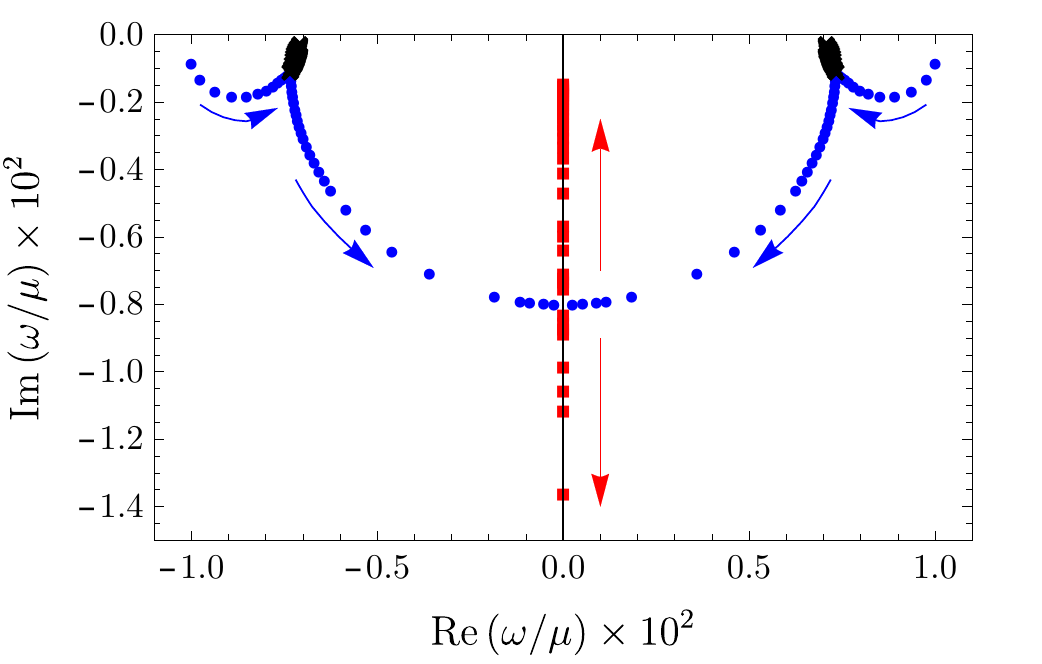}
\caption{\label{figa01}$\tau=10^{-4}$, $\talpha=0.1$, $T/\mu = 5 \times 10^{-4}$ to $2\times10^{-2}$.}
\end{subfigure}
\begin{subfigure}{0.49\textwidth}
\includegraphics[width=\textwidth]{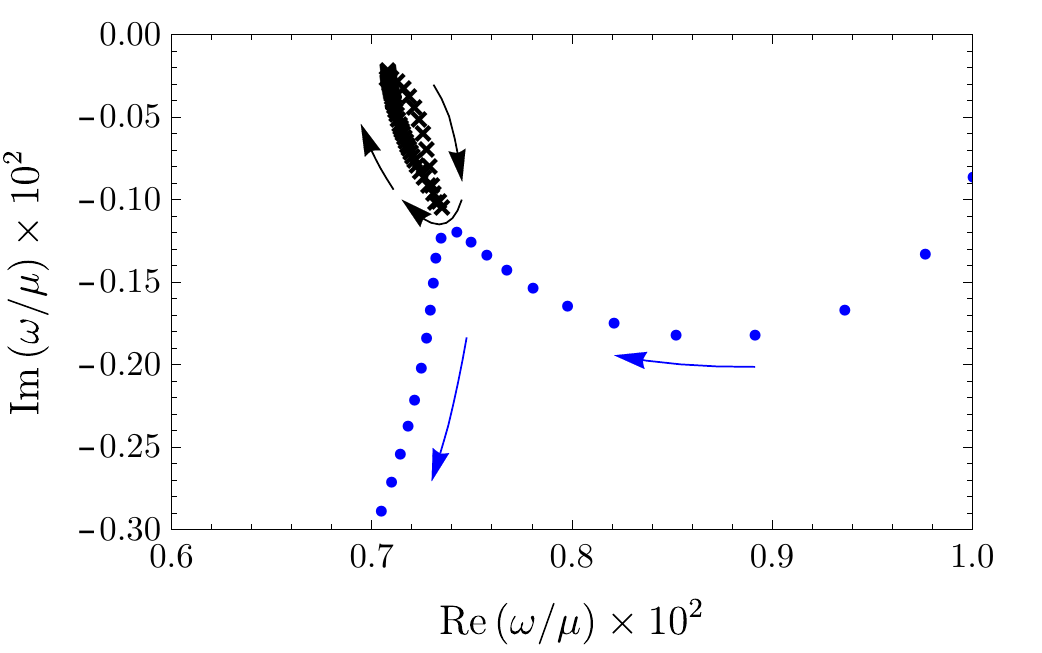}
\caption{\label{figa01cu}Close-up of (a).}
\end{subfigure}
\begin{subfigure}{0.49\textwidth}
\includegraphics[width=\textwidth]{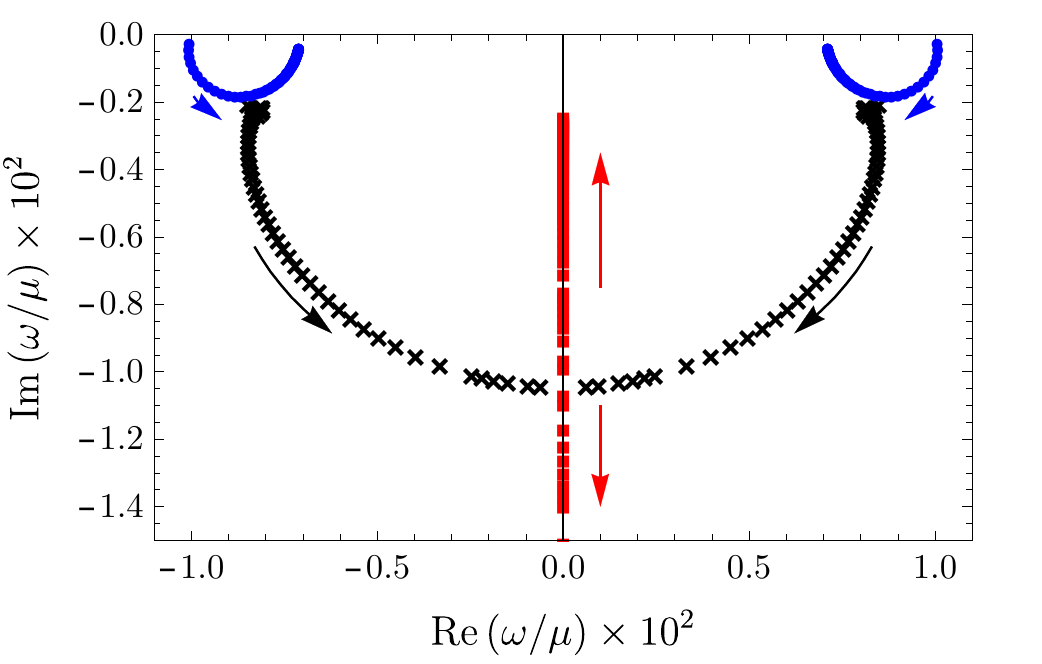}
\caption{\label{figa001}$\tau=10^{-4}$, $\talpha=10^{-2}$, $T/\mu = 2 \times 10^{-4}$ to $10^{-2}$.}
\end{subfigure}
\begin{subfigure}{0.49\textwidth}
\includegraphics[width=\textwidth]{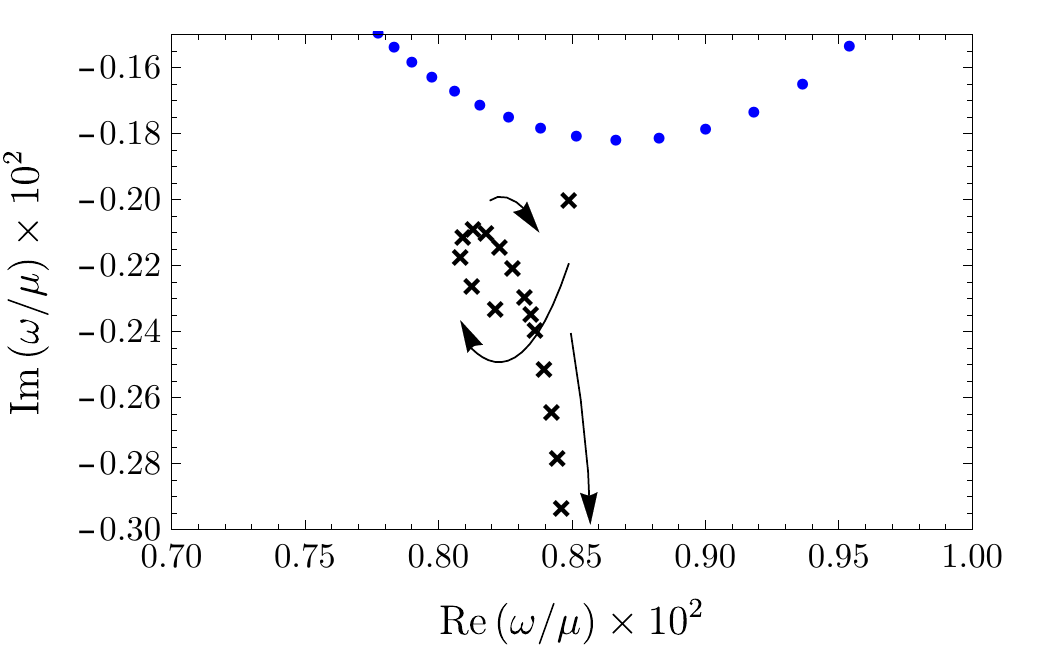}
\caption{\label{figa001cu} Close-up of (c).}
\end{subfigure}
\caption{\label{plane_tau0p0001_changing_alpha} Positions of poles of $G_J$ and $G_{tt}$ in the complex $\omega/\mu$ plane with $\tau=10^{-4}$ for different $\talpha$. We have enhanced $\textrm{Re}\left(\omega/\mu\right)$ and $\textrm{Im}\left(\omega/\mu\right)$ by $10^2$ for clarity. The arrows indicate the movement of poles as $T/\mu$ increases. (a) and (b) $\talpha=0.1$ and $5\times10^{-4} \leq T/\mu \leq 2\times10^{-2}$. At $T/\mu=5\times10^{-4}$ we find four poles, two HZS poles (black crosses) and two relativistic poles (blue dots). As $T/\mu$ increases the black crosses execute a loop-the-loop and eventually become hydrodynamic sound poles, while the blue dots move down, then back up, and then down and towards the imaginary axis, where they collide and split into two imaginary poles, one of which becomes the hydrodynamic charge diffusion pole. (c) and (d) $\talpha=10^{-2}$ and $2\times10^{-4} \leq T/\mu \leq 10^{-2}$. At $T/\mu=2\times10^{-4}$ we again find four poles, two HZS (black crosses) and two relativistic (blue dots). However now as $T/\mu$ increases the blue dots move down and then back up, becoming hydrodynamic sound, while the black crosses execute a loop-the-loop and then move down and towards the imaginary axis, where they collide and split into two imaginary poles, one of which becomes the hydrodynamic charge diffusion pole.}
\end{center}
\end{figure}

\begin{figure}[h!]
\begin{center}
\begin{subfigure}{0.49\textwidth}
\includegraphics[width=\textwidth]{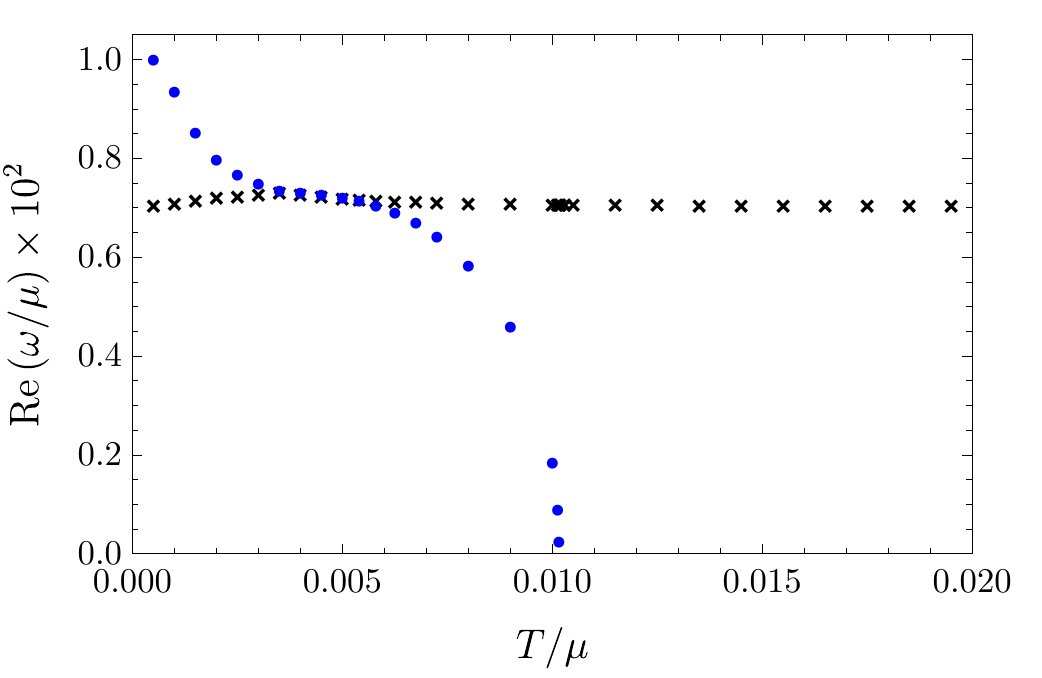}
\caption{\label{figa01re}$\tau=10^{-4}$ and $\talpha=0.1$.}
\end{subfigure}
\begin{subfigure}{0.49\textwidth}
\includegraphics[width=\textwidth]{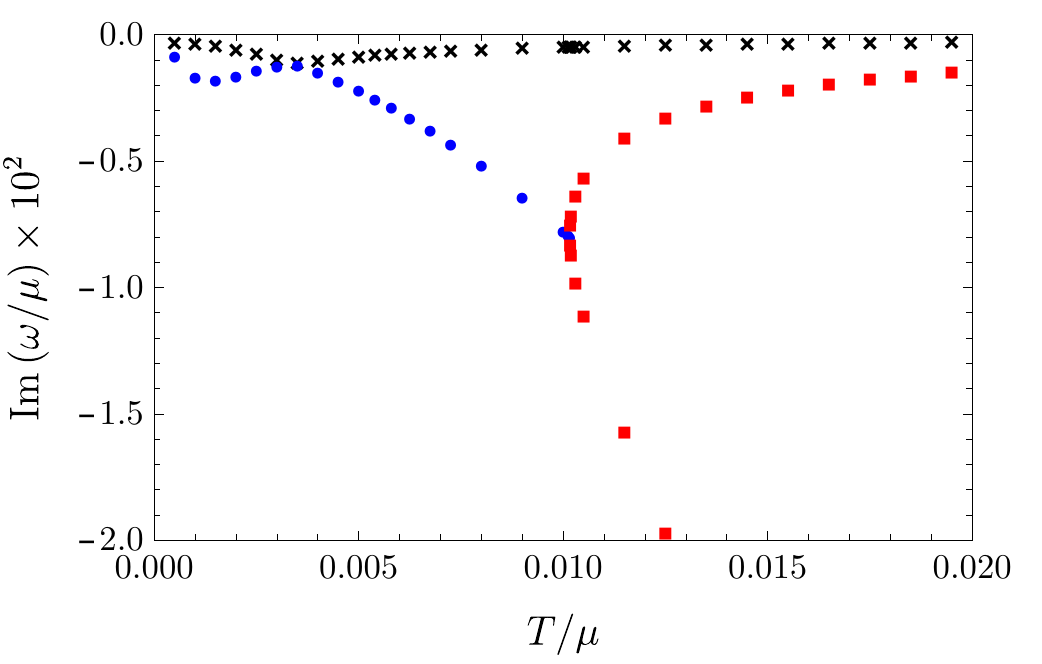}
\caption{\label{figa01im}$\tau=10^{-4}$ and $\talpha=0.1$.}
\end{subfigure}
\begin{subfigure}{0.49\textwidth}
\includegraphics[width=\textwidth]{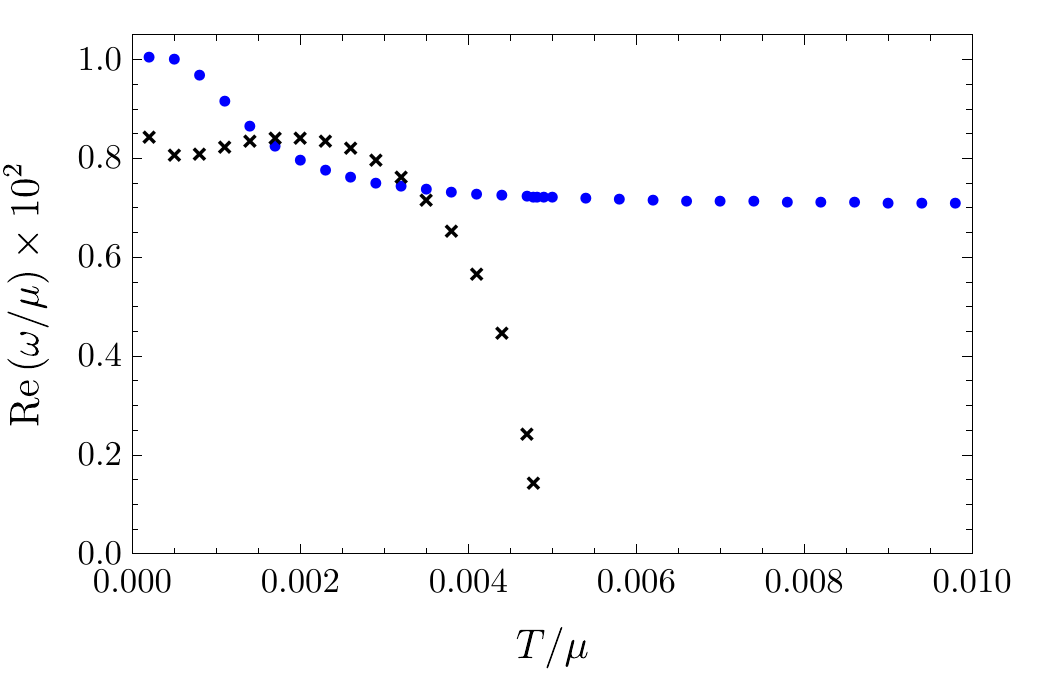}
\caption{\label{figa001re}$\tau=10^{-4}$ and $\talpha=10^{-2}$.}
\end{subfigure}
\begin{subfigure}{0.49\textwidth}
\includegraphics[width=\textwidth]{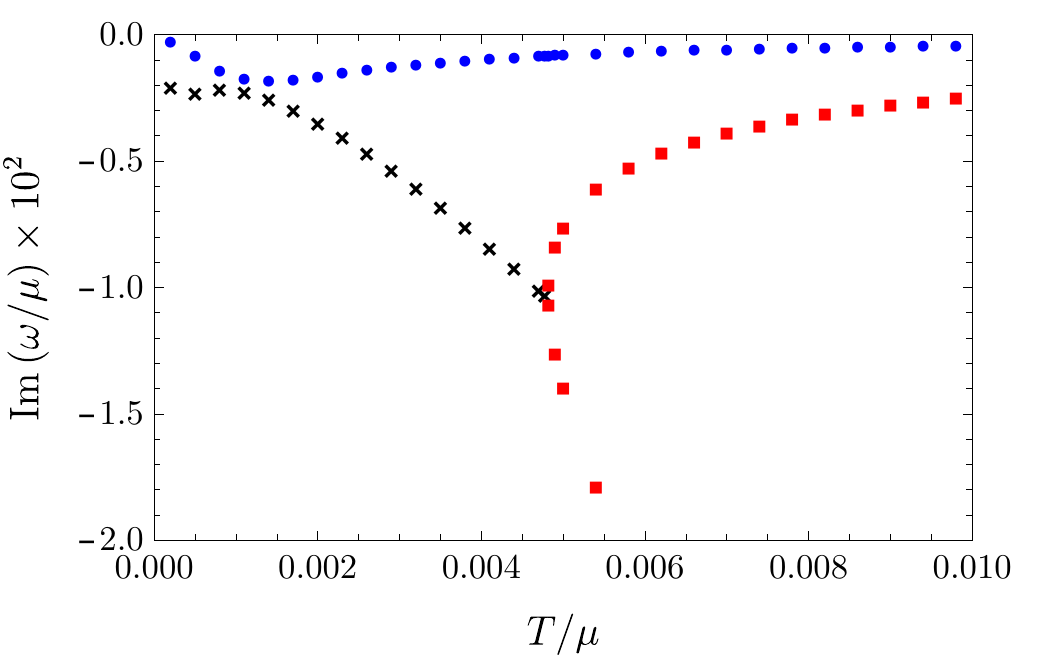}
\caption{\label{figa001im} $\tau=10^{-4}$ and $\talpha=10^{-2}$.}
\end{subfigure}
\caption{\label{plane_tau0p0001_changing_alpha_re_im} The same data as fig.~\ref{plane_tau0p0001_changing_alpha}, but with separate plots for $\textrm{Re}\left(\omega/\mu\right)$ and $\textrm{Im}\left(\omega/\mu\right)$, each enhanced by $10^2$ for clarity, versus $T/\mu$. The color and shape coding are the same as in fig.~\ref{plane_tau0p0001_changing_alpha}. (a) $\textrm{Re}\left(\omega/\mu\right)$ and (b) $\textrm{Im}\left(\omega/\mu\right)$ for $\tau=10^{-4}$ and $\talpha=0.1$. (c) $\textrm{Re}\left(\omega/\mu\right)$ and (d) $\textrm{Im}\left(\omega/\mu\right)$ for $\tau=10^{-4}$ and $\talpha=10^{-2}$.}
\end{center}
\end{figure}

Figs.~\ref{planesmalltau} and~\ref{realimsmalltau} showed our numerical results for the poles in the complex $\omega/\mu$ plane for $\tau=10^{-4}$, $k/\mu=10^{-2}$, and $\talpha=1$. Fig.~\ref{plane_tau0p0001_changing_alpha} shows our numerical results for the same $\tau =10^{-4}$ and $k/\mu=10^{-2}$, but now with $\talpha=0.1$ and $10^{-2}$. Fig.~\ref{plane_tau0p0001_changing_alpha_re_im} shows the same data as fig.~\ref{plane_tau0p0001_changing_alpha} but with $\textrm{Re}\left(\omega/\mu\right)$ and $\textrm{Im}\left(\omega/\mu\right)$ plotted separately versus $T/\mu$ for clarity.

In particular, fig.~\ref{figa01} shows our numerical results for $\talpha=0.1$, $\tau=10^{-4}$, $k/\mu=10^{-2}$, and $5 \times 10^{-4}\leq T/\mu \leq 2\times10^{-2}$. At $T/\mu = 5 \times 10^{-4}$ the four highest poles include two HZS poles (black crosses) and two relativistic poles (blue dots).  As $T/\mu$ increases, the black dots move up and then back down in a ``loop-the-loop,'' eventually becoming the hydrodynamic sound poles. Meanwhile, the relativistic poles move down, up, and then down again, all the while moving towards the imaginary axis and eventually colliding there at $T/\mu=0.010$. They then split into two purely imaginary poles (red squares), one moving up the imaginary axis while the other moves down, where the one moving up eventually becomes the hydrodynamic charge diffusion pole. Fig.~\ref{figa01cu} shows a close-up of a black cross's loop-the-loop. Aside from these loop-the-loops, the $\tau=10^{-4}$ crossover with $\talpha=0.1$ is very similar to $\talpha=1$ in fig.~\ref{plane_10Em4}.

Fig.~\ref{figa001} shows our numerical results for $\talpha=10^{-2}$, $\tau=10^{-4}$, $k/\mu=10^{-2}$, and $2 \times 10^{-4} \leq T/\mu \leq 0.02$. At $T/\mu=2 \times 10^{-4}$ the four highest poles again include two HZS poles (black crosses) and two relativistic poles (blue dots). As $T/\mu$ increases the two relativistic poles move down and then up, all the while moving closer to the imaginary axis, and eventually become the hydrodynamic sound poles. Meanwhile the HZS poles perform a loop-the-loop and then move down and towards the imaginary axis, approximately tracing semi-circles, before colliding on the axis at $T/\mu=4.8 \times 10^{-3}$. They then split into two purely imaginary poles (red squares), one moving up the imaginary axis and one moving down, where the one moving up eventually becomes the hydrodynamic charge diffusion pole. In short, aside from the loop-the-loop, the $\tau=10^{-4}$ crossover with $\talpha=10^{-2}$ is very similar to the probe limit $\tau=0$ with $\talpha=1$ in fig.~\ref{plane_probe} (although now the poles are in both $G_J$ and $G_{tt}$).

For $\tau=10^{-4}$, clearly a change in the crossover occurs as $\talpha$ decreases: when $\talpha=0.1$ HZS crosses over to hydrodynamic sound, whereas when $\talpha=10^{-2}$ the relativistic poles cross over to hydrodynamic sound. A critical value of $\talpha$ thus exists, between $\talpha=0.1$ and $10^{-2}$, where the change in crossover occurs. We have found the critical value to be $\talpha \approx 0.07$.

In short, for fixed $k/\mu$ we find that in general, aside from the loop-the-loops, fixing $\tau$ and decreasing $\talpha$ is similar to fixing $\talpha$ and decreasing $\tau$, as advertised.

\begin{figure}[t!]
\begin{center}
	\begin{subfigure}{0.49\textwidth}
		\includegraphics[width=\textwidth]{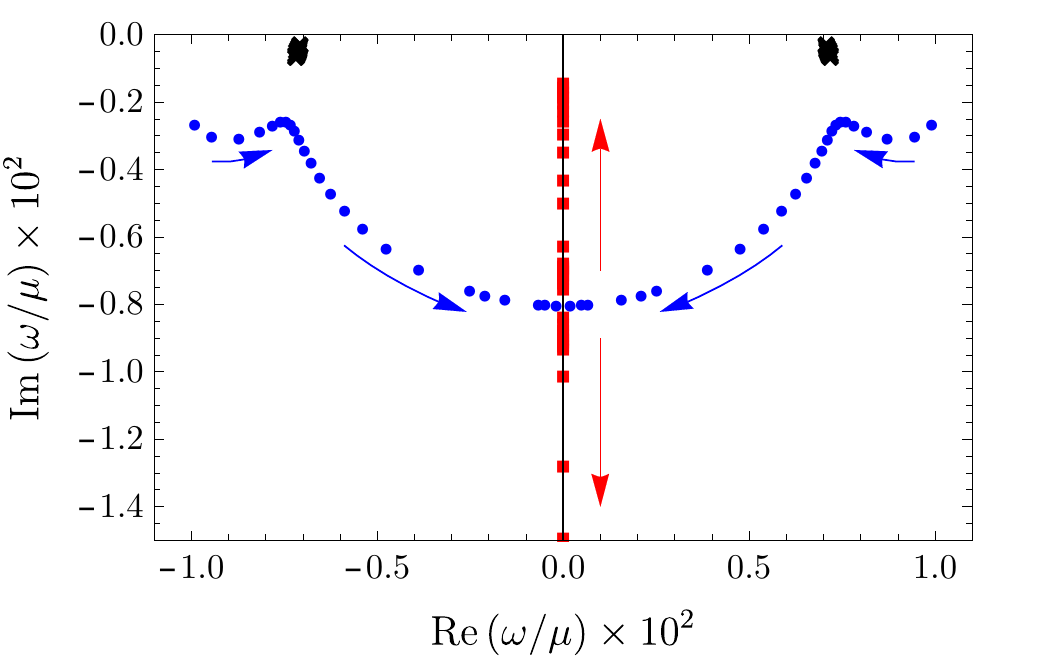}
		\caption{\label{plane_tau_0p01_a0p1} $\tau=10^{-2}$, $\talpha=0.1$, $T/\mu=5\times10^{-4}$ to $2\times10^{-2}$.}
	\end{subfigure}
	\begin{subfigure}{0.49\textwidth}
		\includegraphics[width=\textwidth]{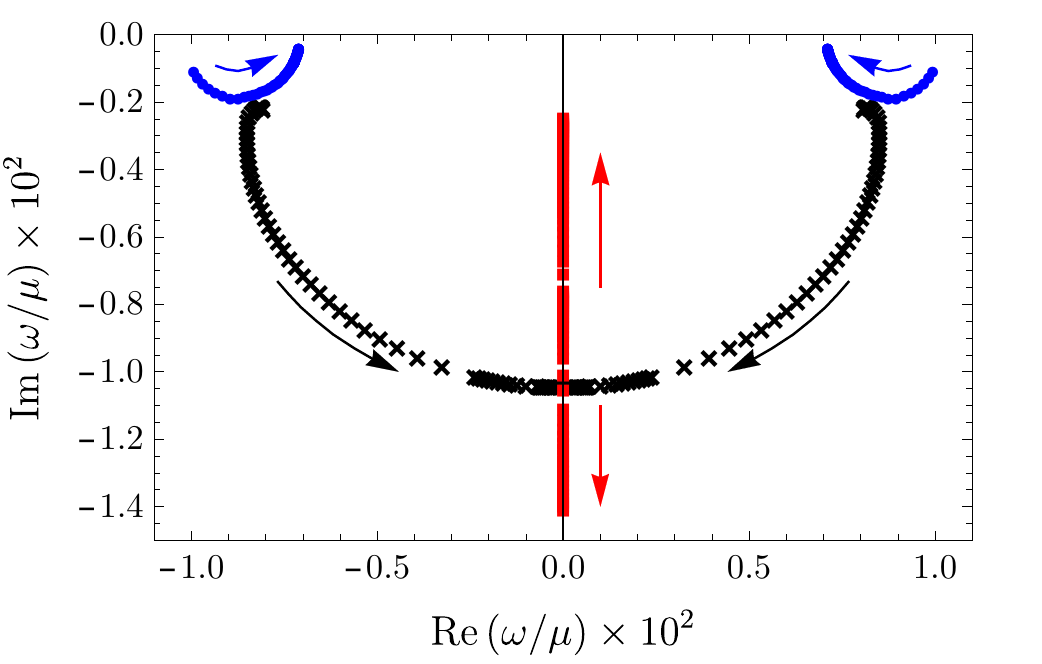}
		\caption{\label{plane_tau_0p01_a0p01} $\tau=10^{-2}$, $\talpha=10^{-2}$, $T/\mu = 6\times10^{-4}$ to $10^{-2}$.}
	\end{subfigure}
	\begin{subfigure}{0.49\textwidth}
		\includegraphics[width=\textwidth]{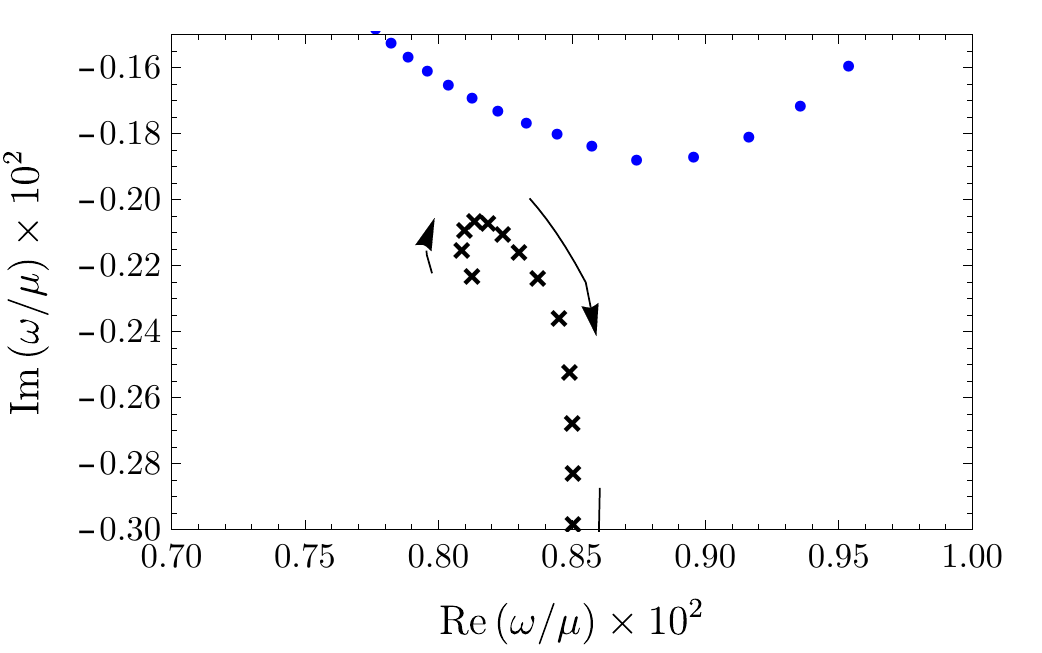}
	\caption{\label{plane_tau_0p01_a0p01_cu} Close-up of (b).}
	\end{subfigure}
	\caption{\label{plane_tau_0p01_a0p1_a0p01} Positions of poles of $G_J$ and $G_{tt}$ in the complex $\omega/\mu$ plane with $\tau=10^{-2}$ for different $\talpha$. We have enhanced $\textrm{Re}\left(\omega/\mu\right)$ and $\textrm{Im}\left(\omega/\mu\right)$ by $10^2$ for clarity. The arrows indicate the movement of poles as $T/\mu$ increases. (a) and (b) $\talpha=0.1$ and $5\times10^{-4} \leq T/\mu \leq 2\times10^{-2}$. At $T/\mu=5\times10^{-4}$ we find four poles, two HZS poles (black crosses) and two relativistic poles (blue dots). As $T/\mu$ increases the black crosses move very little but eventually become hydrodynamic sound poles, while the blue dots move down, then up, and then down and towards the imaginary axis, where they collide and split into two imaginary poles, one of which becomes the hydrodynamic charge diffusion pole. (c) and (d) $\talpha=10^{-2}$ and $6\times10^{-4} \leq T/\mu \leq 10^{-2}$. At $T/\mu=6\times10^{-4}$ we again find four poles, two HZS (black crosses) and two relativistic (blue dots). However now as $T/\mu$ increases the blue dots move down and then back up, becoming hydrodynamic sound, while the black crosses execute a loop-the-loop and then move down and towards the imaginary axis, where they collide and split into two imaginary poles, one of which becomes the hydrodynamic charge diffusion pole.}
\end{center}
\end{figure}

Crucially, for $\tau=10^{-4}$ and both $\talpha=0.1$ and $10^{-2}$, the probe limit definition of the crossover is viable: in both cases poles collide on the imaginary axis, producing the hydrodynamic charge diffusion pole. However, when $\tau \gtrsim 3.2 \times 10^{-3}$ and $\talpha=1$ the probe limit definition of crossover was not viable, so in that case we expect decreasing $\talpha$ will restore the collision of poles and make the probe limit definition viable again. Fig.~\ref{plane_tau_0p01_a0p1} shows our numerical results for $\tau=10^{-2}$, $\talpha=0.1$, $k/\mu=10^{-2}$, and $5\times10^{-4}\leq T/\mu \leq 10^{-2}$, which confirm this expectation. For $T/\mu=5\times10^{-4}$ the four highest poles are two HZS poles (black crosses) and two poles with relativistic dispersion (blue dots). As $T/\mu$ increases, the HZS poles move very little, but eventually cross over to hydrodynamic sound. Meanwhile the blue dots move down, up, and then down again, all while moving closer to the imaginary axis, eventually colliding there at $T/\mu =9.8\times10^{-3}$. They then split into two purely imaginary poles (red squares), one moving up and one moving down, where the latter becomes the hydrodynamic charge diffusion pole.

The behavior is thus qualitatively similar to the $\tau=10^{-4}$ and $\talpha=1$ case in fig.~\ref{plane_10Em4}. In other words, once again, fixing $\tau$ and decreasing $\talpha$ is qualitatively similar to fixing $\talpha$ and decreasing $\tau$. In particular, the probe limit definition of the crossover is viable, in contrast to the $\tau=10^{-2}$ and $\talpha=1$ case in fig.~\ref{plane_tau_0p001}. Indeed, for fixed $\tau=10^{-2}$ and decreasing $\talpha$, clearly a critical $\talpha$ exists where the collision of poles occurs again, making the probe limit definition of the crossover viable. We find the critical value is $\talpha\approx 0.46$. In fact, if we start from $\tau=10^{-2}$ and $\talpha=1$ and then decrease $\talpha$, we find that the second-highest purely imaginary pole (the highest purple triangle in fig.~\ref{complex_plane_tau0p01_a1_q0p01_t0p0083_to_0p01}) reaches a higher and higher maximum, and eventually collides with the charge diffusion pole (red square). As we continue to decrease $\talpha$, this collision leads to complicated pole movement similar to the $\tau=10^{-3}$ and $\talpha=1$ case in fig.~\ref{plane_tau_0p001}: after the two purely imaginary poles collide, they ``pop off'' the axis, moving out and up, becoming propagating modes, but then stop, turn around, and return to the imaginary axis where they split into two purely imaginary poles again. Decreasing $\talpha$ further still leads to a transition similar to that for fixed $\talpha=1$ and decreasing from $\tau=10^{-3}$ to $10^{-2}$, leading to a transition similar to that from fig.~\ref{plane_tau_0p001} to fig.~\ref{plane_10Em4}. We thus find yet again, in still greater detail, that fixing $\tau$ and decreasing $\talpha$ is qualitatively similar to fixing $\talpha$ and decreasing $\tau$.

This theme continues in fig.~\ref{plane_tau_0p01_a0p01}, which shows our numerical results for $\tau=10^{-2}$, $\talpha=10^{-2}$, $k/\mu=10^{-2}$, and $6\times10^{-4}\leq T/\mu \leq 10^{-2}$. At $T/\mu=6\times10^{-4}$ the four highest poles are two HZS poles (black crosses) and two poles with relativistic dispersion (blue dots). As $T/\mu$ increases, the relativistic poles move down and then up, all while moving closer to the imaginary axis, and eventually becoming the hydrodynamic sound poles. The HZS poles execute part of a loop-the-loop, shown in detail in fig.~\ref{plane_tau_0p01_a0p01_cu}, and then move down and towards the imaginary axis, eventually colliding there at $T/\mu=4.8\times10^{-3}$, and then splitting into two purely imaginary poles (red squares), one moving up the axis and one down, where the one moving up eventually becomes the charge diffusion pole. These results are similar to those of the probe limit, $\tau=0$ and $\talpha=1$ in fig.~\ref{plane_probe}, so yet again we find that fixing $\tau$ and decreasing $\talpha$ is qualitatively similar to fixing $\talpha$ and decreasing $\tau$. We also have a second critical $\talpha$ value: for $\tau=10^{-2}$ and $\talpha=0.1$, the HZS crosses over the hydrodynamic sound, while for $\tau=10^{-2}$ and $\talpha=10^{-2}$ the relativistic poles cross over. We find the critical value is $\talpha\approx0.014$.

In summary, for fixed $k/\mu$, while the pole movement depends sensitively on $\talpha$ and $\tau$, in general fixing $\talpha$ and increasing $\tau$, or fixing $\tau$ and increasing $\talpha$, causes poles to move up the imaginary axis and begin ``interfering'' with the movement of the highest poles, eventually changing the crossover qualitatively, so that the probe limit definition is no longer viable.

Crucially, the loop-the-loops in figs.~\ref{plane_tau0p0001_changing_alpha} and~\ref{plane_tau_0p01_a0p1_a0p01}, i.e. the sound poles' changing $\textrm{Re}\left(\omega\right)$ at fixed $k/\mu$, suggests that the sound speed does not remain $v = 1/\sqrt{2}$ as $T/\mu$ changes. However, as mentioned above, the gravity theory's scaling symmetry implies that fixing $k/\mu$ and decreasing $\talpha$ is equivalent to fixing $\talpha$ and increasing $k/\mu$, so in fact we can interpret the loop-the-loops as \text{high momentum} effects. In other words, we are in effect fixing $\talpha=1$ and increasing $k/\mu$, so that higher powers of $k/\mu$ grow in $\textrm{Re}\left(\omega\right)$, obscuring the sound poles' linear in $k/\mu$ behavior. However, in all cases, for fixed $\talpha$ and sufficiently small $k/\mu$ we recover the sound dispersion $\omega = \pm v \, k + \ldots$ with $v=1/\sqrt{2}$.

Such a perspective also reveals that for a given $\tau$, fixing $\talpha$ and increasing $k/\mu$ can produce qualitative changes at critical values of $k/\mu$. Since the combination $k/(\talpha \mu)$ is invariant under the scaling symmetry, and for fixed $k/\mu$ we know the critical $\talpha$ values, if we instead fix $\talpha$ then we can immediately infer the critical $k/\mu$ values. For example, for $\tau=10^{-4}$ and fixed $k/\mu=10^{-2}$, for $\talpha$ below the critical value $\talpha\approx 0.07$ the relativistic poles instead of the HZS crossed over to hydrodynamic sound, as shown in fig.~\ref{figa001}. The critical value of the invariant combination is thus $k/(\talpha \mu)\approx 0.14$, so if instead we fix $\talpha=1$ and increase $k/\mu$, then the critical value will be $k/\mu \approx 0.14$.

\subsection{Spectral Functions}
\label{sec:spectral_functions}

In this section we present our numerical results for the charge and energy spectral functions, $\rho_J$ and $\rho_{tt}$, respectively, obtained via eqs.~\eqref{eq:spectraldef} and~\eqref{eq:retgreen}. We will compare our numerical results to an approximation in which the Green's function matrix is simply a sum of poles,
\beq
\label{eq:mero}
G_{ij}(\omega,k) \approx \sum_{n}\frac{{\cal R}_{ij}^{(n)}(k)}{\omega-\omega_*^{(n)}(k)},
\eeq
where $\omega_*^{(n)}(k)$ are our numerical results for the highest poles, specifically the sound poles and the next highest pole, or pair of poles, and ${\cal R}_{ij}^{(n)}(k)$ is a matrix of pole residues, which are generically complex-valued. In the appendix we explain how we compute the matrix of residues, using the techniques of ref.~\cite{Kaminski:2009dh}.

To our knowledge, in principle nothing requires $G_{ij}(\omega,k)$ to be simply a sum of poles, i.e. nothing forbids either additional terms analytic in $\omega$ or terms more singular in $\omega$, such as branch cuts. Indeed, via the Mittag-Leffler theorem, a partial fraction expansion would provide a more accurate approximation, by including additional terms that, among other things, would capture the large-$\omega$ asymptotics. (For a recent example of such an expansion in holography, see ref.~\cite{Solana:2018pbk}.) However, in the region of small, real-valued $\omega$ we expect many of these terms to be negligible. Indeed, in the following our sum of poles approximations for $\rho_J$ and $\rho_{tt}$ will agree very well with our numerical results for many, but not all, values of $\tau$, $\talpha$, and $T/\mu$, indicating that the great majority of spectral weight comes only from the few highest poles---and indeed primarily from the sound and charge diffusion poles. We fix $k/\mu=10^{-2}$ throughout this subsection.

\begin{figure}[t!]
\includegraphics[width=0.98\textwidth]{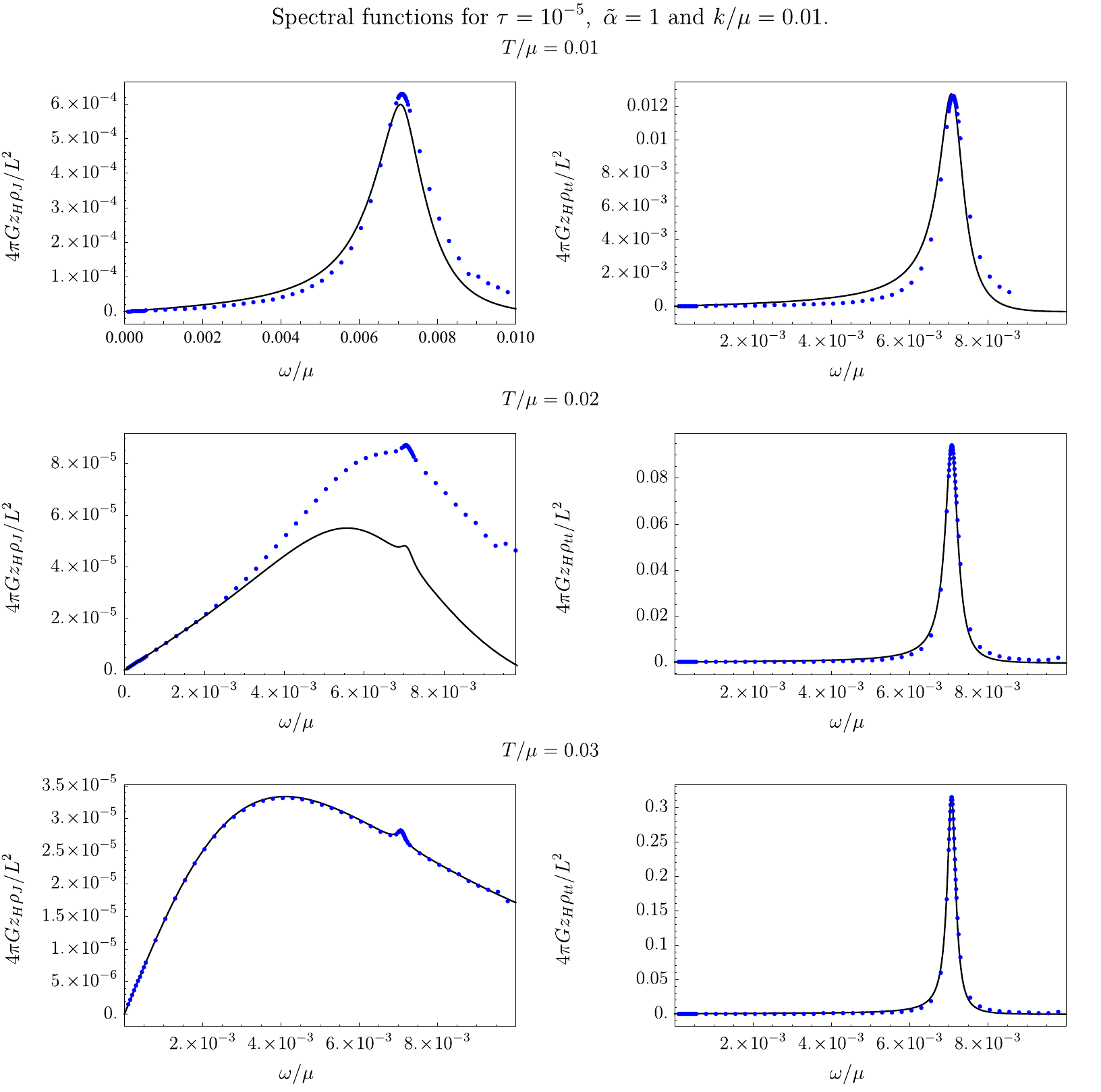}
\caption{\label{tau_0p00001_spectral_functions} Our numerical results (blue dots) for the charge spectral function, $\rho_J$ (left column) and energy spectral function, $\rho_{tt}$ (right column), each made dimensionless by a factor of $4\pi G z_H/L^2$, as functions of $\omega/\mu$ for $\tau=10^{-5}$, $\talpha=1$, $k/\mu =10^{-2}$ and $T/\mu = 10^{-2}$ (top row), $2\times10^{-2}$ (middle row), and $3\times10^{-2}$ (bottom row). The solid black lines come from the sum-of-poles approximation to the Green's functions in eq.~\eqref{eq:mero}. Both $\rho_J$ and $\rho_{tt}$ exhibit a peak from the sound pole (HZS or hydrodynamic) at $\omega/\mu \approx v \, k/\mu\approx 7.1\times 10^{-3}$. As $T/\mu$ increases the sound peak's height decreases in $\rho_J$ but increases in $\rho_{tt}$. Simultaneously, in $\rho_J$ a second peak rises closer to $\omega/\mu = 0$, from the charge diffusion pole, while $\rho_{tt}$ exhibits no other significant features. The crossover can be defined as the value of $T/\mu$ where the two peaks in $\rho_J$ have equal height~\cite{Davison:2011uk}.}
\end{figure}

Fig.~\ref{tau_0p00001_spectral_functions} shows our numerical results for $\rho_J$ and $\rho_{tt}$ for $\tau=10^{-5}$, $\talpha=1$, $k/\mu=10^{-2}$ and $T/\mu=10^{-2}$, $2\times10^{-2}$, and $3\times10^{-2}$. In fig.~\ref{tau_0p00001_spectral_functions} the blue dots are our numerical data while the solid black line comes from the sum-of-poles approximation to the Green's functions in eq.~\eqref{eq:mero}. This approximation is excellent over most of the regime shown, except for one curious outlier, namely $\rho_J$ at $T/\mu=2\times10^{-2}$, where the sum of poles roughly captures some key features of the shape, but otherwise is clearly a poor approximation. We have not found any other poles that provide a significant contribution to the spectral functions in the plotted regimes, suggesting that this is a genuine breakdown of the approximation. The same is true in later examples where the sum-of-poles approximation is poor.

\begin{figure}[t!]
\includegraphics[width=0.96\textwidth]{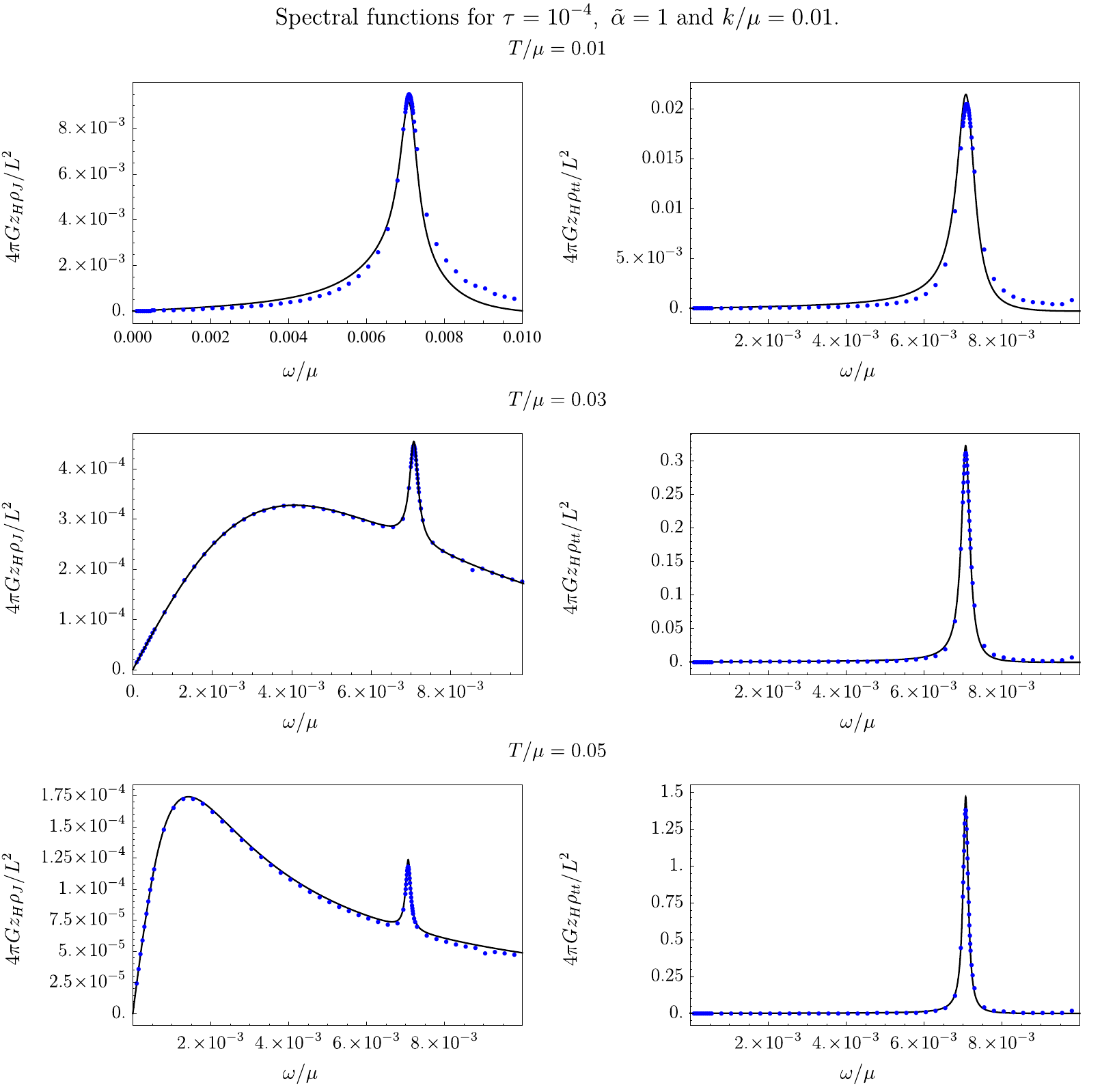}
\caption{\label{tau_0p0001_spectral_functions} Our numerical results (blue dots) for the charge spectral function, $\rho_J$ (left column) and energy spectral function, $\rho_{tt}$ (right column), each made dimensionless by a factor of $4\pi G z_H/L^2$, as functions of $\omega/\mu$ for $\tau=10^{-4}$, $\talpha=1$, $k/\mu =10^{-2}$ and $T/\mu = 10^{-2}$ (top row), $0.03$ (middle row), and $0.05$ (bottom row). The solid black lines come from the sum-of-poles approximation to the Green's functions in eq.~\eqref{eq:mero}. Both $\rho_J$ and $\rho_{tt}$ exhibit a peak from the sound pole (HZS or hydrodynamic) at $\omega \approx k/\sqrt{2}$. As $T/\mu$ increases the sound peak's height decreases in $\rho_J$ but increases in $\rho_{tt}$. Simultaneously, in $\rho_J$ a second peak rises near $\omega/\mu = 0$, from the charge diffusion pole, while $\rho_{tt}$ exhibits no other significant features. The crossover can be defined as the value of $T/\mu$ where the two peaks in $\rho_J$ have equal height~\cite{Davison:2011uk}, which gives $T/\mu=0.039$. (Animated versions of these figures are available on this paper's arxiv page.)}
\end{figure}

\begin{figure}[t!]
\includegraphics[width=\textwidth]{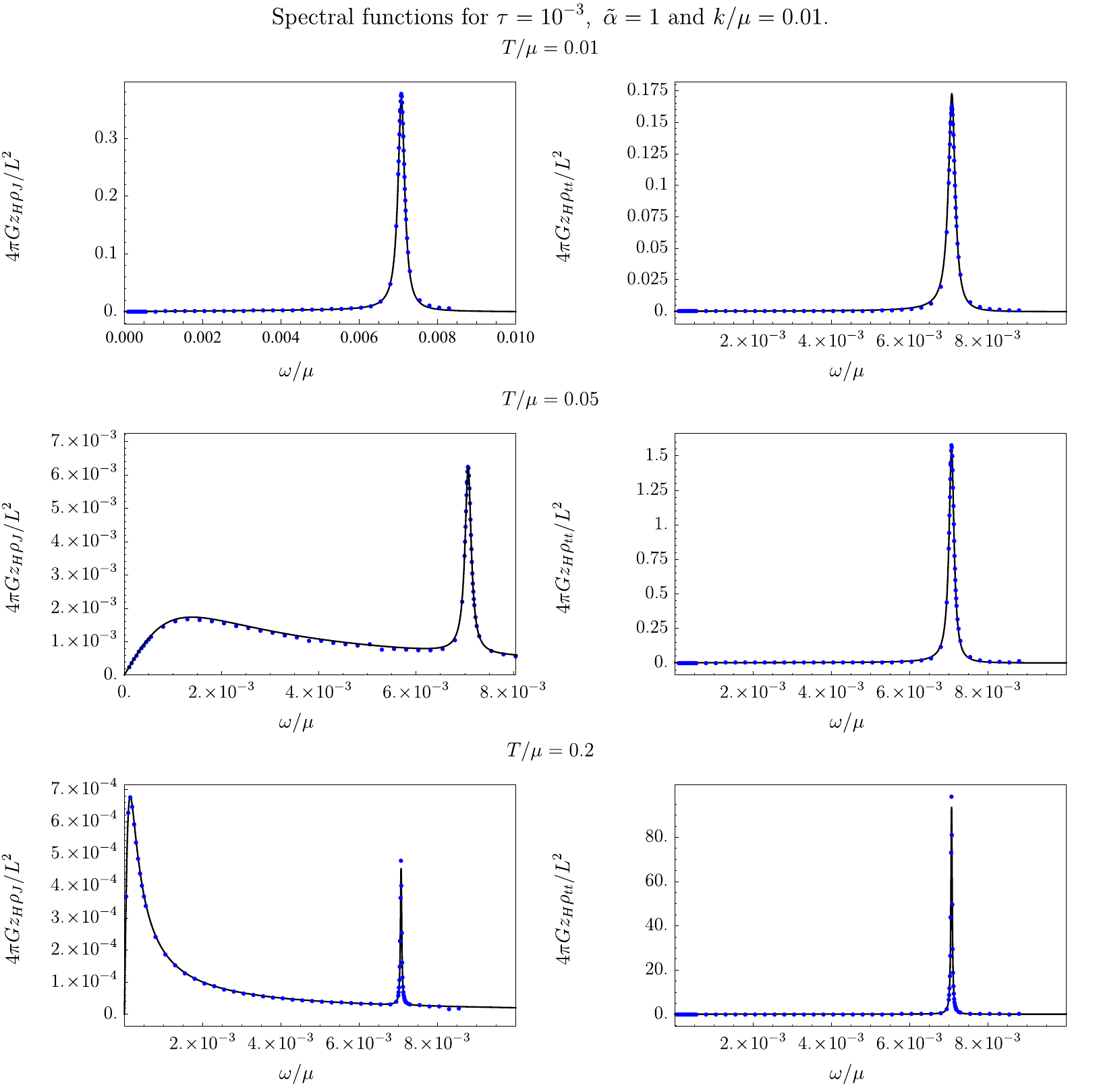}
\caption{\label{tau_0p001_spectral_functions} Our numerical results (blue dots) for the charge spectral function, $\rho_J$ (left column) and energy spectral function, $\rho_{tt}$ (right column), each made dimensionless by a factor of $4\pi G z_H/L^2$, as functions of $\omega/\mu$ for $\tau=10^{-3}$, $\talpha=1$, $k/\mu =0.01$ and $T/\mu = 0.01$ (top row), $0.05$ (middle row), and $0.2$ (bottom row). The solid black lines come from the sum-of-poles approximation to the Green's functions in eq.~\eqref{eq:mero}. As $T/\mu$ increases, the behaviors of both $\rho_J$ and $\rho_{tt}$ are similar to the $\tau=10^{-4}$ case in fig.~\ref{tau_0p0001_spectral_functions}: in $\rho_J$ the sound peak shrinks while the charge diffusion peak grows, and in $\rho_{tt}$ the only significant feature is a sound peak that grows. The crossover can be defined as the value of $T/\mu$ where the two peaks in $\rho_J$ have equal height~\cite{Davison:2011uk}, which gives $T/\mu=0.136$}
\end{figure}

\begin{figure}[t!]
\includegraphics[width=\textwidth]{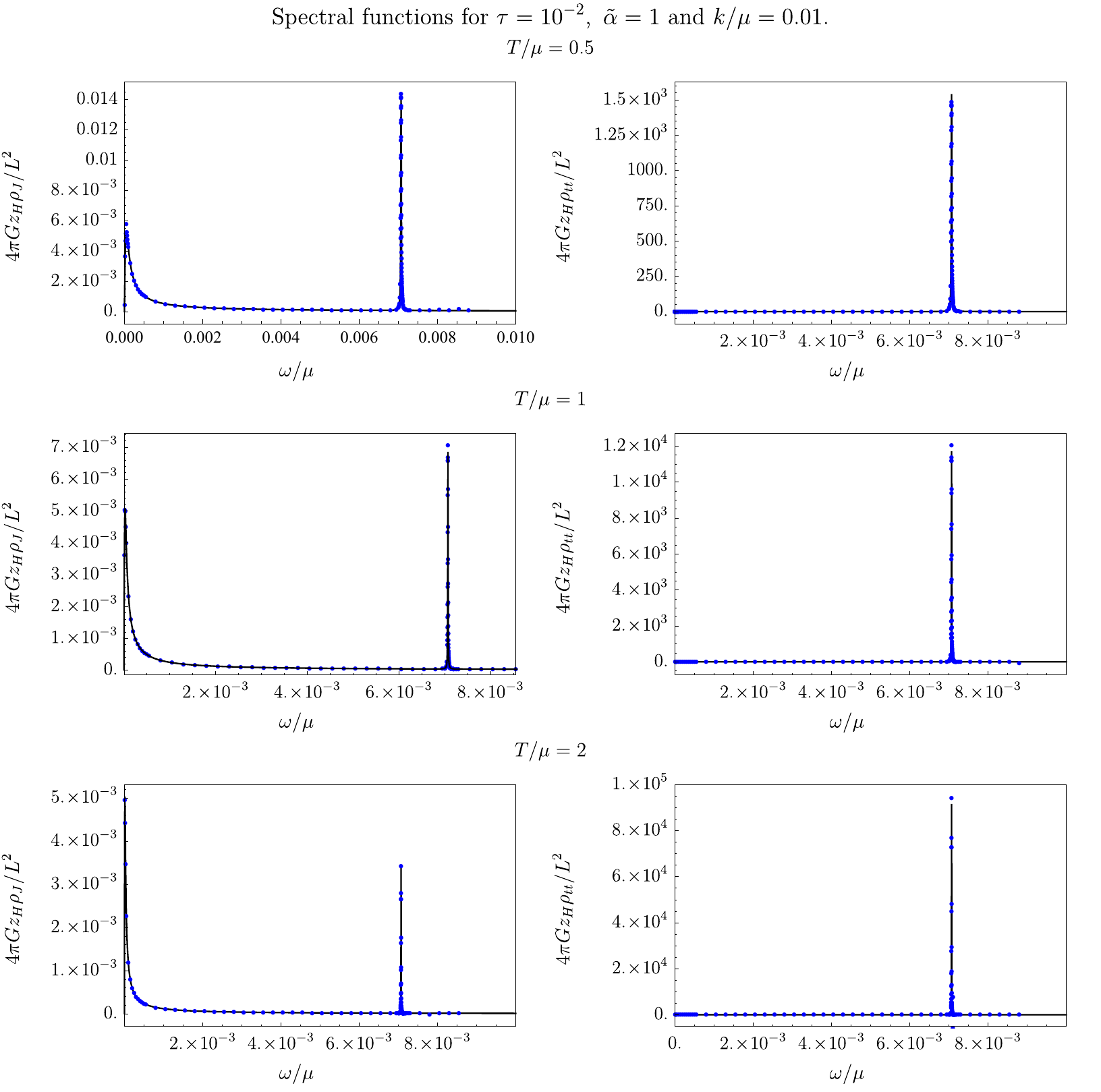}
\caption{\label{tau_0p01_spectral_functions} Our numerical results (blue dots) for the charge spectral function, $\rho_J$ (left column) and energy spectral function, $\rho_{tt}$ (right column), each made dimensionless by a factor of $4\pi G z_H/L^2$, as functions of $\omega/\mu$ for $\tau=10^{-2}$, $\talpha=1$, $k/\mu =10^{-2}$ and $T/\mu = 0.5$ (top row), $1$ (middle row), and $2$ (bottom row). The solid black lines come from the sum-of-poles approximation to the Green's functions in eq.~\eqref{eq:mero}. As $T/\mu$ increases, the behaviors of both $\rho_J$ and $\rho_{tt}$ are similar to the $\tau=10^{-4}$ and $10^{-3}$ cases in figs.~\ref{tau_0p0001_spectral_functions} and~\ref{tau_0p001_spectral_functions}: in $\rho_J$ the sound peak shrinks while the charge diffusion peak grows, and in $\rho_{tt}$ the only significant feature is a sound peak that grows. The crossover can be defined as the value of $T/\mu$ where the two peaks in $\rho_J$ have equal height~\cite{Davison:2011uk}, which gives $T/\mu=1.45$}
\end{figure}

In both $\rho_J$ and $\rho_{tt}$ at $T/\mu = 10^{-2}$ we find a peak from the sound pole at $\omega \approx k/\sqrt{2}$. As $T/\mu$ increases through the values shown, in $\rho_J$ the sound peak's height decreases by a factor of $\approx 20$, while in $\rho_{tt}$ the height increases by a factor of $\approx 25$, indicating that as $T/\mu$ increases the sound pole's residue decreases in $G_J$ but increases in $G_{tt}$. In both cases the sound peak's width decreases as $T/\mu$ increases. These features are consistent with our results for the pole positions, which are similar to those at $\tau=10^{-4}$ and $\talpha=1$ in figs.~\ref{plane_10Em4},~\ref{realimsmalltau}, and~\ref{fig:dispersionsmalltau}. In particular, as $T/\mu$ increased the HZS poles (black crosses) cross over to the hydrodynamic sound poles, with constant real part $\approx k/\sqrt{2}$ and decreasing imaginary part.

Crucially, aside from the sound peak no other significant features are visible in $\rho_{tt}$. Our numerical results from eq.~\eqref{eq:residue} indicate that in $G_{tt}$ the charge diffusion pole does generically have non-zero residue, however at the $T/\mu$ shown in fig.~\ref{tau_0p00001_spectral_functions} the sound pole's residue is $\approx 10$ times larger, explaining why no charge diffusion peak is visible in $\rho_{tt}$ in fig.~\ref{tau_0p00001_spectral_functions}.

However, in $\rho_J$ a dramatic new feature appears as $T/\mu$ increases, namely a charge diffusion peak rises closer to $\omega/\mu = 0$. Indeed, while the sound peak shrinks the charge diffusion peak grows and eventually dominates the spectral weight. Such behavior is qualitatively similar to that of AdS-RN~\cite{Davison:2011uk}, despite the more complicated motion of poles, which is similar to that in fig.~\ref{plane_10Em4}. Indeed, following ref.~\cite{Davison:2011uk}, in principle we could define a precise moment of crossover as the $T/\mu$ where the charge diffusion and sound peaks have equal height, which occurs between $T/\mu=2 \times 10^{-2}$ and $3 \times 10^{-2}$. In practice, however, given how small the sound peak was and how broad the charge diffusion peak was, we struggled to extract a more precise crossover value of $T/\mu$ from our numerics.

Fig.~\ref{tau_0p0001_spectral_functions} shows our numerical results for $\rho_J$ and $\rho_{tt}$ for $\tau=10^{-4}$, $\talpha=1$, $k/\mu=10^{-2}$ and $T/\mu=10^{-2}$, $0.03$, and $0.05$, with the same color coding as in fig.~\ref{tau_0p00001_spectral_functions}. (An animated version of fig.~\ref{tau_0p0001_spectral_functions} is available on this paper's arxiv page.) Unlike the previous $\tau=10^{-5}$ case, now the sum-of-poles approximation in eq.~\eqref{eq:mero} is clearly excellent over most of the regime shown. In general, the results are similar to the previous case. In both $\rho_J$ and $\rho_{tt}$ at $T/\mu = 10^{-2}$ we find a peak from the sound pole at $\omega \approx k/\sqrt{2}$. As $T/\mu$ increases through the values shown, in $\rho_J$ the sound peak's height decreases by a factor of $\approx 10^{2}$, while increasing in $\rho_{tt}$ by a factor of $\approx 75$. In both cases the sound peak's width decreases, though only slightly, as $T/\mu$ increases. These features are consistent with our results for the pole positions at $\tau=10^{-4}$ and $\talpha=1$ in figs.~\ref{plane_10Em4},~\ref{realimsmalltau}, and~\ref{fig:dispersionsmalltau}. Aside from the sound peak no other significant features are visible in $\rho_{tt}$. Our numerical results from eq.~\eqref{eq:residue} indicate that in $G_{tt}$ the charge diffusion pole does generically have non-zero residue, however at the $T/\mu$ shown in fig.~\ref{tau_0p0001_spectral_functions} the sound pole's residue is $\approx 20$ times larger. Again in $\rho_J$ as $T/\mu$ increases a charge diffusion peak rises near $\omega/\mu = 0$. Defining the precise moment of crossover as the $T/\mu$ where the charge diffusion and sound peaks have equal height gives $T/\mu=0.039$. In contrast, the definition based on the collision of poles in fig.~\ref{plane_10Em4} gave the smaller value $T/\mu\approx0.029$.

Fig.~\ref{tau_0p001_spectral_functions} shows our numerical results for $\rho_J$ and $\rho_{tt}$ for $\tau=10^{-3}$, $\talpha=1$, $k/\mu=10^{-2}$ and $T/\mu=0.01$, $0.05$, and $0.2$. These results are qualitatively similar to the $\tau=10^{-5}$ and $10^{-4}$ cases in figs.~\ref{tau_0p00001_spectral_functions} and~\ref{tau_0p0001_spectral_functions}. As $T/\mu$ increases, in $\rho_J$ the sound peak shrinks by a factor of $\approx 10^3$ for the $T/\mu$ shown, while a charge diffusion peak rises at $\omega/\mu = 0$ and eventually dominates the spectral weight. In $\rho_{tt}$ the only significant feature is the sound peak, which grows by a factor of $\approx 10^3$ for the $T/\mu$ shown. All peaks are narrower than in the $\tau=10^{-4}$ case. Again, these features are consistent with our results for the pole positions in fig.~\ref{plane_tau_0p001}. In fact, the complicated motion of poles lower in the complex $\omega/\mu$ plane has little or no apparent effect on $\rho_J$ and $\rho_{tt}$, which are extremely well-approximated by our sum of highest poles in eq.~\eqref{eq:mero}, i.e. the solid black lines in fig.~\ref{tau_0p001_spectral_functions}. Defining the crossover when the two peaks in $\rho_J$ have equal height gives $T/\mu \approx 0.136$. In contrast, defining the crossover by the collision of poles that produces the charge diffusion pole in fig.~\ref{plane_tau_0p001} gave $T/\mu\approx0.027$. 

Fig.~\ref{tau_0p01_spectral_functions} shows our numerical results for $\rho_J$ and $\rho_{tt}$ for $\tau=10^{-2}$, $\talpha=1$, $k/\mu=10^{-2}$, and $T/\mu=0.5$, $1$, and $2$. Again the results are similar to the previous cases. As $T/\mu$ increases, in $\rho_J$ the sound peak shrinks by a factor of $\approx5$ for the $T/\mu$ shown, while the charge diffusion peak rises at $\omega/\mu = 0$ and eventually dominates the spectral weight. In $\rho_{tt}$ the only significant visible feature is a sound peak which grows by a factor of $\approx 1.5$ for the $T/\mu$ shown. All peaks are narrower than the previous cases, and moreover the sound peak is now taller in $\rho_{tt}$ than in $\rho_J$ by a relative factor of $\approx 10^6$, unlike the previous cases where the sound peak was roughly the same height in both spectral functions. Again, these features are consistent with our results for the positions of poles in fig.~\ref{plane_tau_0p01}, and again, the spectral functions are well approximated by the sum of highest poles in eq.~\eqref{eq:mero}. In particular, the complicated pole motion in fig.~\ref{plane_tau_0p01} occurs at much smaller $T/\mu$ than those shown in fig.~\ref{tau_0p01_spectral_functions}. The changes shown in fig.~\ref{tau_0p01_spectral_functions} come only from the three highest poles, and in fact must come primarily from their residues, since those highest poles move very little for the $T/\mu$ shown.  Most importantly, unlike $\tau=10^{-5}$, $10^{-4}$, and $10^{-3}$, when $\tau=10^{-2}$ no collisions of poles producing a charge diffusion pole occurs, so the only definition for a precise moment of crossover is via the exchange of dominance of poles in $\rho_J$, which gives $T/\mu \approx 1.45$.

In short, for $k/\mu=10^{-2}$ and $\talpha=1$, for all $\tau$ we considered the definition of crossover via a transfer of dominance in $\rho_J$ from the sound peak to the charge diffusion peak, remains viable. However, as $\tau \to 0$, we may expect to recover the probe limit result for $\rho_J$, where no transfer of dominance occurs~\cite{Davison:2011ek}. Instead, in the strict probe limit $\rho_J$ exhibits only a single peak at all $T/\mu$, which at low $T/\mu$ comes from HZS and at high $T/\mu$ comes from the charge diffusion pole. More specifically, as shown in fig.~\ref{plane_probe}, as $T/\mu$ increases the HZS poles collide on the imaginary axis and split, producing the charge diffusion pole, and correspondingly in $\rho_J$, the single peak simply moves towards $\omega/\mu=0$ and shrinks in height~\cite{Davison:2011ek}. Apparently $\tau=10^{-5}$ is not small enough to reproduce the probe result, when $k/\mu=10^{-2}$ and $\talpha=1$.

\begin{figure}[t!]
\includegraphics[width=0.98\textwidth]{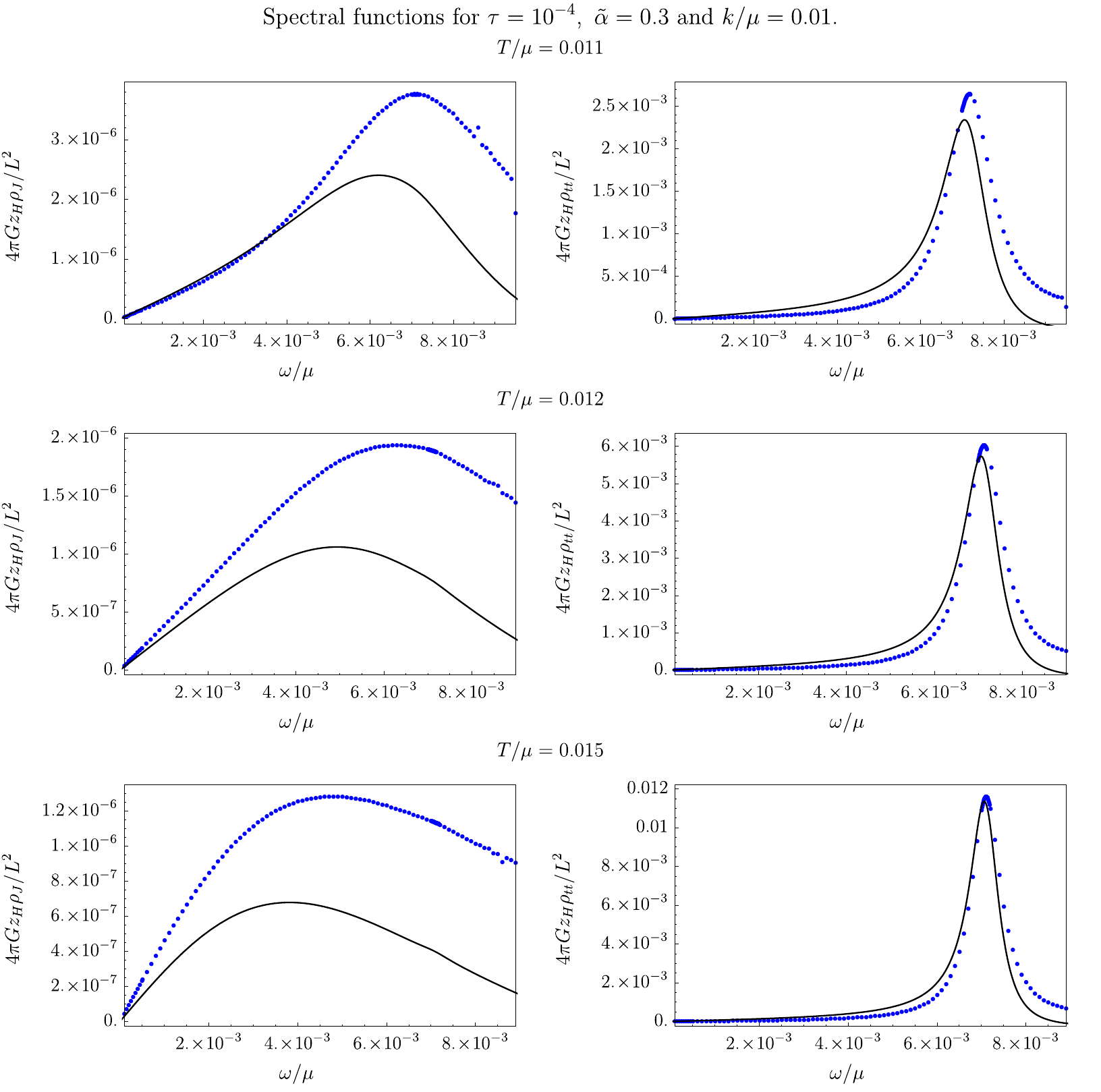}
\caption{\label{a_0p3_spectral_functions} Our numerical results (blue dots) for the charge spectral function, $\rho_J$ (left column) and energy spectral function, $\rho_{tt}$ (right column), each made dimensionless by a factor of $4\pi G z_H/L^2$, as functions of $\omega/\mu$ for $\tau=10^{-4}$, $\talpha=0.3$, $k/\mu =10^{-2}$ and $T/\mu = 1.1\times10^{-2}$ (top row), $1.2\times10^{-2}$ (middle row), and $1.5\times10^{-2}$ (bottom row). The solid black lines come from the sum-of-poles approximation to the Green's functions in eq.~\eqref{eq:mero}. As $T/\mu$ increases, the behaviors of both $\rho_J$ and $\rho_{tt}$ are qualitatively the same as the $\talpha=1$ cases in figs.~\ref{tau_0p00001_spectral_functions} to~\ref{tau_0p01_spectral_functions}: in $\rho_J$ the sound peak shrinks while the charge diffusion peak grows, and in $\rho_{tt}$ the only significant feature is a sound peak that grows. Most importantly, for this smaller $\talpha$ the crossover can still be defined as the value of $T/\mu$ where the charge diffusion and sound peaks in $\rho_J$ have equal height~\cite{Davison:2011uk}.}
\end{figure}

\begin{figure}[t!]
\includegraphics[width=0.95\textwidth]{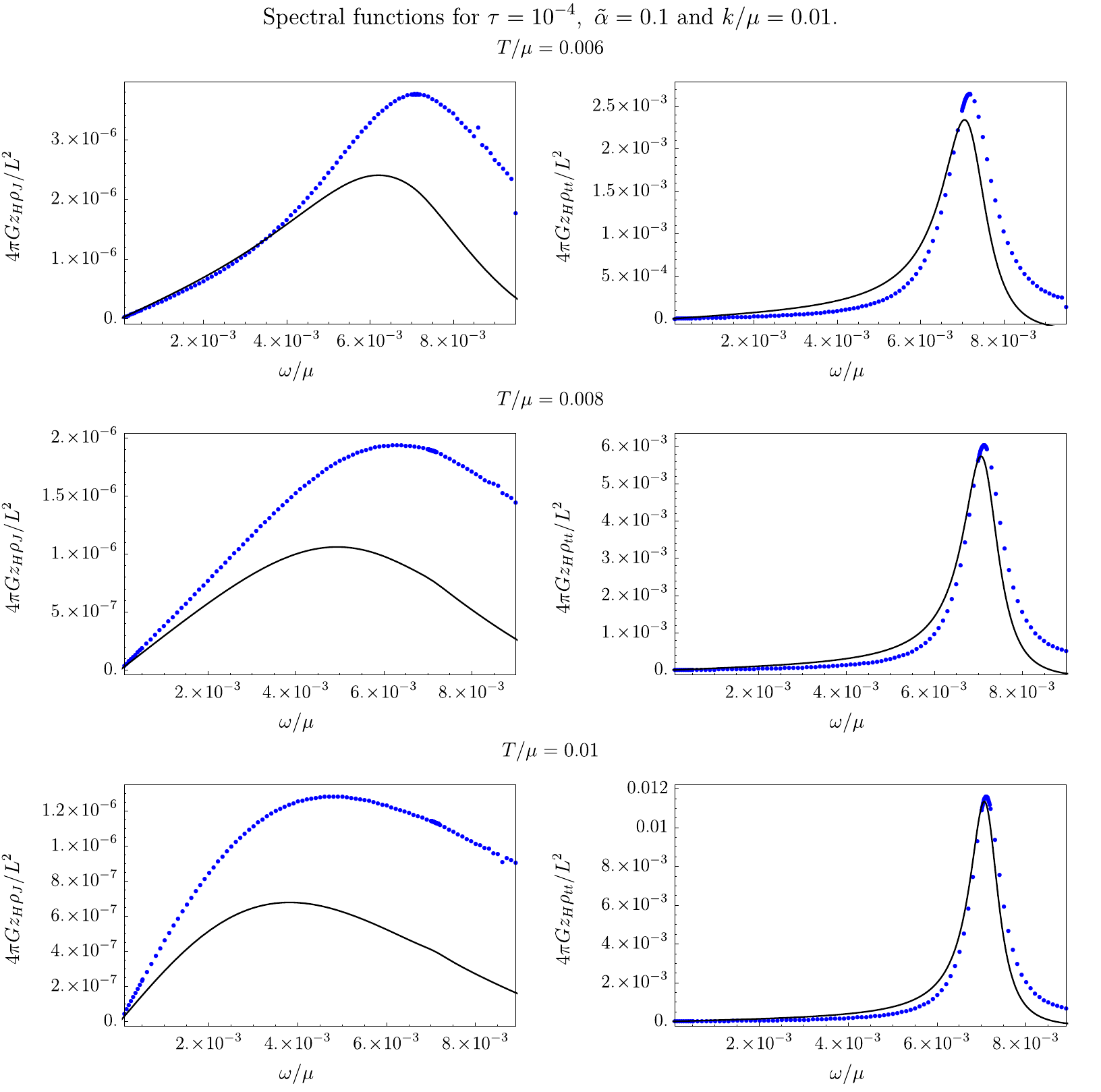}
\caption{\label{a_0p1_spectral_functions} Our numerical results (blue dots) for the charge spectral function, $\rho_J$ (left column) and energy spectral function, $\rho_{tt}$ (right column), each made dimensionless by a factor of $4\pi G z_H/L^2$, as functions of $\omega/\mu$ for $\tau=10^{-4}$, $\talpha=0.1$, $k/\mu =10^{-2}$ and $T/\mu = 6\times10^{-3}$ (top row), $8\times10^{-3}$ (middle row), and $10^{-2}$ (bottom row). The solid black lines come from the sum-of-poles approximation to the Green's functions in eq.~\eqref{eq:mero}. As $T/\mu$ increases, the behavior of $\rho_{tt}$ is qualitatively the same as all previous cases: the only significant feature is a sound peak. However, $\rho_J$ now exhibits only a single peak for all $T/\mu$, which moves towards $\omega/\mu=0$ and shrinks by a factor of $\approx 2.5$ as $T/\mu$ decreases. At $T/\mu=1.1\times10^{-2}$ the peak comes from the sound pole, but by $T/\mu=1.5\times10^{-2}$ it comes from the charge diffusion pole. Only one peak ever appears, so clearly in this case the crossover cannot be defined as the value of $T/\mu$ where two peaks in $\rho_J$ have equal height~\cite{Davison:2011uk}.}
\end{figure}

However as we saw in sec.~\ref{sec:poles}, for fixed $k/\mu$, fixing $\tau$ and decreasing $\talpha$ produces qualitatively similar results to fixing $\talpha$ and decreasing $\tau$. We may thus expect that fixing $\tau$ and decreasing $\talpha$ will produce $\rho_J$ qualitatively similar to the probe limit, and in particular some critical $\talpha$ may exist for which a transfer of dominance no longer occurs. Figs.~\ref{a_0p3_spectral_functions} and~\ref{a_0p1_spectral_functions} confirm that expectation. Fig.~\ref{a_0p3_spectral_functions} shows our numerical results for $\rho_J$ and $\rho_{tt}$ for $\tau=10^{-4}$, $\talpha=0.3$, $k/\mu=10^{-2}$, and $T/\mu=1.1\times10^{-2}$, $1.2\times10^{-2}$, and $1.5\times10^{-2}$. The results are similar to the previous cases. As $T/\mu$ increases, in $\rho_J$ the sound peak shrinks while the charge diffusion peak grows, and a transfer of dominance occurs somewhere between $T/\mu=1.2\times10^{-2}$ and $1.5\times10^{-2}$. In $\rho_{tt}$ the only significant feature is the sound peak, which grows by a factor of $\approx 2.6$ for the $T/\mu$ shown, and is taller than that in $\rho_J$ by a factor of $\approx 10^3$. The sum-of-poles approximation eq.~\eqref{eq:mero} is very good for $\rho_{tt}$, but unlike most previous cases is consistently poor for $\rho_J$, capturing gross features of the shape but not the details or overall size.

In contrast, fig.~\ref{a_0p1_spectral_functions} shows our numerical results for $\rho_J$ and $\rho_{tt}$ for $\tau=10^{-4}$, $\talpha=0.1$, $k/\mu=10^{-2}$, and $T/\mu=6\times10^{-3}$, $8\times10^{-3}$, and $10^{-2}$. At this smaller $\talpha$, the results for $\rho_J$ are qualitatively similar to those of the probe limit: only a single peak appears, which as $T/\mu$ increases moves towards $\omega/\mu=0$ and shrinks by a factor of $\approx 2.5$. In $\rho_{tt}$, again the only significant feature is the sound peak, which grows by a factor of $\approx 4.8$ for the $T/\mu$ shown, and is taller than that in $\rho_J$ by a factor of $\approx 10^3$. The sum-of-poles approximation is again very good for $\rho_{tt}$ but very poor for $\rho_J$.

Clearly for $k/\mu=10^{-2}$ and $\tau=10^{-4}$, a critical $\talpha$ exists where the transfer of dominance in $\rho_J$ no longer occurs. We estimate the critical value as $\talpha \approx0.19$. We also studied $\tau=10^{-2}$ and decreasing $\talpha$, and observed qualitatively similar behavior.

As in previous cases, due to the gravity theory's scaling symmetry $\alpha \to \lambda \, \alpha$ and $F_{MN} \to \lambda \, F_{MN}$, fixing $k/\mu$ and decreasing $\talpha$ is equivalent to fixing $\talpha$ and increasing $k/\mu$, so we may interpret the results above as the effect of increasing momentum. In AdS-RN increasing $k/\mu$ indeed had the effect of merging two peaks in $\rho_J$ into a single peak~\cite{Davison:2011uk}, similar to the transition from fig.~\ref{a_0p3_spectral_functions} to fig.~\ref{a_0p1_spectral_functions}.

In short, for fixed $k/\mu$, our results suggest that the AdS-RN definition of crossover as a transfer in dominance in $\rho_J$ from sound peak to charge diffusion peak, is viable only sufficiently far from the probe limit, meaning fixed $\talpha$ and sufficiently large $\tau$, or fixed $\tau$ and sufficiently large $\talpha$. Additionally, we have shown that the retarded Green's functions are often, but not always, well-approximated simply by the sum of their few highest poles, eq.~\eqref{eq:mero}.

\subsection{Sound Attenuation}
\label{sec:soundatt}

In this section we present our results for the sound attenuation, meaning $\textrm{Im}\left(\omega\right)$ of the sound pole, whether HZS or hydrodynamic sound, as a function of $\tau$, $\talpha$, and $T/\mu$.

As reviewed in sec.~\ref{intro}, in a LFL sound dispersion is typically expressed as complex-valued $k(\omega)$ with real-valued $\omega$. As $T/\mu$ increases, sound exhibits three regimes: quantum collisionless, $0 \leq \pi T/\mu < \omega/\mu$, where $|\textrm{Im}\left(k\right)| \propto \omega^2/\mu$, thermal collisionless, $\omega/\mu < \pi T/\mu < \sqrt{\omega/\mu}$, where $|\textrm{Im}\left(k\right)|\propto \left(\pi T\right)^2/\mu$, and hydrodynamic, $\pi T/\mu > \sqrt{\omega/\mu}$, where $|\textrm{Im}\left(k\right)| \propto \mu \omega^2/T^2$. In other words, in terms of powers of $T$, in a LFL $|\textrm{Im}\left(k\right)|$ scales as $T^0$ in the quantum collisionless regime, $T^2$ in the thermal collisionless regime, and $T^{-2}$ in the hydrodynamic regime. The collisionless-to-hydrodynamic crossover is thus characterized by a maximum in the sound attenuation where the $T^2$ scaling transitions to $T^{-2}$.

In our holographic system, we express the sound dispersion as complex-valued $\omega(k)$ with real-valued $k$. Translating the LFL regimes to that form is easy: simply use the leading small-$\omega$ behavior, $|\omega| = v \, k$, to replace $\omega$ with $k$. For example, the quantum collisionless regime is $0 \leq \pi T/\mu < v\,k/\mu$, where $|\textrm{Im}\left(\omega\right)| \propto \left(v\,k\right)^2/\mu$.

In probe brane models, as $T/\mu$ increases $|\textrm{Im}\left(\omega\right)|$ exhibits $T^0$ scaling followed by $T^2$ scaling, similar to the quantum and thermal collisionless regimes of a LFL, but in the hydrodynamic regime crosses over to charge diffusion, rather than hydrodynamic sound~\cite{Davison:2011ek}. In contrast, in AdS-RN $|\textrm{Im}\left(\omega\right)|$ exhibits $T^0$ scaling at low $T/\mu$, like a LFL, followed by a power of $T$ smaller than $T^2$, unlike a LFL, and then $T^{-1}$ scaling in the hydrodynamic regime, unlike a LFL's $T^{-2}$, but expected for a CFT. In AdS-RN, for sufficiently small $k/\mu$ the sound attenuation exhibits a (very small) maximum at $\pi T/\mu \approx \sqrt{v \, k /\mu}$, signaling the onset of the hydrodynamic regime, as in a LFL. In terms of the pole movement in fig.~\ref{fig:rn_cartoon}, as $T/\mu$ increases the poles are practically stationary at low $T/\mu$ and then start moving up at approximately the $T/\mu$ where $|\textrm{Im}\left(\omega\right)|$ has a small maximum.

\begin{figure}[t!]
\begin{center}
	\includegraphics{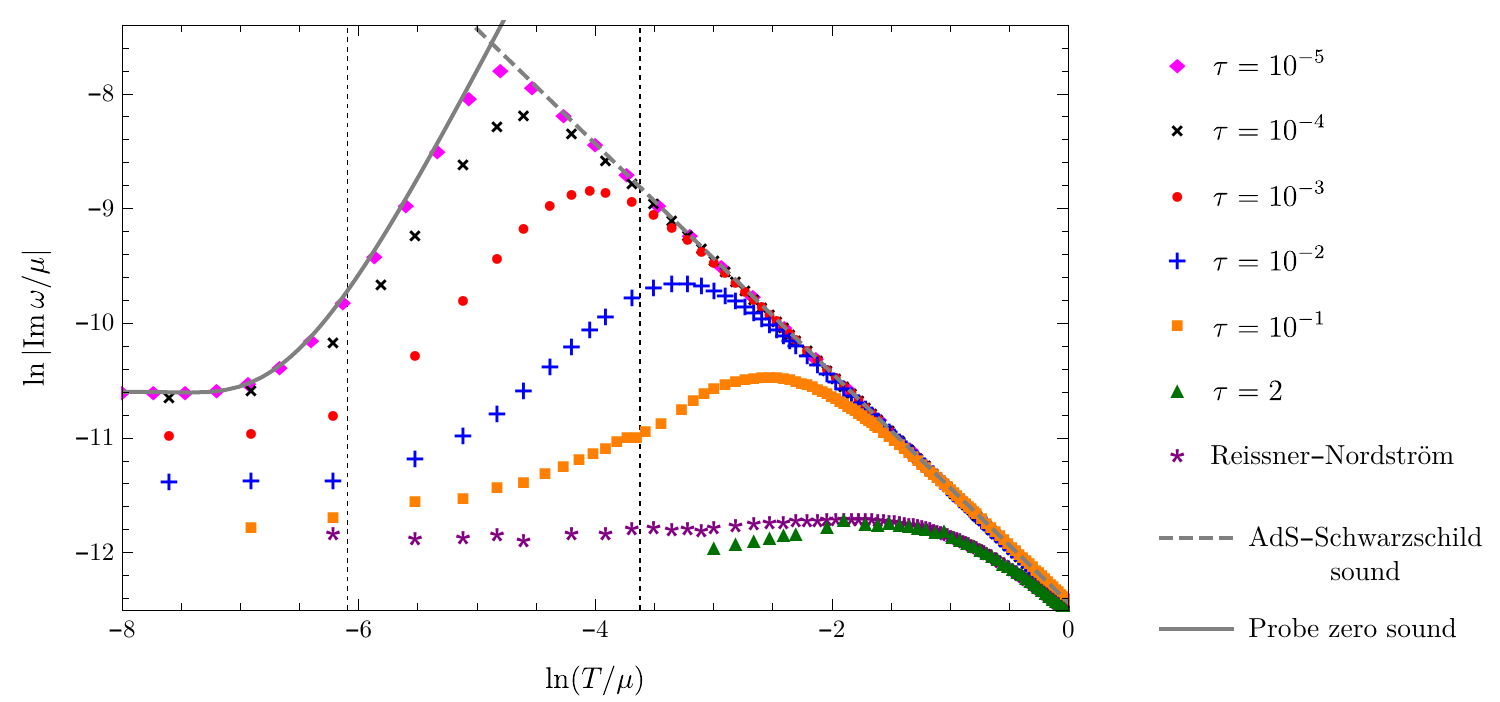}
\caption{\label{fig:sound_attenuation} Our numerical results for $\ln\left|\textrm{Im}\left(\omega/\mu\right)\right|$ versus $\ln \left(T/\mu\right)$ for $\talpha=1$, $k/\mu=10^{-2}$, and $\tau=10^{-5}$ (pink diamonds), $10^{-4}$ (black crosses), $10^{-3}$ (red dots), $10^{-2}$ (blue plus signs), $10^{-1}$ (orange squares), and $2$ (green triangles), as well as the AdS-RN result (purple stars). The solid gray line is the numerical result in the probe limit, while the dashed gray line comes from $\textrm{Im}\left(\omega\right)=-\Gamma k^2$ with the $d=3$ AdS-SCH value $\Gamma=1/(8\pi T)$ (dashed gray). The vertical dashed black lines indicate the LFL definitions of the boundaries between quantum and thermal collisionless regimes, $\ln\left(T/\mu\right)\approx-6.09$, and between thermal collisionless and hydrodynamic regimes, $\ln\left(T/\mu\right)\approx -3.62$. For $\ln\left(T/\mu\right)\lesssim-6.09$, all cases exhibit $|\textrm{Im}\left(\omega\right)|\propto T^0$, similar to the LFL quantum collisionless regime. For $\ln\left(T/\mu\right)\gtrsim-6.09$, $|\textrm{Im}\left(\omega\right)|$ exhibits scaling with a power of $T$ that decreases as $\tau$ increases, from $T^2$ down to, but not exactly to, $T^0$. As $\ln\left(T/\mu\right)$ increases, in all cases such scaling eventually ends in a maximum, followed by $|\textrm{Im}\left(\omega\right)|\propto T^{-1}$, as expected for a CFT in the hydrodynamic regime. As $\tau$ increases the maximum's position moves beyond the LFL value, $\ln\left(T/\mu\right)\approx -3.62$, and its height decreases. Nevertheless, all cases have a maximum, so the LFL definition of the crossover is viable.}
\end{center}
\end{figure}

We start by fixing $k/\mu=10^{-2}$ and $\talpha=1$ and increasing $\tau$. Fig.~\ref{fig:sound_attenuation} shows our numerical results for $\ln\left|\textrm{Im}\left(\omega/\mu\right)\right|$ versus $\ln \left(T/\mu\right)$ for $\talpha=1$, $k/\mu=10^{-2}$, and increasing values of $\tau$ from $\tau= 10^{-5}$ (pink diamonds) to $\tau= 2$ (green triangles), and also the AdS-RN result (purple stars). The solid gray line is the numerical result for $\ln|\textrm{Im}\left(\omega/\mu\right)|$ in the probe limit, while the dashed gray line comes from $\textrm{Im}\left(\omega\right)=-\Gamma k^2$ with the AdS-SCH result $\Gamma=1/(8 \pi T)$~\cite{Herzog:2003ke,Kovtun:2005ev}. The vertical dotted black lines represent the LFL boundaries between quantum and thermal collisionless regimes, $\pi T/\mu = v\,k/\mu$, which for $v=1/\sqrt{2}$ and $k/\mu=10^{-2}$ gives $\ln\left(T/\mu\right)\approx-6.09$, and between thermal collisionless and hydrodynamic regimes, $\pi T/\mu = \sqrt{v\,k/\mu}$, which gives $\ln\left(T/\mu\right) \approx-3.62$. LFL sound attenuation exhibits a maximum at the latter boundary.

In fig.~\ref{fig:sound_attenuation}, when $\tau=10^{-5}$ (pink diamonds) and $T/\mu$ is small, the sound attenuation closely follows the probe limit (solid black line), exhibiting $T^0$ scaling when $\ln\left(T/\mu\right)\lesssim-6.09$ and $T^2$ scaling when $\ln\left(T/\mu\right)\gtrsim-6.09$. Such behavior is practically identical to a LFL. However, as $T/\mu$ increases the sound attenuation exhibits a maximum and transitions to the $T^{-1}$ scaling of a CFT in the hydrodynamic regime. Such behavior is not possible in the probe limit. Moreover, the maximum occurs at $\ln\left(T/\mu\right)\approx-4.75<-3.62$, in contrast to a LFL.

Fig.~\ref{fig:sound_attenuation} also shows that the quantum collisionless type scaling $T^0$ for $\ln\left(T/\mu\right)\lesssim-6.09$ persists to higher $\tau$. In contrast, in the LFL thermal collisionless regime, $\ln\left(T/\mu\right)\gtrsim-6.09$, the power of $T$ clearly decreases as $\tau$ increases, from $T^2$ down to, but not exactly to, $T^0$. At sufficiently high $T/\mu$ the CFT hydrodynamic scaling $T^{-1}$ always emerges, hence a maximum appears in all cases, including AdS-RN. However, as $\tau$ increases the maximum's position drifts to higher and higher $\ln\left(T/\mu\right)$, blithely moving past the LFL value $\ln\left(T/\mu\right)\approx-3.62$.

Additionally, as $\tau$ increases the maximum's height decreases. As discussed in sec.~\ref{results}, such a result is perhaps surprising, if we recall that $\tau$ effectively counts the number of charged fields (such as quark flavors), so that na\"ively we would expect that increasing $\tau$ would cause $\ln\left|\textrm{Im}\left(\omega/\mu\right)\right|$ to increase, i.e. that increasing $\tau$ would \textit{dampen} sound. Instead we find the opposite: in our holographic model, sound becomes \textit{less} damped as we increase $\tau$.

In any case, our results suggest that with $k/\mu=10^{-2}$ and $\talpha=1$, for all $\tau$ a maximum always appears in $|\textrm{Im}\left(\omega\right)|$, and hence the LFL definition of crossover is viable. \textit{Indeed, the shape of all our sound attenuation curves is qualitatively similar to that of a LFL in fig.~\ref{fig:attenuation_cartoon}.}

\begin{figure}[t!]
\begin{center}
	\includegraphics{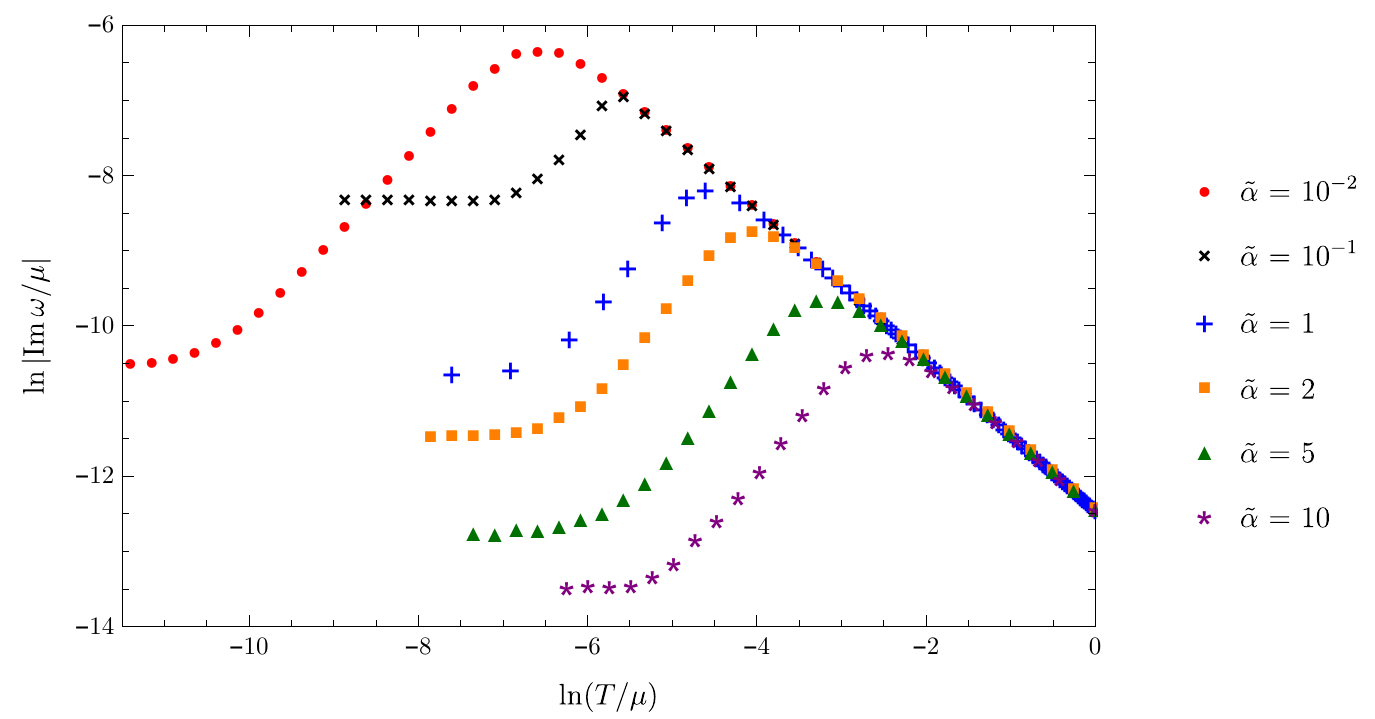}
	\caption{\label{sound_attenuation_tau_0p0001} Our numerical results for $\ln\left|\textrm{Im}\left(\omega/\mu\right)\right|$ versus $\ln \left(T/\mu\right)$ for $\tau=10^{-4}$, $k/\mu=10^{-2}$, and $\talpha=10^{-2}$ (red dots), $0.1$ (black crosses), $1$ (blue plus signs), $2$ (orange squares), $5$ (green triangles), and $10$ (purple stars). The $\talpha=10^{-2}$ curve differs from the others at small $T/\mu$ because in that case relativistic poles rather than HZS cross over to hydrodynamic sound (figs.~\ref{figa001}, ~\ref{figa001cu}, ~\ref{figa001re}, and~\ref{figa001im}). For all $\talpha$, including $\talpha=10^{-2}$, as $\ln\left(T/\mu\right)$ increases, first $|\textrm{Im}\left(\omega\right)|$ scales as $T^0$, then as a power of $T$ slightly less than $T^2$, then has a maximum, and finally scales as $T^{-1}$. For $\talpha \gtrsim 0.07$, increasing $\talpha$ simply re-scales $\ln\left|\textrm{Im}\left(\omega/\mu\right)\right|$ to smaller total value, without changing its shape, and shifts the maximum to higher $T/\mu$.}
\end{center}
\end{figure}

We next fix $k/\mu=10^{-2}$ and $\tau=10^{-4}$ and change $\talpha$. Fig.~\ref{sound_attenuation_tau_0p0001} shows our numerical results for $\ln\left|\textrm{Im}\left(\omega/\mu\right)\right|$ versus $\ln \left(T/\mu\right)$ for $\tau=10^{-4}$, $k/\mu=10^{-2}$, and increasing $\talpha$ from $\talpha=10^{-2}$ (red dots) to $\talpha=10$ (purple stars). As mentioned in sec.~\ref{sec:poles}, for $\tau=10^{-4}$ and $\talpha \gtrsim 0.07$ the HZS poles cross over to hydrodynamic sound poles (figs.~\ref{figa01},~\ref{figa01cu}, ~\ref{figa01re}, and~\ref{figa01im}), but when $\talpha \lesssim 0.07$ the \textit{relativistic} poles cross over to hydrodynamic sound (figs.~\ref{figa001}, ~\ref{figa001cu}, ~\ref{figa001re}, and~\ref{figa001im}). In fig.~\ref{sound_attenuation_tau_0p0001} as we change $T/\mu$ we always follow the poles that cross over to hydrodynamic sound, so for $\talpha= 10^{-2}< 0.07$ (red dots) the poles become relativistic at low $T/\mu$, rather than HZS. Nevertheless, for all $\talpha$, including $\talpha< 0.07$, fig.~\ref{sound_attenuation_tau_0p0001} shows that as $\ln\left(T/\mu\right)$ increases, $|\textrm{Im}\left(\omega\right)|$ first scales as $T^0$, then as a power of $T$ slightly less than $T^2$, then has a maximum, and finally scales as $T^{-1}$. In other words, the behavior is similar to $\tau=10^{-4}$ in fig.~\ref{fig:sound_attenuation}. In particular, increasing $\talpha$ appears to have two main effects, an overall re-scaling of $\ln\left|\textrm{Im}\left(\omega/\mu\right)\right|$ to smaller total value, without changing its shape, and shifting the maximum to higher $T/\mu$.

The fact that changing $\talpha$ appears to re-scale the sound attenuation sounds suspiciously like an effect of the gravity theory's scaling symmetry, $\alpha \to \lambda \, \alpha$ and $F_{MN} \to \lambda^{-1} \, F_{MN}$. However, that symmetry acts as $\talpha \to \lambda \, \talpha$ and $T/\mu \to \lambda \, T/\mu$, and similarly for $\omega/\mu$ and $k/\mu$, and will thus not only re-scale the axes of fig.~\ref{sound_attenuation_tau_0p0001}, but also re-scale $k/\mu$. The results of fig.~\ref{sound_attenuation_tau_0p0001} thus cannot be determined by the scaling symmetry alone.

\begin{figure}[t!]
\begin{center}
\includegraphics{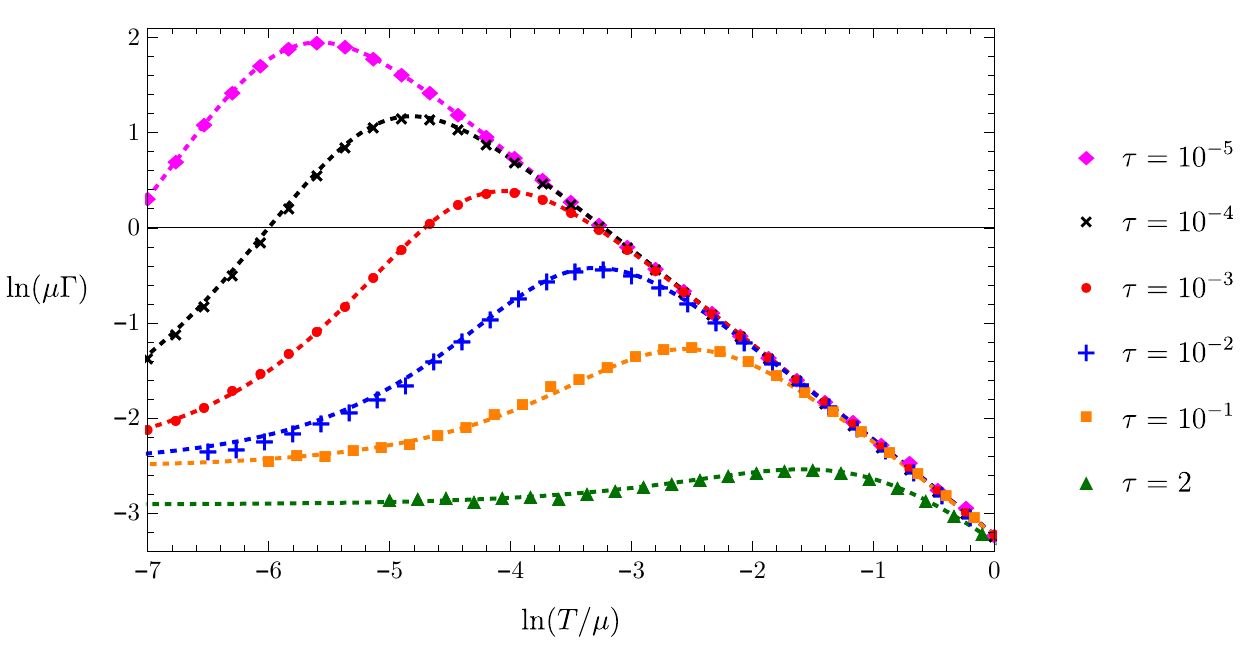}
\caption{\label{hydro_prediction} Our numerical results for $\ln\left(\m\G\right)$ versus $\ln \left(T/\mu\right)$ for $\talpha=1$, $k/\mu=10^{-2}$, and $\tau= 10^{-5}$ (pink diamonds), $10^{-4}$ (black crosses), $10^{-3}$ (red dots), $10^{-2}$ (blue plus signs), $10^{-1}$ (orange squares), and $2$ (green triangles). We obtain $\Gamma$ by numerically fitting $-\G \, k^2 + \delta k\,^4$ to the sound pole's  $\textrm{Im}\left(\omega\right)$ over a range of small $k/\mu$. The dashed lines show the corresponding results using the hydrodynamic form of eq.~\eqref{eq:sound_mode_hydro} with $d=3$, namely $\Gamma=\frac{1}{2}\frac{\eta}{\varepsilon+P}$. Clearly our numerical results for $\Gamma$ agree with the hydrodynamic form for all $\tau$ and $T/\mu$.}
\end{center}
\end{figure}

Nevertheless, we can clarify the role of the scaling symmetry using a key numerical result of refs.~\cite{Davison:2013bxa,Davison:2013uha}: in AdS-RN, for $\omega$ and $k$ sufficiently small compared to $\mu$, the hydrodynamic form of the sound attenuation constant (eq.~\eqref{eq:sound_mode_hydro} with $d=3$), $\Gamma = \frac{1}{2} \frac{\eta}{\varepsilon +P}$, is valid not just in the hydrodynamic regime, but for all $T/\mu$, down to and including $T/\mu=0$. To check whether the same is true in our model, we fit our numerical results for the sound pole's $|\textrm{Im}\left(\omega\right)|$ to a form $\Gamma \, k^2 + \delta \, k^4$ over a range of small $k/\mu$, with fit parameters $\Gamma$ and $\delta$. Fig.~\ref{hydro_prediction} shows the resulting $\ln\left(\mu\Gamma\right)$ versus $\ln\left(T/\mu\right)$, for $\talpha=1$, $k/\mu=10^{-2}$, and increasing $\tau$ from $\tau=10^{-5}$ (pink diamonds) to $\tau=2$ (green triangles). Fig.~\ref{hydro_prediction} also shows the corresponding value of $\Gamma$'s hydrodynamic form for each $\tau$ (dotted lines). The hydrodynamic form indeed agrees precisely with our numerical results for all $\tau$ and $T/\mu$. In short, our results agree with and extend those of refs.~\cite{Davison:2013bxa,Davison:2013uha}: for charged black branes in Einstein-DBI theory, as for AdS-RN, the hydrodynamic form $\Gamma = \frac{1}{2} \frac{\eta}{\varepsilon +P}$ is in fact valid for all $T/\mu$.

In hydrodynamics the shear diffusion constant is also $\propto \eta/(\varepsilon+P)$. A key result of ref.~\cite{Gushterov:2018nht} for the Einstein-DBI charged black brane is that the numerical results for the shear diffusion constant also agree with the hydrodynamic form for all $\tau$ and $T/\mu$.

Our model, like all rotationally-invariant holographic models, has $\eta = s/(4\pi)$~\cite{Policastro:2001yc,Buchel:2003tz,Kovtun:2004de}, so the hydrodynamic form $\Gamma = \frac{1}{2} \frac{\eta}{\varepsilon + P}$ is in fact completely determined by thermodynamics. We can eliminate $s$ from $\Gamma$ using $\eta=s/(4\pi)$, $\varepsilon + P = sT + \mu \Jt$, and as mentioned below eq.~\eqref{jt}, $\Jt =  \tau \talpha^2 Q \, s/(2\pi)$, giving
\beq
\label{eq:thermogamma}
\Gamma = \frac{1}{2} \frac{\eta}{\varepsilon + P} = \frac{1}{8 \pi T + 4 \mu \, \tau \talpha^2 \, Q}.
\eeq
This form of $\Gamma$ makes clear that the probe limit, $\tau\to0$ with $\talpha$ and $T/\mu$ fixed, gives the AdS-SCH result $\Gamma=1/(8 \pi T)$, and that the extremal limit, $T/\mu\to0$ with $\tau$ and $\talpha$ fixed, gives $\Gamma \to (4 \mu \tau \talpha^2 Q_{\textrm{ext}})^{-1}\neq 0$.

The form of $\Gamma$ in eq.~\eqref{eq:thermogamma} also enables us to explain some of our numerical results. For example, to clarify the role of the gravity theory's scaling symmetry, we move a factor of the scaling-invariant product $\talpha \mu$ to the left-hand-side,
\beq
\label{talphamugamma}
\talpha \, \mu \, \Gamma = \frac{1}{8\pi\frac{1}{\talpha} \frac{T}{\mu}+4\tau \, \talpha Q},
\eeq
and observe from eq.~\eqref{tmu} that $\frac{1}{\talpha} \frac{T}{\mu}$ is a function only of the scaling-invariant quantities $\tau$ and $\talpha \, Q$,
\beq
\label{talphatmu}
\frac{1}{\talpha}\frac{T}{\mu} = \frac{3+\tau\left(1-\sqrt{1+\tilde{\alpha}^2 Q^2}\right)}{4\pi \, \talpha Q \,\,_2F_1\left(\frac{1}{2},\frac{1}{4};\frac{5}{4};-\tilde{\alpha}^2 Q^2\right)},
\eeq
which makes clear that $\talpha \, \mu \, \Gamma$ is scaling-invariant. In particular, fixing $\tau$ and changing $\talpha$ does not change the form of $\Gamma$ as a function of $T/\mu$, but rather just acts as a re-scaling. That almost but not quite explains the results of fig.~\ref{sound_attenuation_tau_0p0001}, because that figure shows the full sound attenuation, $|\textrm{Im}\left(\omega\right)|$, not just the order $k^2$ contribution. As a result, the different curves in fig.~\ref{sound_attenuation_tau_0p0001} are not related by re-scalings alone, as mentioned above. Nevertheless, the scaling symmetry does explain the general pattern apparent in fig.~\ref{sound_attenuation_tau_0p0001}.

Ideally we would invert eq.~\eqref{talphatmu} to find $\talpha \, Q$ as a function of $\tau$ and $\frac{1}{\talpha} \frac{T}{\mu}$, but that is impossible to do in full generality, due to the hypergeometric function in the denominator on the right-hand side. However, we can invert eq.~\eqref{talphatmu} in certain limits. For example, suppose $\tau$ is small, such that we can take $\tau=0$ on the right-hand side of eq.~\eqref{talphatmu}. Suppose we then take $T/\mu\ll 1$, which in eq.~\eqref{talphatmu} with $\tau=0$ means $Q \to \infty$. Expanding the hypergeometric function at large argument and solving for $Q$ then gives
\beq
\label{eq:largeq}
Q =\tilde{\alpha}\left(\frac{3}{\sqrt{\pi}\,\Gamma\left(\frac{1}{4}\right)^{2}}\right)^{2}\left(\frac{\mu}{T}\right)^{2}+ {\cal O}\left(\frac{\mu}{T}\right).
\eeq
Dropping all sub-leading terms from eq.~\eqref{eq:largeq} and inserting the result into eq.~\eqref{talphamugamma} gives
\beq
\label{eq:approx}
 \mu \Gamma = \left(\frac{8\pi T}{\mu}+ \frac{36\,\tau \, \talpha^3}{\pi\,\Gamma\left(\frac{1}{4}\right)^4}\left(\frac{\mu}{T}\right)^{2}\right)^{-1}.
\eeq
The approximations leading to eq.~\eqref{eq:approx} are brutal. For example, when $T/\mu\ll1$ the ${\cal O}\left(\frac{\mu}{T}\right)$ term in eq.~\eqref{eq:largeq} is \textit{larger} than the $T/\mu$ term in eq.~\eqref{talphamugamma} and hence should not be dropped. Indeed, eq.~\eqref{eq:approx} fails to capture key features of the actual result, for instance, when $T/\mu \to 0$ eq.~\eqref{eq:approx} gives $\mu\Gamma \to 0$, while the actual limit is non-zero. Eq.~\eqref{eq:approx} nevertheless provides a surprisingly good approximation to certain features. In particular, eq.~\eqref{eq:approx} manifestly describes a transition in $\mu\Gamma$'s scaling from $T^2$ to $T^{-1}$, as expected at small $\tau$, and has a maximum whose position $\left(T/\mu\right)_{\textrm{max}}$ is given by
\beq
\label{eq:max}
\left(\frac{T}{\mu}\right)^{3}_{\textrm{max}}=\frac{9\,\tilde{\alpha}^{3}\tau}{\pi^{2}\Gamma\left(\frac{1}{4}\right)^{4}}.
\eeq
Remarkably, eq.~\eqref{eq:max} describes the actual $\left(T/\mu\right)_{\textrm{max}}$ extremely well---even \textit{away} from small $\tau$. In fact, eq.~\eqref{eq:max} agrees with our numerical results for $\left(T/\mu\right)_{\textrm{max}}$ for all values of $\tau$ and $\talpha$ that we have checked! However, eqs.~\eqref{eq:approx} and~\eqref{eq:max} do not provide a good approximation to the \textit{height} of the maximum, i.e. the \textit{value} of $\mu \Gamma$ at $\left(T/\mu\right)_{\textrm{max}}$, and indeed the approximation to the height grows worse as $\tau$ increases. For example, when $\talpha=1$ and $\tau=10^{-5}$, eqs.~\eqref{eq:approx} and~\eqref{eq:max} suggest a maximum value $\ln\left(\mu\Gamma\right) \approx 1.96$, very close to the actual value in fig.~\ref{hydro_prediction} (pink diamonds), but when $\tau=2$ they suggest a maximum value $\ln\left(\mu\Gamma\right) \approx -2.11$, while the actual value in fig.~\ref{hydro_prediction} (green triangles) is close to $-3$.

As a side comment, Einstein-DBI charged black brane solutions are known for any value of the CFT spacetime dimension $d$~\cite{Dey:2004yt,Cai:2004eh,Pal:2012zn,Tarrio:2013tta}. If the fact that $\Gamma$ takes the hydrodynamic form at all $T/\mu$ persists to all $d$, then using the results for the thermodynamics for arbitrary $d$, and repeating the approximations leading to eq.~\eqref{eq:approx}, gives a transition from $T^{d-1}$ to $T^{-1}$. Apparently in Einstein-DBI models the $T^2$ scaling similar to the LFL thermal collisionless regime may be unique to $d=3$.

In summary, we have three main results for sound attenuation. First is that fixing $\talpha$ and increasing $\tau$ preserves the $T^0$ and $T^{-1}$ scalings at low and high $T/\mu$, respectively, but suppresses the $T^2$ scaling at intermediate $T/\mu$ to a lower (but non-zero) power. Second is that both the full $\textrm{Im}\left(\omega\right)$ and $\Gamma$ are similar in form to that of a LFL for all $\tau$ and $\talpha$ we accessed, including in particular a maximum that can provide a definition for the crossover. Third is that $\Gamma$ assumes the hydrodynamic form, $\Gamma=\frac{1}{2} \frac{\eta}{\varepsilon +P}$, for all $\tau$, $\talpha$, and $T/\mu$ we accessed, which provided us with an excellent approximation for the location of the maximum, eq.~\eqref{eq:max}.

This third result is similar to phenomena observed in other back-reacted models, including AdS-RN~\cite{Bhattacharyya:2007vs,Davison:2013bxa,Davison:2013uha}. The proposal of refs.~\cite{Bhattacharyya:2007vs,Davison:2013bxa,Davison:2013uha} was therefore that hydrodynamics remains reliable even for energies $\gg T/\mu$, outside the usual hydrodynamic regime, as long as $k \ll \mu$ or $T$. In other words, in several holographic models hydrodynamics appears to remain reliable at distances shorter than a mean free path $\propto 1/T$ at high $T/\mu$ but $\propto 1/\mu$ at low $T/\mu$.

\section{Discussion and Outlook}
\label{summary}

For the large-$N$, strongly-coupled CFT states with non-zero $T$ and $\mu$ holographically dual to the Einstein-DBI charged black brane, we studied how the poles of $G_{tt}$ and $G_J$, the associated spectral functions, and the sound dispersion evolved with increasing $T/\mu$, and how that evolution depended on $\tau$ and $\talpha$. For fixed $k/\mu$, we found that the probe limit definition of crossover, as a collision of HZS poles on the imaginary $\omega/\mu$ axis that produces the charge diffusion pole, was viable only for sufficiently small $\tau$ or $\talpha$. The AdS-RN definition of crossover, as a transfer in dominance from sound to charge diffusion peaks in $\rho_J$, was viable only for sufficiently large $\tau$ or $\talpha$. However, outside of the probe limit, the LFL definition of the crossover, as a maximum in the sound attenuation, was always viable. Moreover, the sound attenuation constant, $\Gamma$, took the hydrodynamic form, even outside the hydrodynamic regime, and hence in these holographic models was completely determined by thermodynamics.

\begin{figure}[t!]
	\begin{center}
		\begin{subfigure}{0.58\textwidth}
			\includegraphics[width=\textwidth]{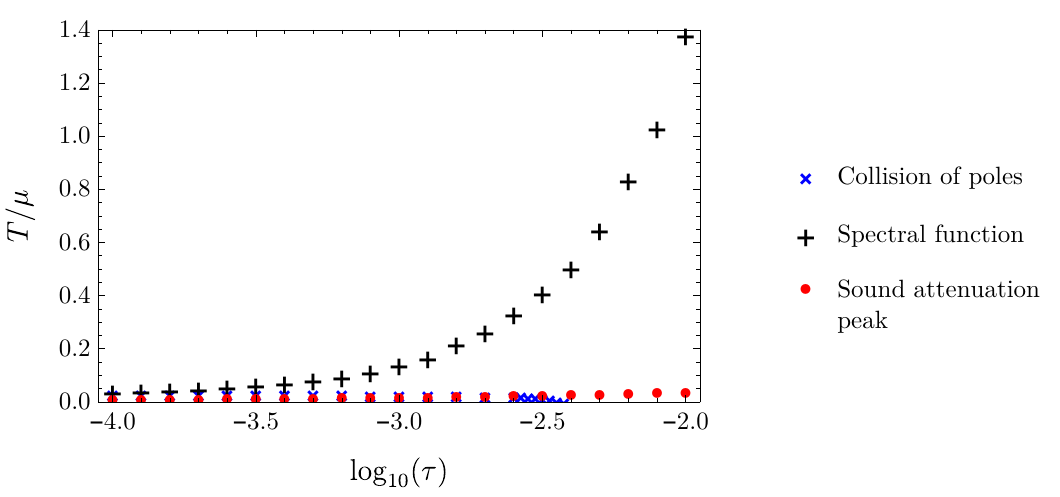}
			\caption{\label{fig:compare} Comparison of crossover definitions.}
		\end{subfigure}
		\begin{subfigure}{0.41\textwidth}
			\includegraphics[width=\textwidth]{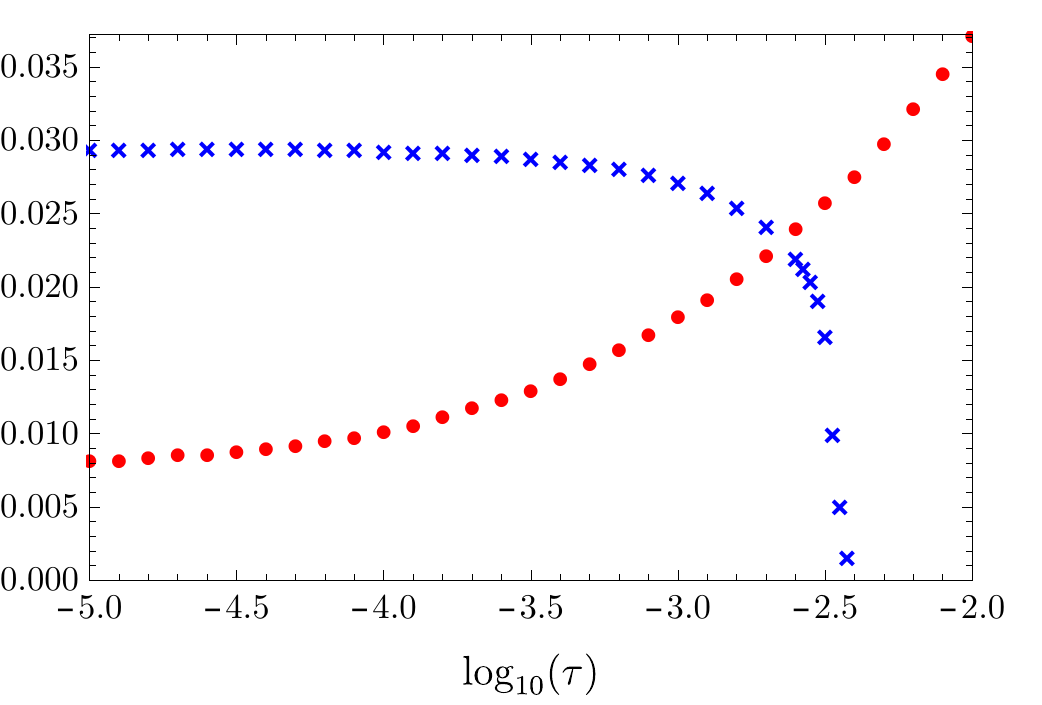}
			\caption{\label{fig:comparecu} Close-up of (a)}
		\end{subfigure}
		\caption{\label{fig:comparedefs} Our numerical results for the crossover value of $T/\mu$ as a function of $\textrm{log}_{10}\left(\tau\right)$, for fixed $\talpha=1$, using the three different definitions: the probe limit definition, via a collision of poles that produces the charge diffusion pole (blue crosses), the AdS-RN definition, via a transfer of dominance from sound to charge diffusion peak in $\rho_J$ (black plus signs), and the LFL definition, via the sound attenuation maximum (red dots).}
	\end{center}
\end{figure}

Fig.~\ref{fig:comparedefs} summarizes our numerical results for the crossover value of $T/\mu$ as a function of $\textrm{log}_{10}\left(\tau\right)$, for fixed $k/\mu=10^{-2}$ and $\talpha=1$, using the three different definitions: the probe limit definition (blue crosses), the AdS-RN definition (black plus signs), and the LFL definition (red dots). The AdS-RN definition gives crossover $T/\mu$ larger than the others by about an order of magnitude. The AdS-RN and LFL definitions both increase without bound as $\tau$ increases, while the probe limit definition instead decreases, eventually dropping to zero at the critical value for $k/\mu=10^{-2}$ and $\talpha=1$, $\tau \approx 3.2 \times 10^{-3}$ or equivalently $\textrm{log}_{10}\left(\tau\right) \approx -2.49$.

Our results motivate the following speculation about the effective description of these strongly-interacting quantum compressible states. In the probe limit $\tau=0$, at high $T/\mu$ the highest poles are the two sound poles in $T^{\mu\nu}$'s two-point function and the charge diffusion pole in $J^{\mu}$'s two-point function. The effective description of long-wavelength excitations is thus hydrodynamics, though with the poles in $T^{\mu\nu}$ and $J^{\mu}$'s two-point functions decoupled due to the probe limit. At low $T/\mu$ the two highest pairs of poles are the propagating HZS and relativistic poles. The effective description of long wavelength excitations thus appears to be uncharged hydrodynamics for the poles of $T^{\mu\nu}$'s two point functions~\cite{Kovtun:2005ev} combined with the effective theory of a weakly-conserved current~\cite{Chen:2017dsy} for the poles of $J^{\mu}$'s two-point functions.

For all non-zero $\tau$, sufficiently small $k/\mu$, and all $T/\mu$, our highest poles are similar to those of AdS-RN, namely the two highest poles are sound and the next highest pole is purely imaginary. This is expected at large $T/\mu$, where hydrodynamics is a reliable effective description, while small $T/\mu$ takes us outside the usual hydrodynamic regime. For AdS-RN such behavior was interpreted as evidence that, for excitations with small $k/\mu$, hydrodynamics remains a valid effective description for \textit{all} $T/\mu$, down to and including $T/\mu=0$~\cite{Davison:2011uk}.

However, in our Einstein-DBI model the range of $k/\mu$ where hydrodynamics is valid decreases as $\tau$ decreases. For example, at $\tau=10^{-2}$ and $k/\mu = 10^{-2}$ the second highest pole is purely imaginary for all $T/\mu$ we considered: see fig.~\ref{plane_tau_0p01}. However for $\tau=10^{-4}$ and $T/\mu=10^{-2}$ the second highest pole is purely imaginary only for $k/\mu \lesssim 2\times10^{-3}$, as shown in fig.~\ref{fig:dispersionsmalltau}. Our results therefore suggest that, outside of the probe limit, for the smallest $k/\mu$ (largest distances) hydrodynamics is indeed a reliable effective description for all $T/\mu$, similar to AdS-RN, but as $k/\mu$ increases (shorter distances) the effective description becomes uncharged hydrodynamics combined with the effective theory of a weakly conserved current~\cite{Chen:2017dsy} for the charged sector. The $k/\mu$ where the transition between effective descriptions occurs increases as $\tau$ increases, where $\tau$ counts the number of charged fields (such as quark flavors). Our results thus reveal how, in a strongly-interacting system, changing the number of charged fields can dramatically change the effective description, and indeed produce non-hydrodynamic modes.

More generally, low-temperature sound modes have been relatively under-explored in holography, especially outside of the probe limit. However, our results, combined with the accumulated body of evidence about low-temperature sound modes in holography, raise many questions relevant to strongly-coupled systems, and worthy of future research.

In our model, one immediate task would be to attempt analytic, rather than numerical, calculations of the leading powers of $k$ in the imaginary parts of correlators at exactly $T/\mu=0$. We expect that, as in the back-reacted models of refs.~\cite{Edalati:2010pn,Davison:2013bxa}, these will be fixed by dimensions of operators in the $(0+1)$-dimensional CFT dual to the near-horizon $AdS_2$. In the Einstein-DBI model, a key question is how these dimensions depend on $\tau$ and $\talpha$.

More generally, however, the primary task at hand is simply to continue searching for low-temperature sound modes in holographic quantum compressible matter. In what cases does HZS appear? Is it universal? If not, then what distinguishes systems with HZS from those without? To date, HZS has appeared in systems with and without extensive entropy at $T/\mu=0$, with heat capacity scaling as various powers of $T$, etc. Indeed, so far only two patterns have emerged. First, HZS appears in systems with some form of non-linearity. In particular, a probe Maxwell action does not produce HZS. To obtain HZS we must introduce non-linearities, either by replacing the probe Maxwell action with the probe DBI action, or by allowing the Maxwell action to back-react, so that we must solve the Einstein equation, which is non-linear. Second, HZS appears in systems with non-zero spectral weight at $\omega=0$ over a finite range of $k$, up to a characteristic value of $k$, in a fashion reminiscent of a Fermi-Dirac distribution~\cite{Hartnoll:2016apf}. Are these patterns universal? Moreover, when HZS does appear, how does it evolve in the crossover to hydrodynamics?

Low-temperature sound modes can be further probed by a variety of deformations. As just one example, how does HZS respond to an external magnetic field? General arguments, such as Kohn's theorem for non-relativistic electrons with pair-wise interactions, and evidence from holographic probe brane models~\cite{Jokela:2012vn,Goykhman:2012vy,Brattan:2012nb,Brattan:2013wya}, suggest that a magnetic field will gap HZS. In the Einstein-DBI model with $d=3$, magnetically-charged solutions are straightforward to obtain via electric-magnetic duality. Does the magnetic field gap HZS in such models, and if so, then how does the gap depend on $\tau$ and $\talpha$?

Of course, the over-arching question is what lessons HZS may teach us about real strongly-coupled systems. Do real quantum compressible systems and non-Fermi liquids, such as graphene, the cuprates, the heavy fermion compounds, etc., support sound modes? Recent evidence suggests that in LFLs in two spatial dimensions described by kinetic theory, both zero sound and hydrodynamic sound are replaced by plasmons~\cite{Lucas:2017idv,PhysRevB.97.115449}. However, the most important question remains: what types of effective theories give rise to low-temperature sound modes, and what do those sound modes, and their crossover to hydrodynamic sound, tell us about the underlying degrees of freedom?

\section*{Acknowledgements}

We thank T.~Andrade, M.~Blake, R.~Davison, S.~Grozdanov, C.~Herzog, N.~Kaplis, A.~Lucas, I.~Papadimitriou, G.~Policastro, A.~Prybitoks, and B.~Robinson for useful conversations and correspondence. We also thank P.~Kovtun and A.~Starinets for useful comments on an early draft of the paper. N.~I.~G. and A.~O'B. were partially supported by the Royal Society research grant ``Strange Metals and String Theory'' (RG130401). N.~I.~G. was also supported by the European Research Council under the European Union's Seventh Framework Programme (ERC Grant agreement 307955). A.~O'B. is a Royal Society University Research Fellow.  R.~R. acknowledges support from STFC through Consolidated Grant ST/L000296/1.

\appendix
\section{Holographic Calculation of Green's Functions}

In this appendix we discuss technical details of our holographic calculations of the retarded Green's functions, their poles, and the spectral functions. We use standard techniques, and in particular the method of ref.~\cite{Kaminski:2009dh}.

We want to study the CFT's response to linearized perturbations about the equilibrium state described holographically by the solution in eq.~\eqref{bgsol}. Specifically, we want to compute the retarded Green's functions of $T^{\mu\nu}$ and $J^{\mu}$ as functions of complex frequency $\omega$ and real momentum $k$. In a retarded Green's function with fixed $k$, a pole in the complex $\o$ plane at position $\o_*$ with $\textrm{Re}(\omega_*)\neq0$ and $\textrm{Im}(\omega_*)\neq0$ represents a propagating excitation, with $|\textrm{Im}(\omega_*)|\propto$ the excitation's decay rate. If $|\textrm{Im}(\omega_*)| < |\textrm{Re}(\omega_*)|$, then the excitation is a long-lived quasi-particle, like a sound wave. If $\textrm{Re}(\omega_*)=0$ then the excitation is dissipative rather than propagating, like a charge diffusion mode. Stability requires $\textrm{Im}(\omega_*)\leq 0$, since $\textrm{Im}(\omega_*)>0$ means the mode grows without bound over time. The mode with smallest $|\textrm{Im}(\omega_*)|$ dominates the late-time response, as all other modes will decay faster. We focus on the ''highest'' poles, \textit{i.e.} those closest to the $\textrm{Re}(\omega)$ axis, with relatively small $|\textrm{Im}(\omega_*)|$.

For a set of operators $\mathcal{O}_i$ with $i=1,2,\ldots$, the matrix of spectral functions, $\rho_{ij}(\o,k)$ is defined as the anti-Hermitian part of the matrix of retarded Green's functions, $G_{ij}(\o,k)$:
\beq
\label{eq:spectraldef}
\rho_{ij}(\o,k) \equiv i \left(G_{ij}(\o,k)-G_{ji}(\o,k)^*\right).
\eeq
In general, a pole in $G_{ij}(\o,k)$ at $\o_*$ produces a peak in $\rho_{ij}(\o,k)$ as a function of $\textrm{Re}(\o)$, with position $\propto \textrm{Re}(\o_*)$, width $\propto 2 |\textrm{Im}(\omega_*)|$, and height $\propto$ the pole's residue divided by $|\textrm{Im}(\omega_*)|$.

In holography, the CFT's generating functional is proportional to the on-shell bulk action~\cite{Gubser:1998bc,Witten:1998qj}. To compute $G_{ij}(\o,k)$ and hence $\rho_{ij}(\o,k)$ holographically, we must thus solve for fluctuations of bulk fields with in-going boundary conditions at the horizon, plug the solutions into the bulk action, renormalize~\cite{deHaro:2000xn,Skenderis:2002wp,Edalati:2010pn}, and take two functional derivatives~\cite{Son:2002sd,Policastro:2002se,Starinets:2002br,Policastro:2002tn,Kovtun:2003wp,Kovtun:2005ev}. However, we can obtain the location $\o_*$ of a pole in $G_{ij}(\o,k)$ simply by solving the bulk linearized equations of motion, without evaluating the on-shell action: $\o_*$ corresponds to an $\o$ value where a linearized, in-going, \textit{normalizable} solution, namely a QNM, exists~\cite{Son:2002sd,Kovtun:2005ev}.

We thus introduce fluctuations around the solutions $g_{MN}(z)$ and $A_M(z)$ in eq.~\eqref{bgsol}, with dependence on $z$, $t$, and without loss of generality due to rotational invariance, $x$ but not $y$,
\beq
g_{MN}(z) \to g_{MN}(z) + \delta g_{MN}(z,t,x), \qquad A_M(z) \to A_M(z) + \delta A_M(z,t,x).
\eeq
We next linearize the equations of motion in $\delta g_{MN}(z,t,x)$ and $\delta A_M(z,t,x)$, and then introduce Fourier transforms in $t$ and $x$,
\beq
\delta g_{MN}(z,t,x) \equiv \int \frac{d\o \, dk}{\left(2\pi\right)^2} \, e^{- i \omega t + i k x} \, \delta g_{MN}(z,\o,k),
\eeq
and similarly for $\delta A_M(z,\omega,k)$. We hence obtain fourteen equations for the ten components of $\delta g_{MN}(z,\o,k)$ and four components of $\delta A_M(z,\o,k)$. However, following refs.~\cite{Edalati:2010pn,Davison:2011uk,Tarrio:2013tta}, for the sound channel we can reduce these to only two equations, in two steps, as follows.

At linearized order fluctuations in different representations of the parity transformation $y \to -y$ decouple. The sound modes appear in the parity-even sector. The first step is thus to set to zero the parity-odd fluctuations, $(\delta g_{zy},\delta g_{ty},\delta g_{xy},\delta A_y)$, leaving ten equations for the parity-even fluctuations, $(\delta g_{zz},\delta g_{tt},\delta g_{xx},\delta g_{yy},\delta g_{zt},\delta g_{zx},\delta g_{tx},\delta A_z,\delta A_t,\delta A_x)$. These ten equations are cumbersome and unilluminating, so we will not write them here. They are special cases of the equations written explicitly in the appendix of ref.~\cite{Tarrio:2013tta}.\footnote{To obtain our equations from those in ref.~\cite{Tarrio:2013tta}, in ref.~\cite{Tarrio:2013tta}'s bulk action send $d \to 3$, $\kappa^2 \to 8 \pi G$, $L\to L_0$, $T_b/(2\kappa^2) \to T_D$, $\lambda \to \a$, and $\alpha \to 0$ and $\beta \to 0$ so that $Z_1 \to 1$ and $Z_2 \to 1$, and in in ref.~\cite{Tarrio:2013tta}'s solution send $r \to z$, $\phi \to 0$, $\varphi \to 0$, $\rho \to \alpha Q/z_H^2$, $h_{\mu\nu} \to \delta g_{MN}$, $a_{\mu} \to \delta A_M$, $\alpha_0 \to \delta A_t$, $\alpha_1 \to \delta A_x$, and $q \to k$.} Six of these equations are second order (dynamical), four from Einstein's equation and two from Maxwell's equation, while the other four equation are first order (constraints), three from the radial components of Einstein's equation and one from the radial component of Maxwell's equation. The second-order equations are in fact linear combinations of derivatives of the first-order equations, hence the latter contain no independent information.

The second step is to form diffeomorphism- and $U(1)$-gauge invariant linear combinations of the fluctuations~\cite{Kovtun:2005ev}. Any sum of diffeomorphism- and gauge-invariant fluctuations is again diffeomorphism- and gauge-invariant, so we must make a choice. For example, one choice is to use Ishibashi-Kodama ``master fields''~\cite{Kodama:2003kk}, involving $z$ derivatives of fluctuations, which have the advantage of producing two \textit{decoupled} equations~\cite{Edalati:2010pn}. However, we will instead use the linear combinations of refs.~\cite{Davison:2011uk,Davison:2013bxa,Tarrio:2013tta}, involving fluctuations with no $z$ index, which ultimately lead to two \textit{coupled} equations. The fields of refs.~\cite{Davison:2011uk,Davison:2013bxa,Tarrio:2013tta} have several advantages over those of Ishibashi-Kodama, for example they make transparent not only the mapping from the fields' boundary values to the dual operator sources~\cite{Michalogiorgakis:2006jc,Edalati:2010pn,Davison:2011uk} but also the fact that the CFT Ward identities for $T_{\mu\nu}$'s and $J^{\mu}$'s Green's functions are satisfied~\cite{Davison:2011uk}. We thus choose the diffeomorphism- and gauge-invariant linear combinations of refs.~\cite{Davison:2011uk,Davison:2013bxa,Tarrio:2013tta},
\begin{subequations}
\label{zdef}
\bea
Z_1 & \equiv & k \, \delta a_t + \omega \, \delta a_x + \frac{1}{2} \, k \, z \, F_{zt} \, \delta g^y_{~y},  \\
Z_2 & \equiv & - k^2 \, f \, \delta g^t_{~t} + \omega^2 \, \delta g^x_{~x} + 2 \, \o k \, \delta g^x_{~t} + \left(-\o^2+ k^2 \, f - \frac{1}{2} \, k^2 \, z \, f'\right) \delta g^y_{~y},
\eea
\end{subequations}
where we raised an index on $\delta g_{MN}$ using the background metric $g_{MN}(z)$ in eq.~\eqref{bgsol}. The equations of motion of $Z_1$ and $Z_2$ are of the form
\begin{subequations}
\label{eoms}
\bea
Z_1'' & + & A_1 Z_1' + A_2 Z_2' + A_3 Z_1 + A_4 Z_2 = 0, \\
Z_2'' & + & B_1 Z_1' + B_2 Z_2' + B_3 Z_1 + B_4 Z_2 = 0,
\eea
\end{subequations}
where if we define the notation
\(
\mathcal{F}(z) \equiv \sqrt{1 - \talpha^2 z^4 F_{tz}^2},
\)
then the coefficients
can be written as
\small{
\begin{align}
	A_1 &= \frac{1}{f \mathcal{F}^2 z \left(f k^2 \mathcal{F}^2-\omega ^2\right) \left(k^2 \left(z f'-4 f\right)+4 \omega ^2\right)}
	\nonumber \\ & \phantom{=}
		\times \biggl\{
			k^4 f \mathcal{F}
		\left[  \tau \left(\mathcal{F}^2-1\right) \left(2 f \left(\mathcal{F}^2+1\right)-z f'\right)-3 z^6 f  \mathcal{F}^2  \mathcal{F}' \left(f/z^4\right)'\right]
		+ 4 \mathcal{F} \omega ^4 z \left(f \mathcal{F}'-\mathcal{F} f'\right)
	\nonumber \\ & \phantom{= \times \biggl[}
		+ k^2 \omega ^2 \mathcal{F}  \left[-4 z f^2  \mathcal{F}'\left(3 \mathcal{F}^2+1\right) +f \left(z^2 f' \mathcal{F}'+4 \mathcal{F} z f'-4 \mathcal{F}^2 \tau +4 \tau \right)-\mathcal{F} z^2
		   f'^2\right]
	\biggr\},
	\\
	A_2 &= \frac{k \mathcal{F}^2}{z^4\talpha^2 F_{tz}\mathcal{F}^3 \left(f k^2 \mathcal{F}^2-\omega ^2\right) \left(k^2 \left(z f'-4 f\right)+4 \omega ^2\right)^2}
	\nonumber \\ & \phantom{=}
	\times \biggl\{
		k^4 \Bigl[
			 \left(1-\mathcal{F}^2\right) \Bigl(-4 f^2 \mathcal{F} \left(\mathcal{F}^4+3
			   \mathcal{F}^2-6\right)+2 f \left(-3 z \mathcal{F} f'+\tau -\tau  \mathcal{F}^4\right)
	\nonumber \\ & \hspace{1.5cm}
				+z f' \left(z \mathcal{F}^3
			   f'+\tau  \left(\mathcal{F}^2-1\right)\right)\Bigr)-f \mathcal{F} \left(\mathcal{F}^2-3\right) \left(z f'-4
			   f\right) \left(z \mathcal{F} \mathcal{F}'-2 \mathcal{F}^2+2\right)
		\Bigr]
	\nonumber \\ & \phantom{= \times \biggl[}
		+ k^2 \omega ^2 \Bigl[
			   -2 \left(\mathcal{F}^2-1\right) \left(\mathcal{F} \left(\left(2 \mathcal{F}^2-1\right) z f'+f \left(6 \mathcal{F}^2+2\right)+2 \mathcal{F} \tau \right)-2 \tau
			      \right)
			  \nonumber \\ & \hspace{6cm}
			  - z \mathcal{F}' \left(\left(\mathcal{F}^2+1\right) z f'+4 f \left(\mathcal{F}^4-4 \mathcal{F}^2-1\right)\right)
		   \Bigr]
	\nonumber \\ & \phantom{= \times \biggl[}
		-4 \omega ^4 \left(\left(\mathcal{F}^2+1\right) z \mathcal{F}'-2 \mathcal{F} \left(\mathcal{F}^2-1\right)\right)
	\biggr\},
\end{align}
\begin{align}
	A_3 &= \frac{1}{f^2 \mathcal{F}^3 z^2 \left(f k^2 \mathcal{F}^2-\omega ^2\right) \left(k^2 \left(z f'-4 f\right)+4 \omega ^2\right)^2}
	\nonumber \\ & \phantom{=}
	\times \biggl\{
		2  k^6  \tau f^2 \mathcal{F} \Bigl[4 f^2 \mathcal{F}^3 \left(\mathcal{F}^4-1\right) +z f' \left(\mathcal{F}^2-1\right) \left(2  z f'\mathcal{F}^3 + \tau\left(\mathcal{F}^2-1\right) \right)
	\nonumber \\ & \hspace{2cm}
		+z f  \mathcal{F}^2  \mathcal{F}' \left(\mathcal{F}^2-3\right) \left(z f'-4 f\right)-2 f \left(\mathcal{F}^2-1\right) \left( z
				   f' \mathcal{F}^3\left(\mathcal{F}^2+4\right)+ \tau \left(\mathcal{F}^4-1\right)\right)\Bigr]
	\nonumber \\ & \phantom{= \times \biggl[}
		-k^8 z^{12}  f^2 \mathcal{F}^7 \left(f/z^4\right)'^2
		+ 2  k^6 \omega ^2 z^7 f \mathcal{F}^5  \left(f/z^4\right)' \left(z f'-4 f \left(\mathcal{F}^2+1\right)\right)
	\nonumber \\ & \phantom{= \times \biggl[}
		+ 2 k^4 \omega ^2  \tau  f  \mathcal{F} \Bigl[z f \mathcal{F}' \Bigl( z f'\left(\mathcal{F}^2+1\right)+4 f \left(\mathcal{F}^4-4
		   \mathcal{F}^2-1\right)\Bigr)
		  \nonumber \\ & \hspace{3.1cm}
		   -\left(\mathcal{F}^2-1\right) \Bigl(-2 z f' f \mathcal{F} \left(7 \mathcal{F}^2+2\right) +\mathcal{F} z^2 f'^2
		   \nonumber \\ & \hspace{5.5cm}
		   +4 f \left\{f
		   \left(2 \mathcal{F}^5+5 \mathcal{F}^3+\mathcal{F}\right)+\tau \left( \mathcal{F}^2 + 1\right) \right\}\Bigr)\Bigr]
	\nonumber \\ & \phantom{= \times \biggl[}
		+ 8 k^2 \omega ^4 \tau  f \mathcal{F}   \left[2 \mathcal{F} \left(\mathcal{F}^2-1\right) \left(f \left(3 \mathcal{F}^2+2\right)-z f'\right)+ z f \mathcal{F}' \left(\mathcal{F}^2+1\right)
		  \right]
	\nonumber \\ & \phantom{= \times \biggl[}
		-k^4  \omega ^4 z^2 \mathcal{F}^3 \left[16 f^2 \left(\mathcal{F}^4+4 \mathcal{F}^2+1\right)-8 z f' f \left(2 \mathcal{F}^2+1\right)+z^2
		   f'^2\right]
	\nonumber \\ & \phantom{= \times \biggl[}
		-32 \omega^6 \tau  f \mathcal{F}^2 \left(\mathcal{F}^2-1\right) 
		+ 8 \omega ^6 k^2 z^2  \mathcal{F}^3 \left[4 f \left(\mathcal{F}^2+1\right)-z f'\right]
		-16 \omega ^8  z^2 \mathcal{F}^3
	\biggr\},
\end{align}
\begin{align}
	A_4 &= \frac{\mathcal{F}^2}{2 \talpha^2 F_{tz} z^5  f k \mathcal{F}^3 \left(\omega ^2-f k^2 \mathcal{F}^2\right) \left(k^2 \left(z f'-4 f\right)+4 \omega ^2\right)^2}
	\nonumber \\ & \phantom{=}
		\times \biggl\{
			k^6 f\Bigl[
				4 z f'\left(\mathcal{F}^2-1\right)  \left\{f\mathcal{F}^3(3\mathcal{F}^2-1) - \tau(\mathcal{F}^2-1) \right\}-2 z^2 f'^2 \mathcal{F}^3 \left(\mathcal{F}^4-1\right)
				 \nonumber \\ & \hspace{1.6cm} 
				  + z^6 \mathcal{F}^2 \mathcal{F}' \left(f/z^4\right)' \left(\mathcal{F}^2 z f'+f \left(12-8 \mathcal{F}^2\right)\right)
				 +8 f
				   \left(\mathcal{F}^2-1\right) \left(4 f \mathcal{F}^3+\left(\mathcal{F}^4-1\right) \tau \right)
			\Bigr]
		\nonumber \\ & \phantom{= \times \biggl\{}
			+ 2 k^8 z^7 f \mathcal{F}^3 \left(\mathcal{F}^2-1\right)^2 \left(f/z^4\right)'
			- k^4  \omega ^4 z^2 \mathcal{F} \left(\mathcal{F}^2-1\right)^2
						+ 16\omega ^6 \left(-2 \mathcal{F}^3+z \mathcal{F}'+2 \mathcal{F}\right)
		\nonumber \\ & \phantom{= \times \biggl\{}
			-k^4 \omega ^2 \Bigl[
				16 f^2 \Bigl(2 \mathcal{F} \left(-2 \mathcal{F}^4+\mathcal{F}^2+1\right)+ z \mathcal{F}' \left(-4 \mathcal{F}^4+6 \mathcal{F}^2+1\right) \Bigr)
				\nonumber \\ & \hspace{2.2cm}
				-4 f \mathcal{F} \left(\mathcal{F}^2-1\right) 
				   \Bigl( z f' \left(2 \mathcal{F}^4+7 \mathcal{F}^2-5\right)+2 \tau \mathcal{F} \left(\mathcal{F}^2+2\right)  \Bigr)- z^2 f' \mathcal{F}' \mathcal{F} \left(3 \mathcal{F}^2+1\right)
				   \nonumber \\ & \hspace{2.8cm}
				    -6 \tau \mathcal{F}+z f' \left(2 z f' \left(3 \mathcal{F}^5-4 \mathcal{F}^3+\mathcal{F}\right) -\mathcal{F}^2 z^2 f' \mathcal{F}'+4 \tau \left(\mathcal{F}^2-1\right)^2 \right)\Bigr]
		\nonumber \\ & \phantom{= \times \biggl\{}
			+2 k^6 \omega^2 z^2 \mathcal{F} \left(\mathcal{F}^2-1\right)^2  \left[4 f \left(\mathcal{F}^2+1\right)-z f'\right]
		\nonumber \\ & \phantom{= \times \biggl\{}
			+ 4 k^2 \omega ^4 \Bigl[
				2 \left(\mathcal{F}^2-1\right) \left\{\mathcal{F} \left(\left(3 \mathcal{F}^2-2\right) z f'+4 f \left(\mathcal{F}^2+2\right)+2 \mathcal{F} \tau \right)-2 \tau
				   \right\}
				\nonumber \\ & \hspace{3.1cm}
				+z \mathcal{F}' \left\{4 f \left(\mathcal{F}^2-2\right) \left(2 \mathcal{F}^2+1\right)- z f'\left(\mathcal{F}^2-1\right)\right\}
			\Bigr]
		\biggr\},
\end{align}
\begin{align}
	B_1 &= \frac{ \tau k  \tilde{\alpha}^2 z^2 F_{t z} \left(k^2 \left(z f'-2 f \left(\mathcal{F}^2+1\right)\right)+4 \omega ^2\right)}{\mathcal{F}(f
	   k^2 \mathcal{F}^2- \omega ^2)},
\end{align}
\begin{align}
	B_2 &= \frac{1}{f \mathcal{F} z \left(f k^2 \mathcal{F}^2-\omega ^2\right) \left(k^2 \left(z f'-4 f\right)+4 \omega ^2\right)}
	\nonumber \\ & \phantom{=}
	\times \biggl[
		k^4 f \left(8 f^2 \mathcal{F}^3-2 f \left(\mathcal{F}^3 z^2 f''(z)+\left(\mathcal{F}^4-1\right) \tau \right)+z f' \left(\mathcal{F}^3 z f'+\left(\mathcal{F}^2-1\right)
		   \tau \right)\right)
		+ 4  \omega ^4 \mathcal{F}  \left(2 f-z f'\right)
	\nonumber \\ & \phantom{= \times \biggl[}
		+k^2 \omega ^2 \left(-8 f^2 \left(\mathcal{F}^3+\mathcal{F}\right)-\mathcal{F} z^2 \left(f'\right)^2+2 f \left(\mathcal{F} \left(z^2 f''(z)+2 \mathcal{F} \left(\mathcal{F} z
		   f'+\tau \right)\right)-2 \tau \right)\right)
	\biggr],
\end{align}
\begin{align}
	B_3 &= \frac{\tau k \talpha^2 z F_{tz}}{f \mathcal{F}^2 \left(\omega ^2-f k^2 \mathcal{F}^2\right) \left(k^2 \left(z f'-4 f\right)+4 \omega ^2\right)}
	\nonumber \\ & \phantom{=}
	\times \biggl\{
		2 k^4 f  \left[z f' \Bigl(2 \mathcal{F}^3 z f'+ \tau \left(\mathcal{F}^2-1\right) \Bigr)-2 f \Bigl(\mathcal{F}^3 z
		   \left(z f''+f'\right)+ \tau (\mathcal{F}^4 - 1)\Bigr)\right]
	\nonumber \\ &\phantom{= \times \biggl[}
		+ 4 k^2\omega ^2   \left[f \Bigl(4 \mathcal{F}^3 z f'+\mathcal{F} z \left(z f''+f'\right)+2  \tau (\mathcal{F}^2 - 1) \Bigr)-z^2 f'^2 \mathcal{F} \right]
		-16 \omega ^4  z f' \mathcal{F}
	\biggr\},
\end{align}
\begin{align}
	B_4 &= \frac{1}{f^2 \mathcal{F} z^2 \left(f k^2 \mathcal{F}^2-\omega ^2\right) \left(k^2 \left(z f'-4 f\right)+4 \omega ^2\right)}
	\nonumber \\ & \phantom{=}
	\times \biggl\{
		k^4 f^2  \left[4 z f \left(z^3  \mathcal{F}^3 \left(f'/z^2\right)'+ \tau (\mathcal{F}^4-1)\right)-z f' \Bigl(z f' \mathcal{F}^3 +2 \tau 
		   \left(\mathcal{F}^2-1\right) \Bigr)\right]
	\nonumber \\ & \phantom{= \biggl[ \times}
		+ k^6 z^7  f^2  \mathcal{F}^3\left(f/z^4\right)'
		+ k^4 \omega ^2 z^2 f \mathcal{F} \left[z f'\left(\mathcal{F}^2+1\right) -4 f \left(2 \mathcal{F}^2+1\right)\right]
	\nonumber \\ & \phantom{= \biggl[ \times}
		+k^2 \omega ^2 f \Bigl[z f' \Bigl(z f' \mathcal{F}+2 \tau \left(\mathcal{F}^2-1\right)  \Bigr)
	\nonumber \\ & \hspace{3cm} -4 f \left(z^2 f''\left(\mathcal{F}^3+\mathcal{F}\right) -2
		   z f' \Bigl(\mathcal{F}^3+\mathcal{F}\right) + \tau \left(\mathcal{F}^2+2\right) \mathcal{F}^2  -3 \tau \Bigr)\Bigr]
	\nonumber \\ & \phantom{= \biggl[ \times}
		+ 4 f \omega ^4 \left[z^4 \mathcal{F} \left(f'/z^2\right)'+2 \tau \mathcal{F}^2  -2 \tau \right]
	\nonumber \\ & \phantom{= \biggl[ \times}
		+ k^2 \mathcal{F} \omega ^4 z^2 \left[4 f \left(\mathcal{F}^2+2\right)-z f'\right]
		-4 \omega ^6 \mathcal{F} z^2
	\biggr\}.
\end{align}
}

A key property is $B_1 \propto \tau$ and $B_3 \propto \tau$, so that in the probe limit $\tau \to 0$, $Z_1$ drops out of $Z_2$'s equation of motion. In that case we can consistently set the metric fluctuations to zero, so that $Z_2=0$ and $Z_1 = k \, \delta a_t + \omega \, \delta a_x$, and then solve $Z_1$'s equation of motion in the AdS-SCH background. In that way we can reproduce the probe calculation of ref.~\cite{Davison:2011ek}.

The gravity theory's scaling symmetry $\alpha \to \lambda \, \alpha$ and $F_{MN} \to \lambda^{-1} \, F_{MN}$ implies that $Z_1 \to \lambda^{-1} \, Z_1$ and $Z_2 \to Z_2$. The coupled equations for $Z_1$ and $Z_2$ are invariant under the scaling symmetry. To be explicit: $A_2$ and $A_4$ are each $\propto \left(\talpha^2 F_{tz}\right)^{-1}$ and hence $A_2\to \lambda^{-1}\, A_2$ and similarly for $A_4$, with $A_1$ and $A_3$ invariant, making $Z_1$'s equation of motion invariant. Correspondingly, $B_1$ and $B_3$ are each $\propto \talpha^2 F_{tz}$ and hence $B_1 \to \lambda \, B_1$ and similarly for $B_3$, with $B_2$ and $B_4$ invariant, making $Z_2$'s equation of motion invariant. As a result, all our results for QNMs, spectral functions, and sound attenuation are invariant under the gravity theory's scaling symmetry (as indeed we have checked numerically).

The values of $A_{\mu}$ and $g^{\mu}_{~\nu}$ at the boundary $z \to 0$ are sources for $J^{\mu}$ and  $T_{\mu}^{~\nu}$, respectively. Using $\lim_{z \to 0} \left(z F_{zt}\right)=0$ and $\lim_{z \to 0} \left(z f' \right)= 0$, the $z \to 0$ limit of eq.~\eqref{zdef} thus reveals that the linear combination of bulk fields $Z_1$ is dual to the linear combination of operators $k \, J^t + \omega \, J^x$, while $Z_2$ is dual to $- k^2 \, T_t^{~t} + \omega^2 \, T_x^{~x} + 2 \o k \, T_x^{~t} + (-\o^2 + k^2) T_y^{~y}$. More precisely, the expansions of $Z_1$ and $Z_2$ about the boundary $z \to 0$ are
\begin{subequations}
\begin{align}
	Z_1 &= Z^{(0)}_1 + Z^{(1)}_1 z + \mathcal{O}(z^2),
	\\
	Z_2 &= Z^{(0)}_2 - \frac{1}{2} Z^{(0)}_2(k^2-\omega^2) z^2 + Z^{(3)}_2 z^3 + \mathcal{O}(z^4),
\end{align}
\end{subequations}
where $Z_1^{(0)}$ and $Z_2^{(0)}$ are the sources for these dual operators.

The expansions of $Z_1$ and $Z_2$ about the horizon $z = z_H$ are
\begin{subequations}
\begin{align}
	Z_1 &= (z_H-z)^{-i\omega/4 \pi T} \, \zeta^\mathrm{in}_1(z)+(z_H-z)^{i\omega/4 \pi T} \, \zeta^\mathrm{out}_1(z),
	\\
	Z_2 &= (z_H-z)^{-1-i\omega/4 \pi T} \, \zeta^\mathrm{in}_2(z)+(z_H-z)^{-1+i\omega/4 \pi T} \, \zeta^\mathrm{out}_2(z),
\end{align}
\end{subequations}
where $\zeta^{\mathrm{in}}_1(z)$, $\zeta^{\mathrm{out}}_1(z)$, $\zeta^{\mathrm{in}}_2(z)$, and $\zeta^{\mathrm{out}}_2(z)$ are regular at $z = z_H$. We want to compute retarded Green's functions, which are dual to purely in-going solutions~\cite{Son:2002sd}, so we will impose $\zeta^{\mathrm{out}}_1(z_H)=0$ and $\zeta^{\mathrm{out}}_2(z_H)=0$. QNMs are in-going solutions that are furthermore normalizable, meaning they also have $Z_1^{(0)}=0$ and $Z_2^{(0)}=0$. The values of $\o$ at which such solutions exist are dual to the positions of poles in the retarded Green's functions. Crucially, $Z_1$ and $Z_2$ are coupled, hence the dual Green's functions will mix, and in particular will have poles at the same positions. However, the \textit{residues} of these poles may differ, and hence the spectral functions may differ. Indeed, in our system, as in AdS-RN~\cite{Davison:2011uk}, the spectral functions differ in important ways, as we discuss in sec.~\ref{sec:spectral_functions}.

To compute the QNMs and Green's functions numerically, we use the method of ref.~\cite{Kaminski:2009dh}. For given $\o$ and $k$ we form two linearly independent in-going solutions specified by
\beq\label{eq:independent_solutions}
	\begin{pmatrix}\zeta_1^\mathrm{in}(z_H) \\ \zeta_2^\mathrm{in}(z_H) \end{pmatrix}= \begin{pmatrix}1 \\ \pm 1\end{pmatrix},
\eeq
and then construct a matrix with columns given by these solutions,
\beq
\label{eq:Hdef}
	H_{ia}(z) \equiv \begin{pmatrix}
		Z_1^+(z) & Z_1^-(z)
		\\
		Z_2^+(z) & Z_2^-(z)
	\end{pmatrix}.
\eeq
where the index $a=\pm$ (the superscripts) corresponds to the sign in eq.~\eqref{eq:independent_solutions}. To find QNMs we compute
\beq
\label{eq:qnmmatrix}
\lim_{z \to 0}H_{ia}(z) \equiv \begin{pmatrix} Z_1^{(0)+} & Z_1^{(0)-} \\ Z_2^{(0)+} & Z_2^{(0)-}
\end{pmatrix}.
\eeq
If the determinant of the matrix in eq.~\eqref{eq:qnmmatrix} vanishes, then a normalizable linear combination of our two solutions, that is, a QNM, exists at the given $\o$ and $k$. For the Green's functions we need the on-shell action, which may be written as
\beq \label{eq:on_shell_action}
	S = \int_{\e}^{z_H} dz \int \frac{d \omega d^2 k}{(2\pi)^3} \,  C_{ij}  \partial_z Z_i(z,-\omega,-k)  \partial_z Z_j(z,\omega,k) + \ldots,
\eeq
where $\e$ is a near-boundary cutoff and $\ldots$ represent terms with at most a single $\partial_z$. These terms are both analytic, and so do not affect the poles of $G_{ij}$, and real-valued, and so do not contribute to $\rho_{ij}$. Following ref.~\cite{Davison:2011uk}, we only compute the diagonal components of the Green's functions, $G_{11}$ and $G_{22}$, which we will denote $G_J$ and $G_{tt}$, respectively, since $Z_1$ is dual to a linear combination of $J^{\mu}$ components and $Z_2$ is dual to an operator containing the energy density $T_t^{~t}$. In the main text we somewhat sloppily refer to these as the ``charge'' and ``energy'' Green's functions. The coefficients $C_{ij}$ that we need for $G_{11}$ and $G_{22}$ are
\begin{subequations}
\begin{align}
	C_{11} &= \frac{1}{16 \pi G} \frac{\tau  \tilde{\alpha}^2 \, f}{\mathcal{F} \left(\omega ^2-f \mathcal{F}^2 k^2 \right)},
	\\
	C_{12} &= - C_{21} = - \frac{1}{16 \pi G}\frac{i \tau \talpha^2 L^2 \, z f^2  k \, F_{tz}  }{\mathcal{F} \left(\omega ^2-f \mathcal{F}^2 k^2 \right) \left(k^2 \left(z f'(z)-4
	   f\right)+4 \omega ^2\right)},
	\\
	C_{22} &= \frac{1}{16 \pi G} \frac{f^3 L^2 \left[2 \mathcal{F} (f \mathcal{F}^2 k^2  - \omega^2 ) - k^2 z^4 \tau \talpha^2 F_{tz}^2 \right]}{z^2 \left(f k^2 \mathcal{F}^3-\mathcal{F} \omega^2\right)
	   \left(k^2 \left(z f'(z)-4 f\right)+4 \omega ^2\right)^2}.
\end{align}
\end{subequations}
If we define the matrix
\beq
F_{ij}(z) \equiv H_{ia}(z)H_{aj}^{-1}(\e),
\eeq
then we can write the retarded Green's functions as~\cite{Kaminski:2009dh}
\beq
\label{eq:retgreen}
G_{ij} = - \frac{1}{16\pi G}\lim_{\e\to0} \left[F_{ik}^\dag \left(C_{kl}+C^*_{lk} \right) F_{lj}' + \ldots\right],
\eeq
where $F' \equiv \partial_z F$ and all quantities in the brackets are evaluated at $z = \e$. The $\ldots$ include terms descending from the $\ldots$ in eq.~\eqref{eq:on_shell_action}, as well as the boundary terms, including the holographic renormalization counterterms. We may safely ignore these terms, for the reasons mentioned above. Using the matrices $H_{ia}(z)$ and $C_{ij}$ we can also compute the matrix of pole residues in eq.~\eqref{eq:mero}~\cite{Kaminski:2009dh},
\beq
\label{eq:residue}
{\cal R}_{ij}^{(n)}(k) = - \left . \frac{\textrm{det}\left[H_{ia}(\epsilon)\right]}{\partial_{\omega}\, \textrm{det}\left[H_{ia}(\epsilon)\right]} \left(C_{ik} + C_{ki}^*\right) \, H'_{ka}(\epsilon) H_{aj}^{-1}(\epsilon) \right |_{\omega_*^{(n)}(k)},
\eeq
where $\omega_*^{(n)}(k)$ is the position of the $n^{\textrm{th}}$ pole, computed numerically from the zeroes of the determinant of the matrix in eq.~\eqref{eq:qnmmatrix}, as described above.

\bibliographystyle{JHEP}
\bibliography{dbisound}

\end{document}